\documentclass[11pt,a4paper]{scrartcl}
\pdfoutput=1

\usepackage{CLICdp}

\usepackage{CLICdp_definitions}

\title{All-hadronic HZ production at high energy at 3 TeV CLIC}

\clicdpnote{2019}{005}  

\date{\formatdate{06}{11}{2019}}

\addauthor{Emilia~Leogrande}{\institute{1}}
\addauthor{Philipp~Roloff}{\institute{1}}
\addauthor{Ulrike~Schnoor}{\institute{1}}
\addauthor{Matthias~Weber}{\institute{1}\hcomma\institute{2}}

\addinstitute{1}{CERN, Geneva, Switzerland}
\addinstitute{2}{University of Glasgow, United Kingdom}

\onbehalfof{the CLICdp Collaboration} 

\abstract{In this note the \zhsm production in the all-hadronic final state in \epem collisions at the Compact Linear Collider is studied at the \SI{3}{TeV} stage. At high energies, the events have an experimental signature of back-to-back approximately mono-energetic large jets. Each of these jets contains two sub-jets and substructure compatible with two original objects. The study is based on full simulation including the detector response, as well as the presence of beam-induced background from \gghadrons. Results on the measurement of the total \zhsm cross section are given, and the potential to measure angular asymmetry observables is discussed.}

\titlecomment{This work was carried out in the framework of the CLICdp Collaboration} 

\graphicspath{ {./logos/}{./figures/} }

\begin{document}

\titlepage

\section{Introduction}
\label{sec:intro}

The Compact Linear Collider (CLIC) is a mature option for a future electron-positron collider~\cite{Aicheler:2019dhf}. CLIC will be built in energy stages~\cite{CLIC:2016zwp}, achieving nominal centre-of-mass energies between \SI{380}{GeV} and \SI{3}{TeV}. A comprehensive discussion of the CLIC Higgs physics programme has been published in~\cite{Abramowicz:2016zbo}. At the lowest energy stage of \SI{380}{GeV} Higgsstrahlung ($\Pep\Pem\rightarrow\zhsm$) is the dominant \PH boson production mechanism with a cross section of over \SI{100}{fb}. The total cross section of the Higgsstrahlung can be measured in a model-independent way, i.e. without any additional assumption about the decays of the Higgs boson, using the quantities of the \PZ boson alone. This recoil method relies on precise knowledge of the \roots of the collision and is only possible at lepton colliders. 

At high energy, the cross section for the Higgsstrahlung process decreases as $1/s$, 
where \roots is the centre-of-mass energy, and \PW\PW-fusion ($\Pep\Pem\rightarrow\PH\PGne\PAGne$) dominates the \PH production. However, contributions from some Standard Model Effective Field Theory (EFT) operators grow with energy. Measurements of the Higgsstrahlung production rate at the higher CLIC energy stages enhance the sensitivity to New Physics as illustrated in~\cite{Ellis:2017kfi}. Recent EFT analyses of Higgs and electroweak processes including projections for Higgsstrahlung at the high energy CLIC stages are described in~\cite{deBlas:2019rxi, deBlas:2019wgy}. We present a study of the process $\epem\to\PZ\PH\rightarrow$ with $\PZ\rightarrow\qqbar$ and $\PH\rightarrow\bb$ at the \SI{3}{TeV} CLIC stage based on full detector simulation. This fully hadronic final state is expected to provide the best statistical precision.

Additional sensitivity on EFT operators is provided by angular observables~\cite{Beneke:2014sba}. Projections from a generator-level study using leptonic $\PZ$ boson decays are shown in chapter 2.3 of~\cite{deBlas:2018mhx}. The potential of events with hadronic $\PZ$ decays based on full detector simulation is described here.

Due to the high energy, two large jets are reconstructed and $\PZ\PH$ signal events are identified using substructure information. B-tagging in boosted $\PH$ boson decays at CLIC is investigated for the first time for the analysis presented here. Different methods for the charge reconstruction in subjets are compared

The study uses the updated luminosity numbers and the baseline scenario for luminosity sharing of $\mathrm{L}_{-80\%}=\SI{4}{\abinv}$ and $\mathrm{L}_{+80\%}=\SI{1}{\abinv}$ for -80\% and +80\% polarisation of the electron beam~\cite{Robson:2018zje}. 

This paper is organized as follows: Section~\ref{sec:DetectorSoftware} describes the CLIC detector model, and the software packages in use, while section~\ref{sec:HZSignal} gives an overview of the overall \zhsm topology and signal reconstruction. Section \ref{sec:BTagging} contains considerations for B-tagging in boosted jet topologies. Section~\ref{sec:MCSimulation} describes the Monte Carlo Simulation. The details about the background rejection and signal selection are discussed in section~\ref{sec:variables}. Section~\ref{sec:DifferentialDistributions} presents the cross section and results on the angular distributions, followed by a discussion of systematic uncertainties in section~\ref{sec:systematics}. Section~\ref{sec:summary} concludes with a summary.

\section{Detector model and software chain}
\label{sec:DetectorSoftware}

This study is based on the new detector model CLICdet, which was developed in several optimisation studies~\cite{CLICdet_note_2017,Arominski:2018uuz}. The CLICdet model is designed to cope with experimental conditions at \SI{3}{TeV} CLIC. 
The central feature of CLICdet is a superconducting solenoid with an internal diameter of \SI{7}{m}, providing a magnetic field of \SI{4}{T} in the centre of the detector. Silicon pixel and strip trackers, the electromagnetic (ECAL) and hadronic calorimeters (HCAL) are embedded within the solenoid. Each subdetector is divided into a barrel and two endcap sections. ECAL is a highly granular array of 40 layers of silicon sensors and tungsten plates. HCAL is built from 60 layers of plastic scintillator tiles, read out by silicon photomultipliers, and steel absorber plates. The muon system surrounding the solenoid consists in the endcap of 6, in the barrel of 7 layers of resistive plate chambers interleaved with yoke steel plates. Two smaller electromagnetic calorimeters, LumiCal and BeamCal, cover the very forward region of CLICdet on either side of the interaction point. 

CLICdet uses a right-handed coordinate system, with the origin at the nominal point of interaction. The $z$-axis is along the beam direction, with the electron pointing in the positive direction. The $y$-axis points upwards along the vertical direction. The crossing angle between the electron and positron beams is \SI{20}{mrad}, with electron momentum $p_{x}^{-}>0$ and positron momentum $p_{x}^{+}>0$. The polar angle $\theta$ is measured from the positive $z$-axis.

A new software chain for simulation and reconstruction has been introduced, using the DD4hep detector description toolkit~\cite{Frank:2015ivo,Sailer:2017rnh}. The detector response is simulated using the \geant 10.02.p02 toolkit~\cite{Agostinelli:2002hh}. 
Beam-induced backgrounds from \gghadrons are simulated using the \guineapig program~\cite{Schulte:382453}  and CLIC beam parameters at \SI{3}{TeV}. These background collisions are overlaid on the hard physics event. Tracks are reconstructed using the conformal tracking pattern recognition technique~\cite{Brondolin:2019awm}. Software compensation is applied to hits in HCAL to improve the energy measurement, using local energy density information~\cite{Tran:2017tgrSoftwareCompensation}. Pandora particle flow algorithms~\cite{Marshall:2015rfa,Marshall:2012ry} combine information from tracks, calorimeter clusters and muon hits for particle identification and reconstruction. Jet clustering, jet substructure variables, and the jet resolution parameters ($y_{23}$, $y_{34}$) are calculated using the FastJet 3.3.2~\cite{Cacciari:2011ma} library. The performance of track reconstruction, particle identification, and flavour tagging at CLICdet has been studied with the new software chain in~\cite{Arominski:2018uuz}. Relative jet energy resolution values at \SI{3}{TeV} CLIC are typically around 6--8\% for jet energies around \SI{50}{GeV}, decreasing to 4.5--6\% for jet energies larger than \SI{100}{GeV}, and about 3-4\% for \SI{1}{TeV} jets. In this paper, the relevant jet energies are above \SI{1}{TeV}.

\section{HZ event and signal reconstruction}
\label{sec:HZSignal}

At very high energy the quarks from the Z boson are produced very close to each other, as are the decay products of the H boson. Thus all-hadronic \zhsm events are characterised by two back-to-back ``fat'' jets, each containing two very close-by subjets. Jets are reconstructed with the VLC algorithm~\cite{Boronat:2016tgd} as implemented in the FastJet library~\cite{Cacciari:2011ma} with a radius $R=0.7$ and $\gamma=\beta=1$ in exclusive mode to force the event into two jets.  The VLC algorithm combines a Durham like inter-particle distance  $d_{ij}=2\min(E_{i}^{2\beta},E_{j}^{2\beta})(1-\cos\theta_{ij})/R^{2}$ based on energy and polar angle with a beam distance $d_{i\mathrm{B}}=E_{i}^{2\beta}\sin^{2\gamma}\theta_{i\mathrm{B}}$. The algorithm applies a sequential recombination procedure, providing a robust performance at \epem colliders with non-negligible background. The jet algorithm clusters particle flow objects, identified by the Pandora particle flow algorithms~\cite{Marshall:2015rfa,Marshall:2012ry}. Prior to jet clustering \pT and tight timing cuts are applied to the particle flow objects, to reject hadrons originating from \gghadrons. The timing requirements and the impact of timing cuts at CLIC are described in detail in the CDR~\cite[Section 2.5]{cdrvol2}. The radius of $R=0.7$ is sufficient to catch the energy of both partons in one jet. At energies of around \SI{1}{TeV} and larger, the relative jet energy resolution is between 3\% to 4\% for almost all polar angles. The dominating Higgs boson decay mode in the Standard Model is $\PH\rightarrow b\bar{b}$ with around 58.4\%, at a mass of $m_{\PH}=\SI{125}{GeV}$. Thus the event selection is based on finding events with two high-energy back-to-back jets, each containing two subjets. Each one of the jets are required to have jet masses compatible with \PZ or \PH. Concentrating on the most dominant decay channel and in order to suppress backgrounds, the jet compatible with \PH needs to be B-tagged.

\begin{figure}[htbp!]
\centering
\includegraphics[width=0.6\textwidth]{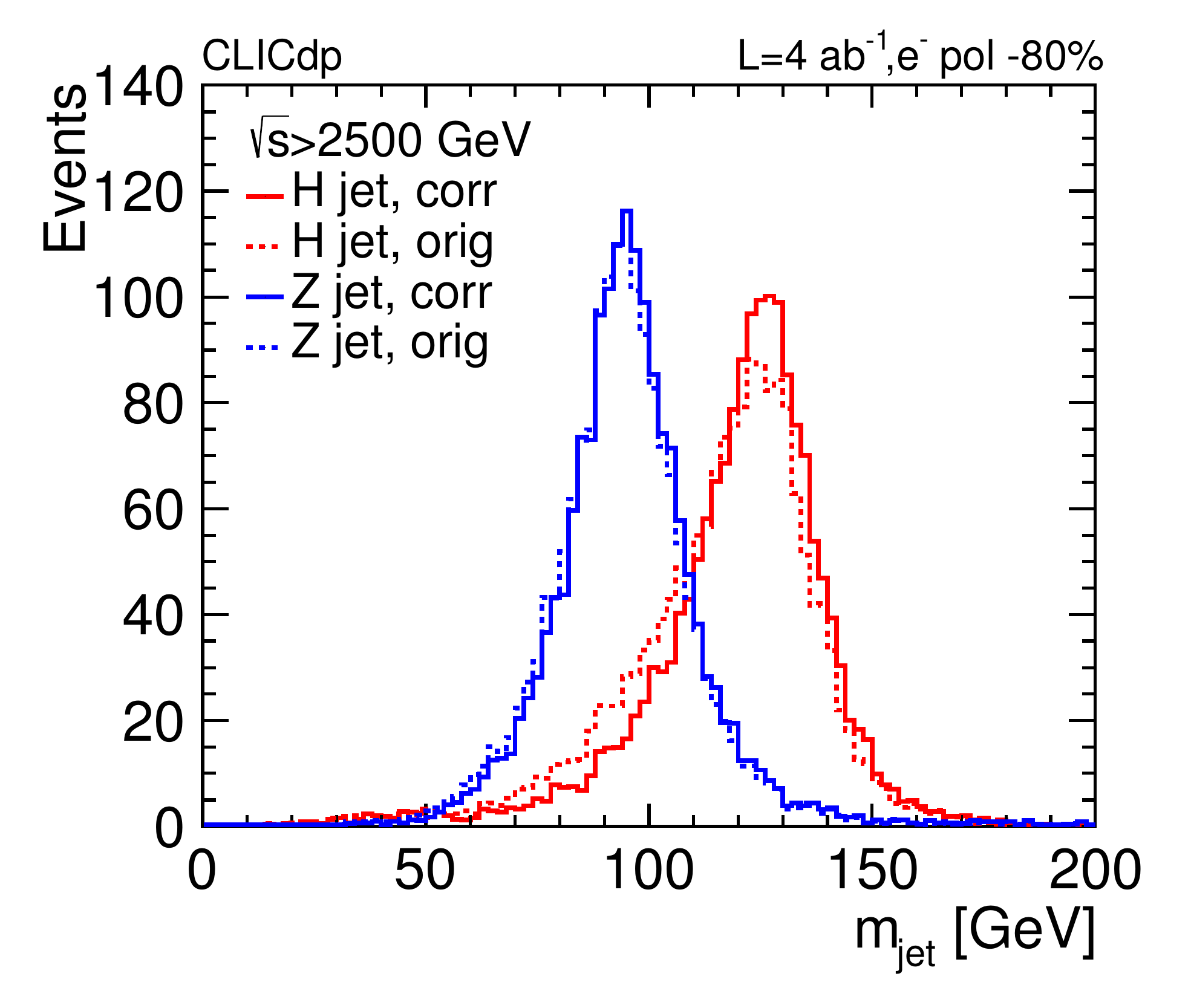}
\caption{The mass distributions of jets, matched to the \PH and \PZ parton direction, for boosted events with $\PH\rightarrow\bb$ and reconstructed $\roots>\SI{2500}{GeV}$ after correcting the jets. The solid lines are the distributions after correcting the jet with the missing transverse momentum vector, the dashed lines represent the reconstructed masses before the correction is applied.}
\label{Fig:jet_mass_H_and_Z_matched}
\end{figure}

\begin{figure}[htbp!]
\centering
\begin{minipage}[l]{0.49\textwidth}
\includegraphics[width=1.0\textwidth]{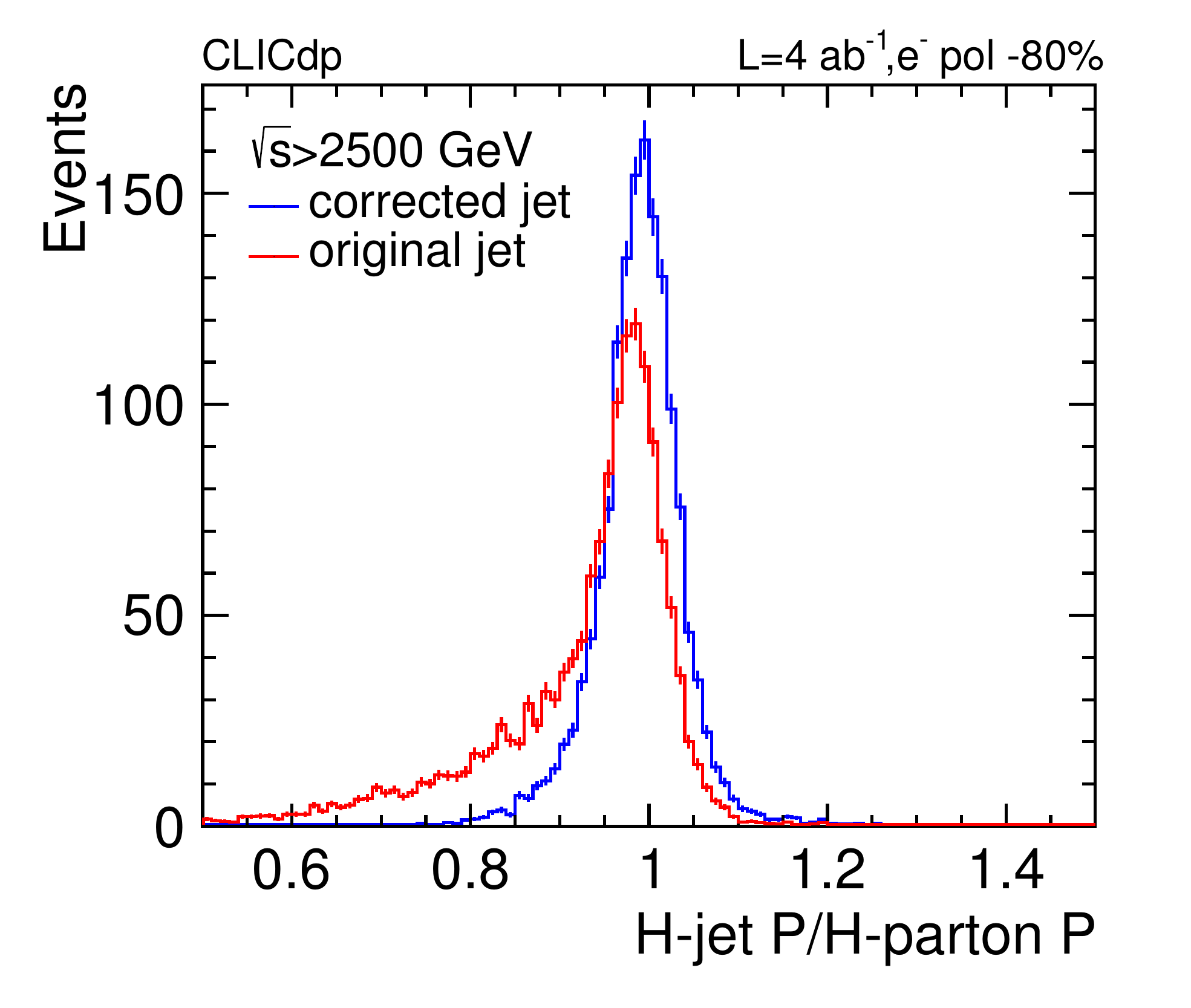}
\end{minipage}
\begin{minipage}[r]{0.49\textwidth}
\includegraphics[width=1.0\textwidth]{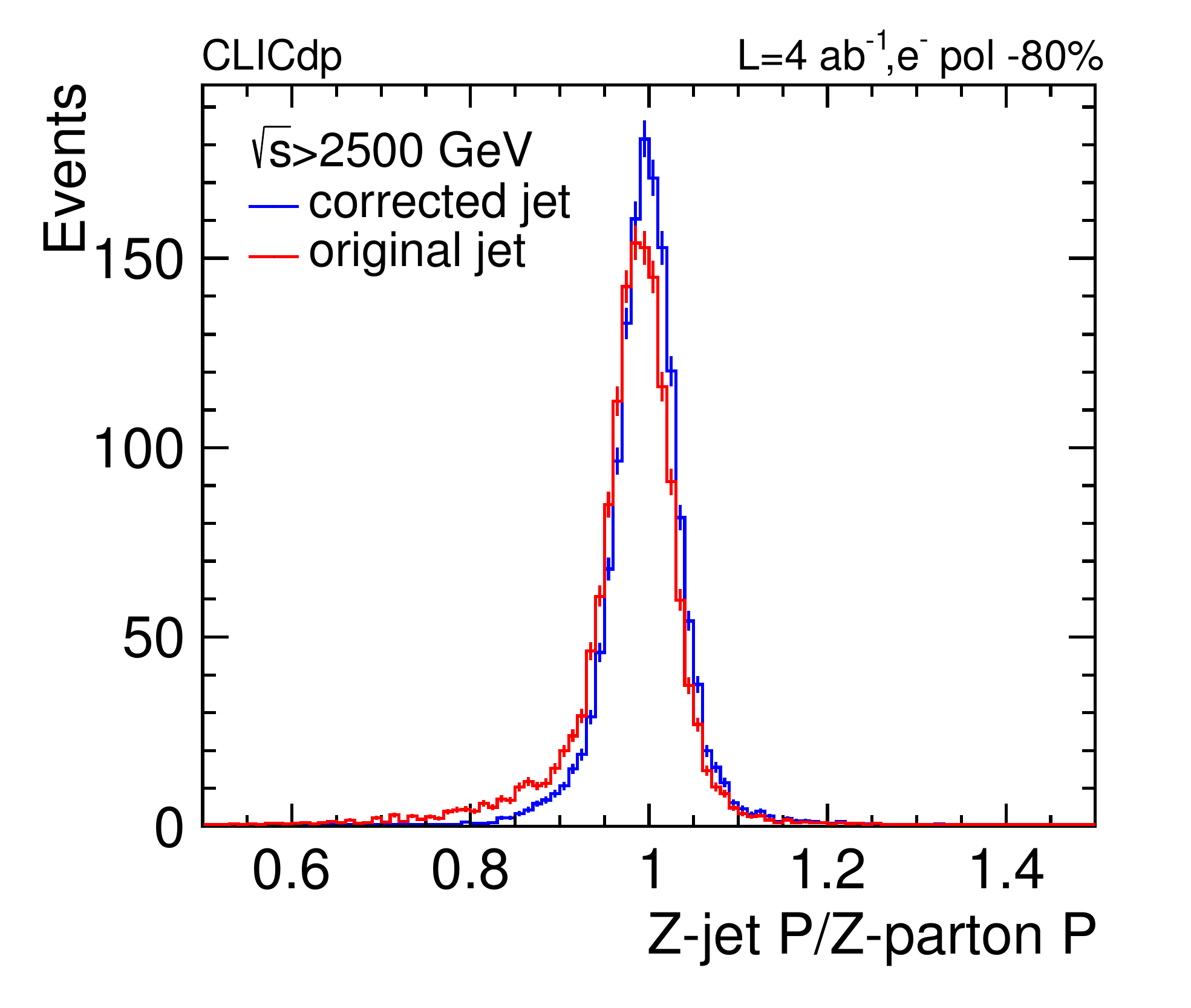}
\end{minipage}
\caption{Impact of the missing transverse momentum correction procedure on the jet momentum for jets matched to the partonic H (left) and Z (right) bosons for boosted events with $\PH\rightarrow\bb$ and reconstructed $\roots>\SI{2500}{GeV}$ after correcting the jets.}
\label{Fig:H_P_METProj}
\end{figure}

\begin{figure}[htbp!]
\centering
\begin{minipage}[l]{0.49\textwidth}
\includegraphics[width=1.0\textwidth]{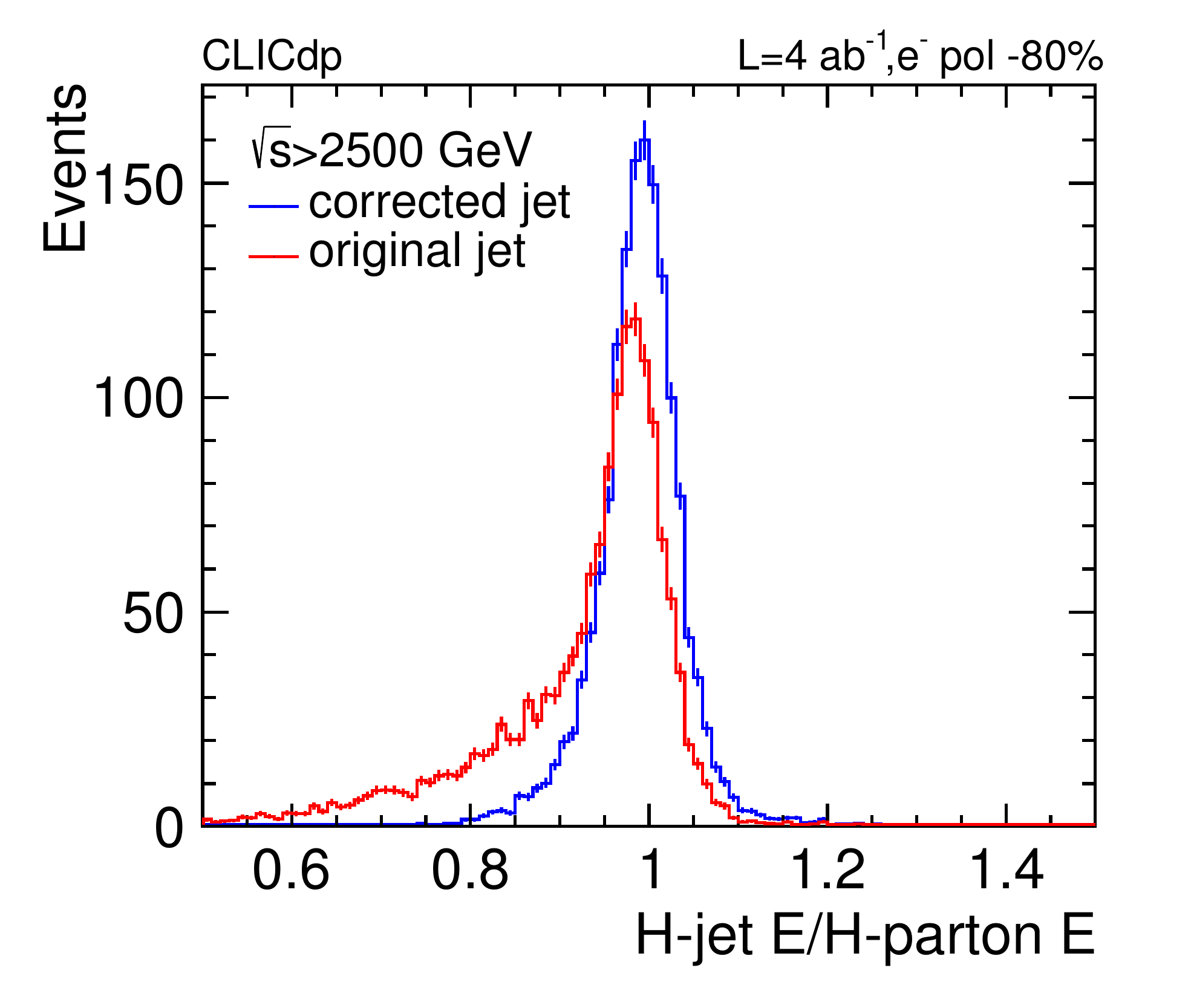}
\end{minipage}
\begin{minipage}[r]{0.49\textwidth}
\includegraphics[width=1.0\textwidth]{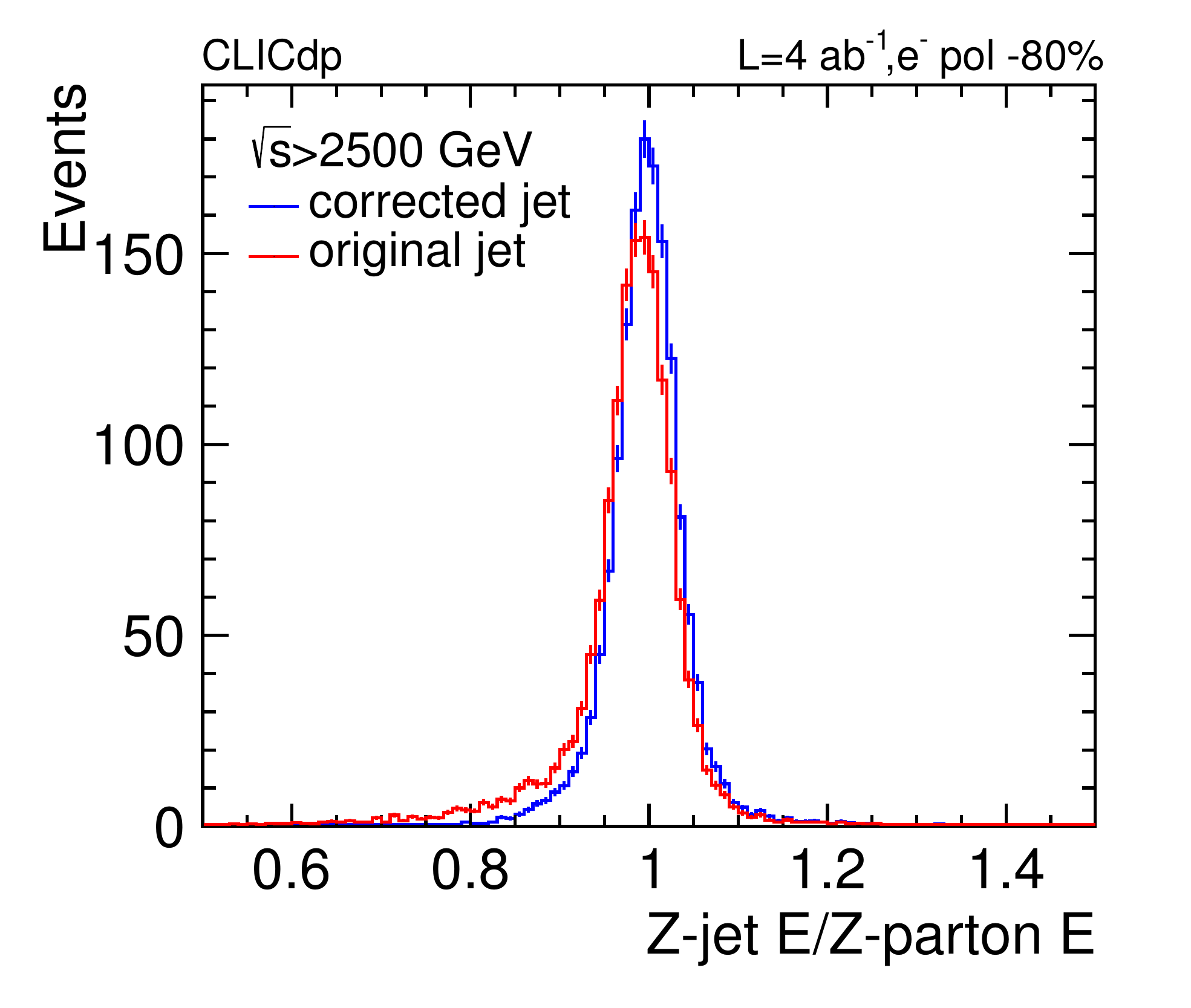}
\end{minipage}
\caption{Impact of the missing transverse momentum correction procedure on the jet energy for jets matched to the partonic H (left) and Z (right) bosons for boosted events with $\PH\rightarrow\bb$ and reconstructed $\roots>\SI{2500}{GeV}$ after correcting the jets.}
\label{Fig:H_E_METProj}
\end{figure}

Reconstructed jets are spatially matched to the Z and H parton directions to evaluate their mass resolution. As Figure~\ref{Fig:jet_mass_H_and_Z_matched} shows, both jet mass distributions peak close to the Z and H masses. While the Z jet mass distribution is symmetric, for the H jet the mass distribution is asymmetric and significantly shifted to lower values. The H jet decays predominantly into b-quarks, and a sizeable fraction of B mesons and baryon decays involves neutrinos, which escape detection. Since genuine missing energy is not present in the hard process of all-hadronic decays of \zhsm, this bias can be partially recovered, using the reconstructed missing transverse momentum. The missing transverse momentum vector is projected onto the transverse momentum vector of the jet, which is in the same hemisphere as the missing transverse momentum $\vec{p}_{\text{T}}^{miss}$. The projected missing transverse momentum $\vec{p}_{\text{T}}^{miss,\mathrm{proj}}$ is added on top of the jet transverse momentum, thus $\vec{\pT}'=\vec{\pT}+\vec{p}_{\text{T}}^{miss,\mathrm{proj}}=f\cdot\vec{\pT}$ with $f>1$. The jet is very boosted and most particles are aligned with the jet axis. Thus the $z$-momentum of the jet is scaled with the same scale-factor and $\vec{p}'=f\cdot{p}$. Since neutrino masses are negligible to the scale of the neutrino momentum, the energy is modified to $E'=E+(f-1)\cdot p$, where $p$ is the original total momentum of the jet. This correction procedure improves the mass resolution of the reconstructed jet matched to the H boson significantly: the distribution is now symmetric, and the mean of the distribution moves closer to the H mass as shown in Fig.~\ref{Fig:jet_mass_H_and_Z_matched}. Besides improving the mass reconstruction, the correction improves both the momentum (see Fig.~\ref{Fig:H_P_METProj}) and the energy of the reconstructed jets (see Fig.~\ref{Fig:H_E_METProj}) separately as well. After applying the missing transverse momentum correction, the leading reconstructed jet in terms of jet mass, referred to as $jet1$, is assigned as \PH and the second leading jet is assigned as \PZ, referred to as $jet2$ in the following. This procedure leads to the correct assignment of jets to the bosons in 85\% of events for high-\roots events. After a final signal selection, the selection by mass is accurate to about 98\%.

\begin{figure}[htbp!]
\centering
\includegraphics[width=0.6\textwidth]{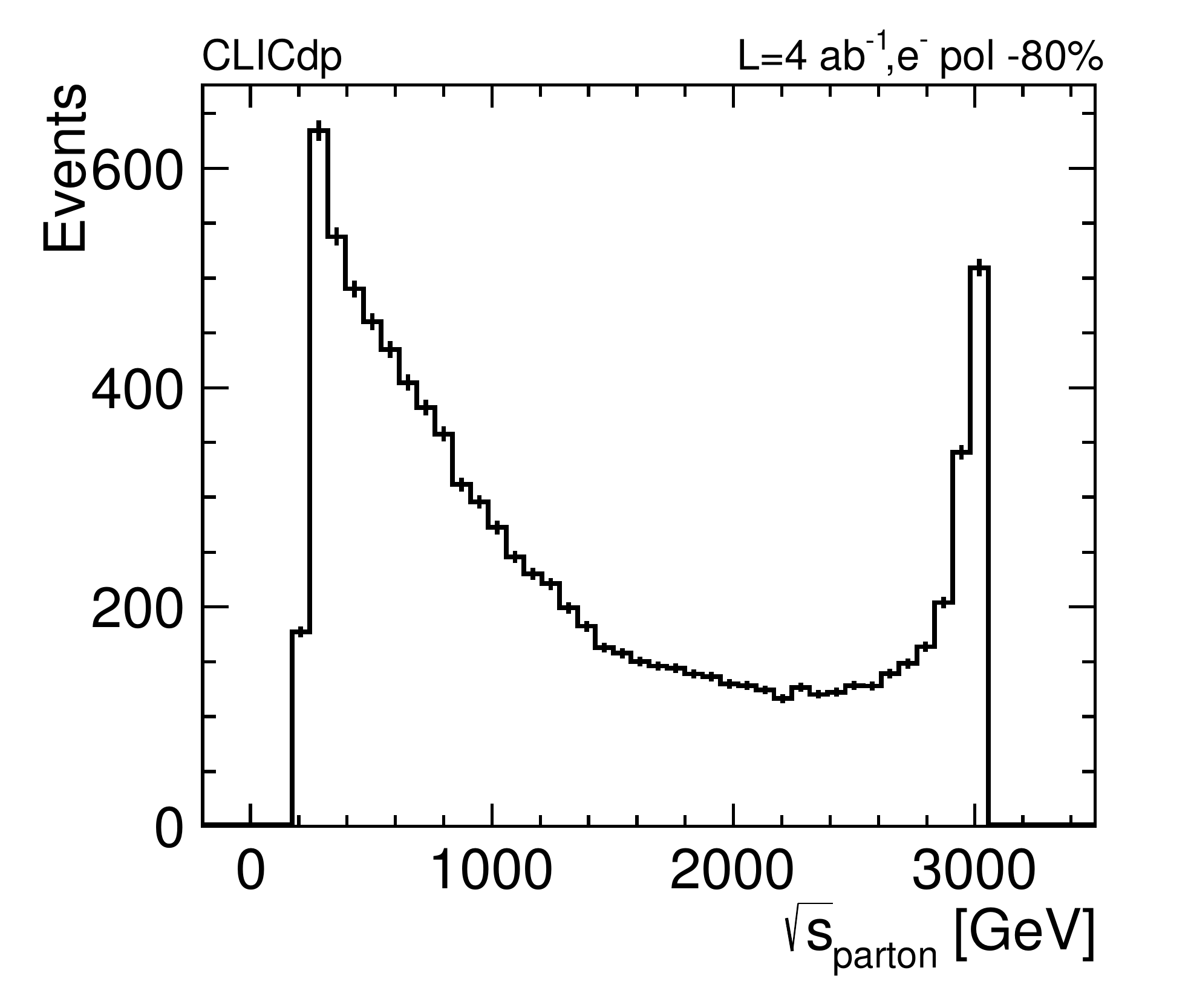}
\caption{The parton level centre-of-mass energy \roots.}
\label{Fig:sqrtS_part_3TeV}
\end{figure}

This study focuses on high energy \roots events. The \roots spectrum of all \zhsm events on parton level is shown in Fig.~\ref{Fig:sqrtS_part_3TeV}. It combines the effects of the luminosity spectrum, which peaks at \SI{3}{TeV} with the steeply falling cross section for \zhsm as function of \roots. In this note the interest is on the high-energy part of the \roots spectrum.

\begin{figure}[htbp!]
\centering
\begin{minipage}[l]{0.49\textwidth}
\includegraphics[width=1.0\textwidth]{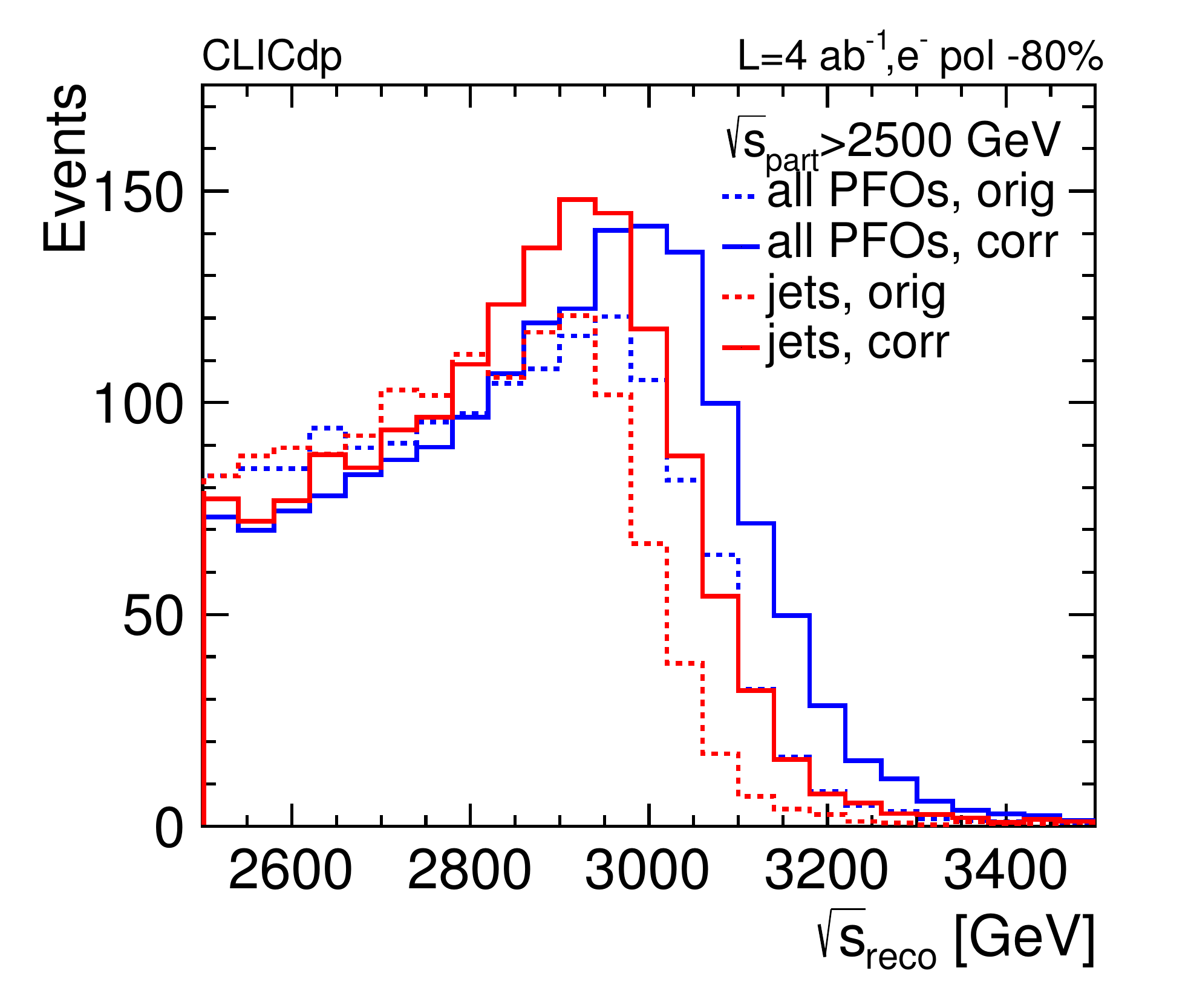}
\end{minipage}
\begin{minipage}[r]{0.49\textwidth}
\includegraphics[width=1.0\textwidth]{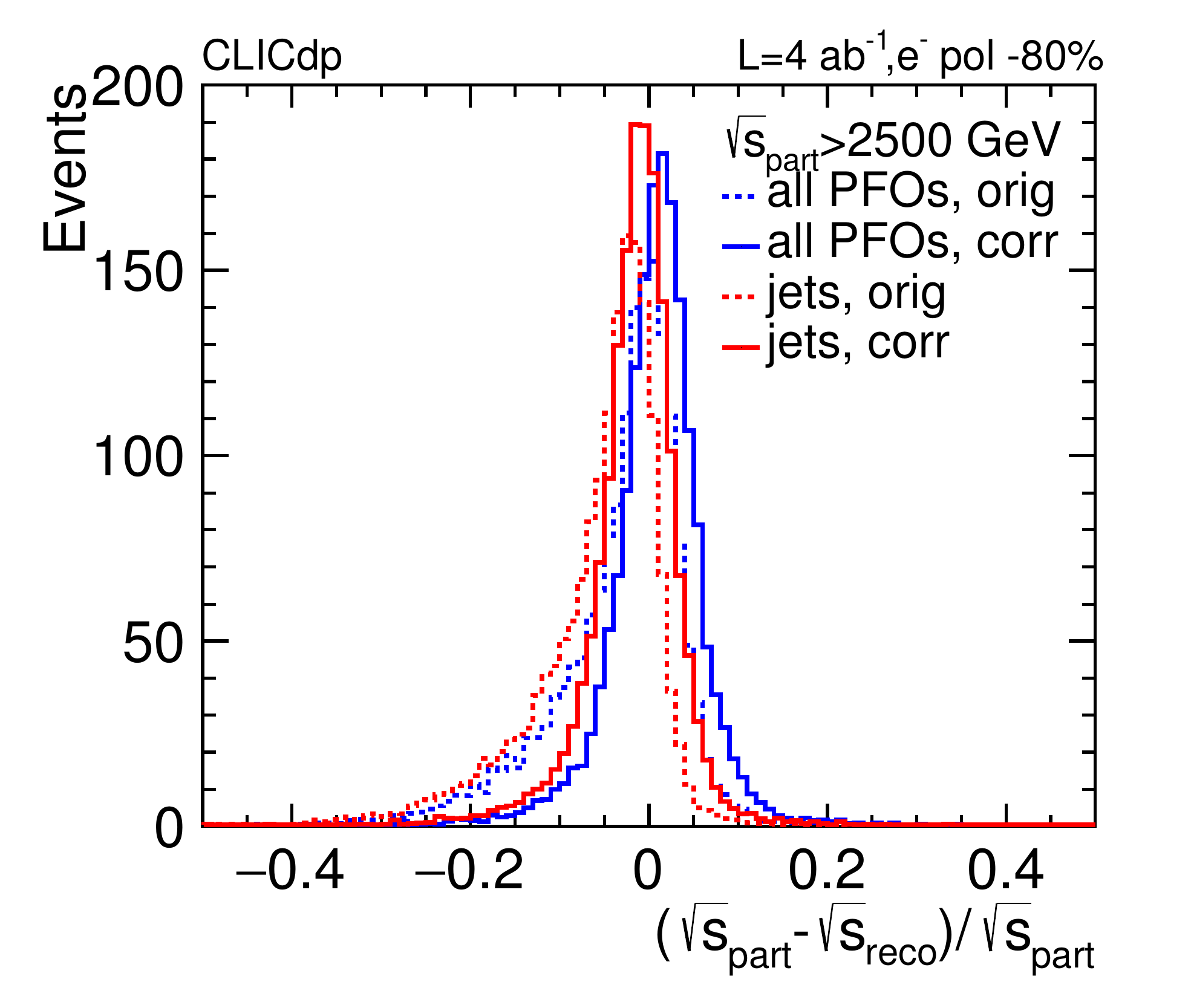}
\end{minipage}
\caption{Comparison of four methods to reconstruct the centre-of-mass energy \roots: the reconstructed distributions at high-energy are shown on the left, the relative ratio with respect to the partonic \roots are shown on the right. Events with $\PH\rightarrow\bb$ and partonic $\roots_{\mathrm{part}}>\SI{2500}{GeV}$ are considered.}
\label{Fig:sqrtS_methods}
\end{figure}

 On reconstructed level four different methods are investigated to determine \roots. The first method uses the four-momentum vector sum of all reconstructed particle flow objects after subtracting isolated photons under the assumption that these originate from beam radiation. The second method corrects this sum by the missing transverse momentum correction vector. The third method uses the four-momentum vector sum of the reconstructed jets to derive \roots. The fourth method uses the four-momentum vector sum of the reconstructed jets after applying the transverse momentum projection correction. The performance of the four methods is shown in Fig.~\ref{Fig:sqrtS_methods} comparing the ratio of the reconstructed \roots and the real partonic \roots for $\roots>\SI{2500}{GeV}$. The fourth method -- using jets after the transverse momentum projection correction -- performs best, the mean is closest to zero and the width of the distribution is the smallest. Thus this method is used in the following. The first and second method suffer from the impact of beam-induced background events from \gghadrons. These hadrons tend to be forward and increase the recorded event energy by order of \SI{100}{GeV}. The jet properties are only mildly affected by these hadrons~\cite{Arominski:2018uuz}. Due to the large boost at high energies, the unclustered energy outside of the jet cone is negligible.

\begin{figure}[htbp!]
\centering
\begin{minipage}[l]{0.49\textwidth}
\includegraphics[width=1.0\textwidth]{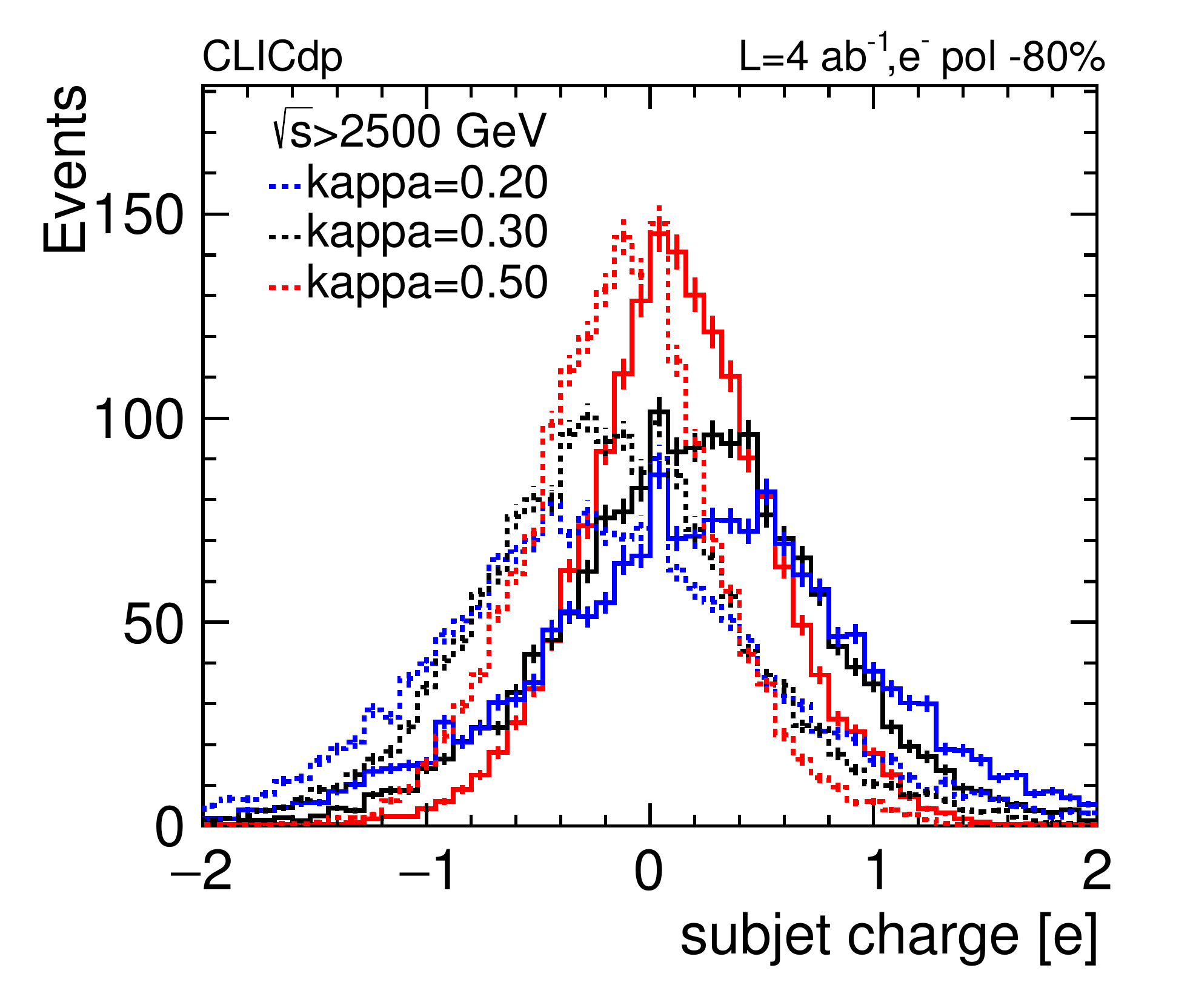}
\end{minipage}
\begin{minipage}[r]{0.49\textwidth}
\includegraphics[width=1.0\textwidth]{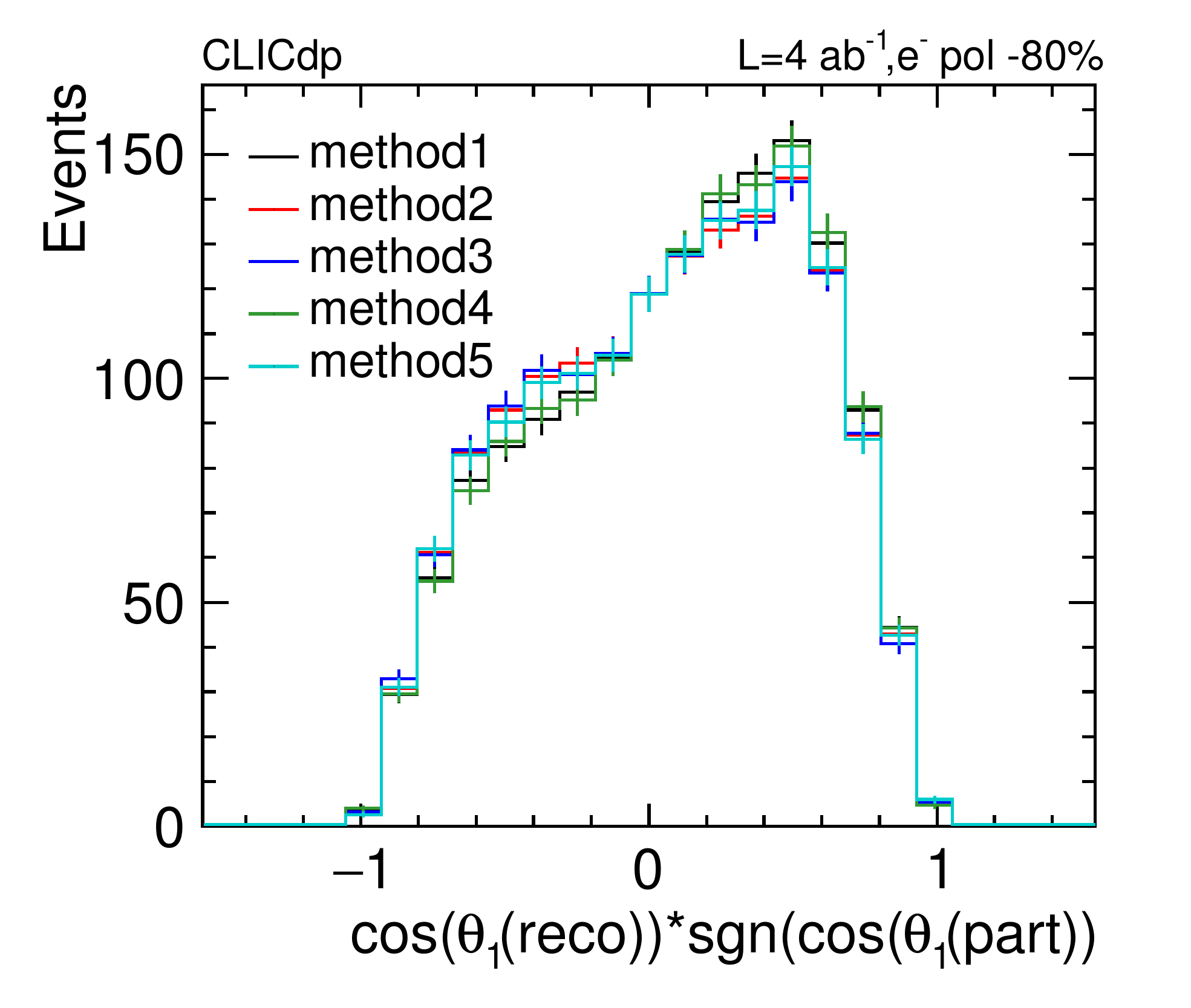}
\end{minipage}
\caption{Impact of the weighting parameter $\kappa$ on the subjet charge distribution (left) for matched positively and negatively quarks from Z decays in \zhsm signal events. On the right the distribution of reconstructed $\cos\theta_{1}$, multiplied by the $\mathrm{sgn}(\cos\theta)$ of the positively charged quark from the Z-decay using five different selection methods. Events with the correctly reconstructed hemisphere will populate positive values.}
\label{Fig:jetChargeSelections}
\end{figure}

The final goal is the measurement of angular asymmetries, which have been found to be particularly sensitive to subtle effects which can point towards an extension of the SM. The asymmetries ${A}_{c\theta_{1},c\theta_{2}}$, ${A}_{\theta_{1}}$, ${A}^{(1)}_{\phi}$, ${A}^{(2)}_{\phi}$, ${A}^{(3)}_{\phi}$, and ${A}^{(4)}_{\phi}$ are defined in terms of three angles~\cite{Craig:2015wwr}. The angle $\theta_{1}$ is defined as the angle between the positively charged fermion $f^{+}$ of the \PZ boson decay in the \PZ centre-of-mass-system and the flight direction of the \PZ boson. The angle $\theta_{2}$ is defined as the angle between the \PH boson and the incoming \Pep direction in the \zhsm centre-of-mass-system. The third angle $\phi$ is defined as the angle between the \PH-\Pep and $f^{+}$-\PZ planes. Previous studies concentrated on \zhsm events with leptonic Z-decays $\PZ\rightarrow\Plp\Plm$~\cite{Craig:2015wwr,Durieux:2017rsg} (with $\Pl=\Pe,\PGm$) where backgrounds are small, and the positively charged lepton can be identified with high accuracy. In this note the possibility to measure \zhsm in an all-hadronic signature is investigated. This branching ratio is about a factor of ten larger than \PZ decays into muons or electrons.  
Following the strategy for \zhsm with $\PZ\rightarrow\Plp\Plm$ for events with $\PZ\rightarrow\qqbar$ the positively charged quark of the \PZ boson decay has to be identified. The two sub-jets of the second jet are identified using FastJet in the exclusive subjet clustering mode. This effectively undoes the last clustering step of the VLC jet clustering of the second jet. Both the direction and energy sharing between both sub-jets replicate the energy ratios and the direction of the quarks on parton level satisfactory. The sub-jet charge $Q_{sj}$ is used to identify the sub-jet originating from the positively charged quark. The sub-jet charge is defined by a weighted charge sum of sub-jet constituents:
\begin{equation}
Q_{sj}^{(\kappa)} = \sum_{i\in sj}Q_{i}(E^{i}/E_{\mathrm{sj}})^{\kappa},
\end{equation}
where $Q_{i}$ and $E_{i}$ are the charge and energy of subjet constituent $i$, and the freely chosen parameter $\kappa>0$ is an exponent.
A slightly modified definition based on a weighting by transverse momentum \pT is given by
\begin{equation}
Q_{sj}^{(\kappa)} = \frac{1}{(\pT^{\mathrm{subjet}})^{\kappa}}\sum_{i\in sj}Q_{i}(\pT^{i})^{\kappa}.
\end{equation}
This definition is widely used by LHC experiments~\cite{Sirunyan:2017tyr,Aad:2015cua}. Both definitions lead to very similar results in identifying the quark charge of the sub-jets from $jet2$. The first definition using energy based weighting has been used as default. Figure~\ref{Fig:jetChargeSelections} (left) shows the subjet charge distributions for different values of $\kappa$, matching the subjets to the positively and negatively charged quark. For lower $\kappa$ values the peaks of the two distributions move apart, at the cost of larger tails. For values of $\kappa$ larger than 0.5 the discriminating power diminishes significantly, while for $\kappa$ values below 0.20 the overlaps between both distributions are above 40\%. For $\kappa=0.3$ the overlap of the $Q_{sj}$ distributions is the smallest with 36.5\% for both definitions. Thus $\kappa=0.3$ is used as default in the following when calculating the jet charges. Five different methods have been evaluated in order to select the positively charged quark:
\begin{enumerate}
\item Consider the sub-jet with the largest absolute jet charge value.
\item Consider the sub-jet with the largest energy.
\item Give preference to the sub-jet with the largest multiplicity of charged particles (pions, muons, electrons). If both sub-jets have the same multiplicity, then choose the sub-jet with the largest absolute jet charge value.
\item Consider the sub-jet with the largest charged energy fraction, defined as the energy sum carried by the charged particles divided by the total sub-jet energy.
\item Consider the sub-jet with the largest energy carried by charged particles.
\end{enumerate}
 If $Q_{sj}>0$, the sub-jet is considered as positively charged quark, otherwise the other sub-jet is chosen. Figure~\ref{Fig:jetChargeSelections} (right) shows the angle $\theta_{1}$ between the chosen subjet and the Z-boson multiplied by the sign of the $\cos\theta$ of the positively charged quark. In about 57\% of events the correct hemisphere is identified for all five methods, with the first method providing the best correct assignment at a value slightly above 60\%; it is subsequently used as default.

\begin{figure}[htbp!]
\centering
\begin{minipage}[l]{0.32\textwidth}
\includegraphics[width=1.0\textwidth]{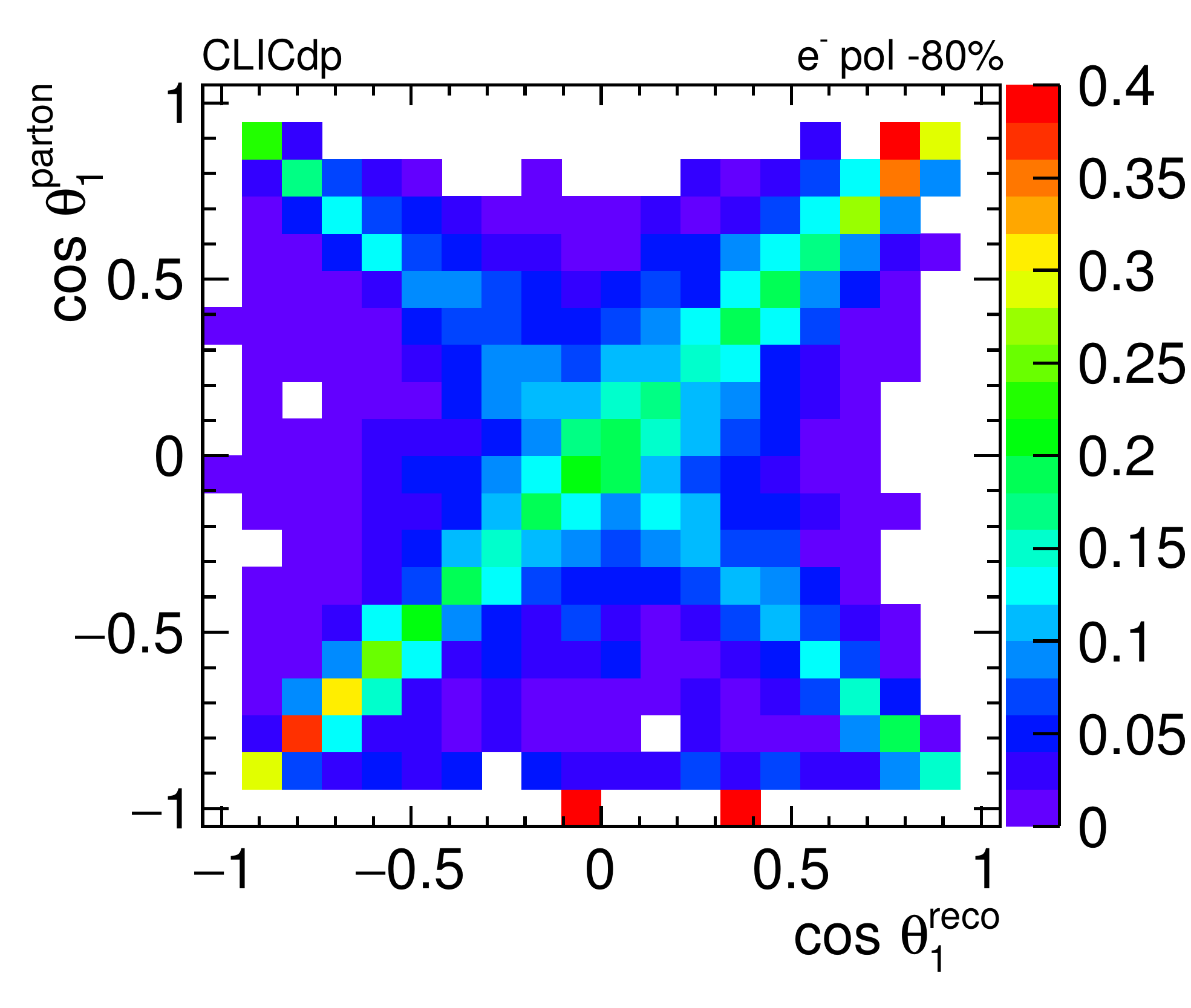}
\end{minipage}
\begin{minipage}[c]{0.32\textwidth}
\includegraphics[width=1.0\textwidth]{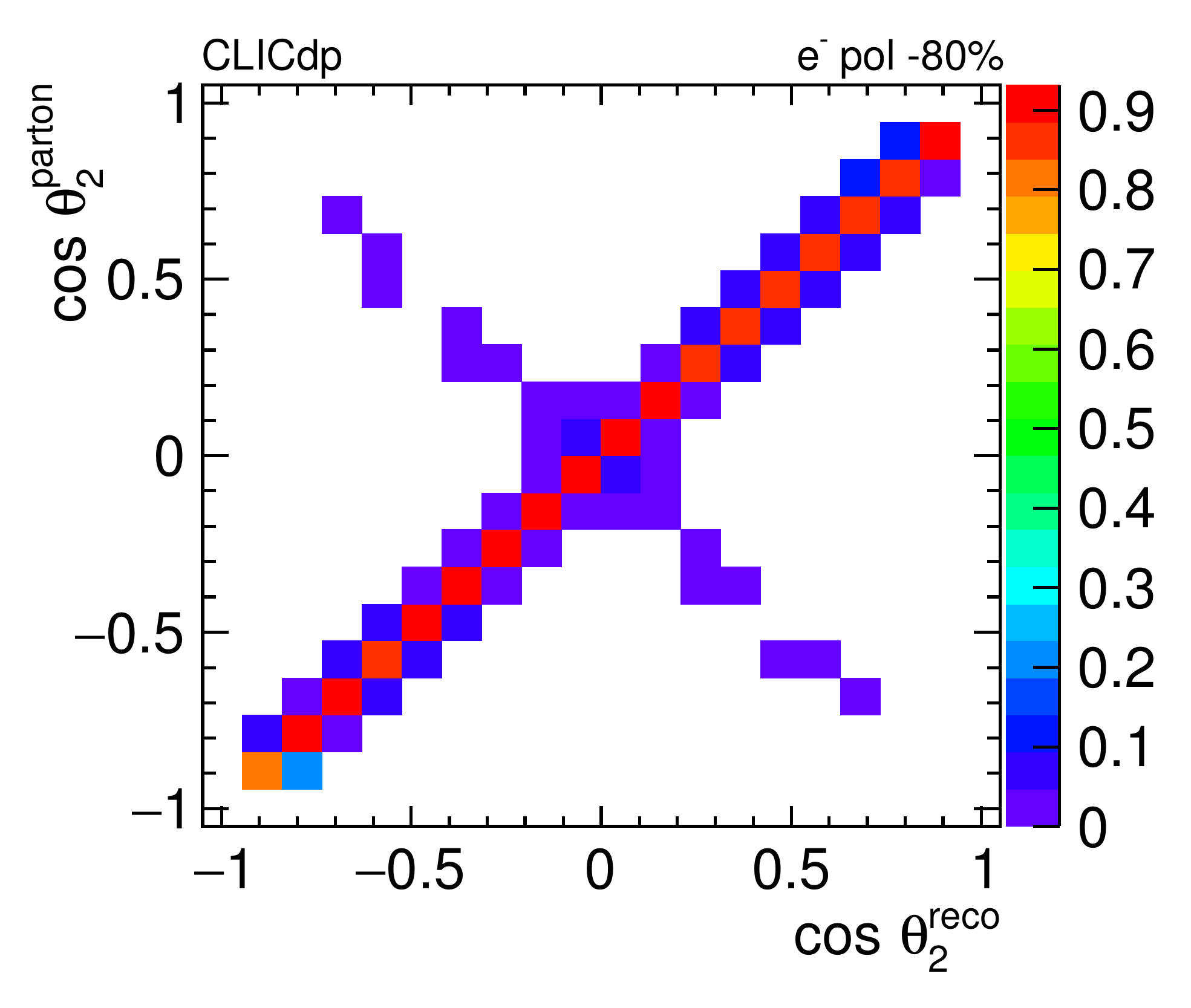}
\end{minipage}
\begin{minipage}[l]{0.32\textwidth}
\includegraphics[width=1.0\textwidth]{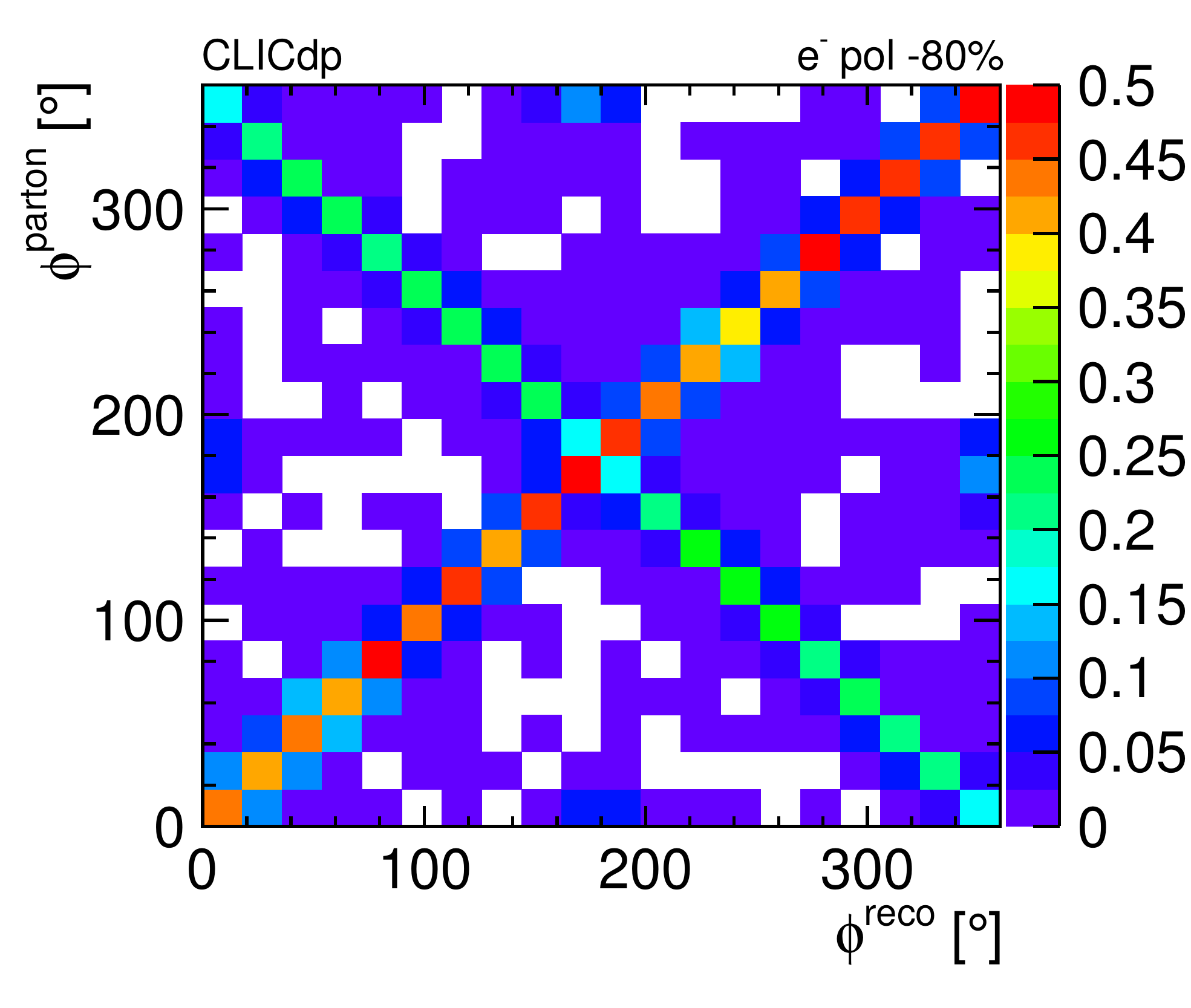}
\end{minipage}
\caption{The normalised correlation matrices between the parton level and reconstructed level of the three angles of $\cos\theta_{1}$ (left), $\cos\theta_{2}$ (centre), and $\phi$ (right) for \zhsm signal in events with negative electron beam polarisation.}
\label{Fig:Angles_parton_vs_reco}
\end{figure}

 The normalised correlation matrices between the reconstructed and parton level for all three angles $\cos\theta_{1}$, $\cos\theta_{2}$, and  $\phi$  are displayed in Fig.~\ref{Fig:Angles_parton_vs_reco}. In the case of $\cos\theta_{1}$ diagonal elements dominate the matrix, illustrating that the subjet clustering of FastJet provides a suitable representation of the underlying quarks. A sizeable fraction of events populate the anti-diagonal, populated by events where the wrong sub-jet was chosen as positively charged quark. This anti-correlation has no impact on the derived values for the asymmetries as long as the transfer matrix is symmetric around an angle $\theta_{1}$ of $90^{\circ}$. The definition of the asymmetry handles these anti-correlated cases in a similar manner as elements on the diagonal, as the positive and negative contributions of the asymmetry are determined by $\mathrm{sgn}(\cos (2\cdot\theta_{1}))$. As long as the transfer of events is symmetric with respect to $\cos \theta_{1}$ the anti-correlation will pose no issue in the unfolding step. The spread around the (anti-) diagonal reflects the confusion term of assigning the reconstructed particles to the correct sub-jet. For $\cos\theta_{2}$ in the overwhelming number of events the correct jet is chosen as direction for the H-jet, and as a consequence the matrix is almost diagonal. Since the angle $\phi$ depends on the correct assignment of both the \PH jet and the positively charged quark direction, a similar behavior is observed as for the $\cos\theta_{1}$ distribution.

\section{B-tagging in boosted H decays}
\label{sec:BTagging}

The linear collider flavour identification (LCFIPlus) tool~\cite{Suehara:2015ura} is used to perform b-jet identification. The tools have been developed in the context of future \epem linear colliders. The first step of LCFIPlus is the primary vertex finder, followed by the secondary vertex finder to identify $b$ and $c$ hadron decays, which are typically embedded in jets. In the second step the secondary vertices are assigned to jets, and isolated leptons are identified within the jet. These isolated leptons can originate from semileptonic decays of heavy flavour hadrons. Tracks from identified secondary vertices and leptons matched to these vertices are used as seeds for the third step of LCFIPlus, the so-called refined jet clustering.
The input particles of the vertex finding and the subsequent jet finding step are the same. In case more particles are considered in the vertex clustering step than for the subsequent refined jet clustering, it can happen that the vertex itself creates its own jet without clustering any further particle specified as input for the refined clustering step. Thus it is not advised to perform secondary vertex finding using all PandoraPFOs clustered in both jets, and subsequently trying to assign the particles of the subjets to the identified vertices. Three different LCFIPlus settings have been investigated:
\begin{itemize}
\item use all PandoraPFOs clustered in both VLC jets as input for secondary vertexing and refined jet clustering, request two output jets. LCFIPlus provides in this case b-tagging information for each VLC jet.
\item use all PandoraPFOs clustered in both VLC jets as input for secondary vertexing and refined jet clustering, request four output jets. LCFIPlus creates for each VLC jet at least one refined jet with b-tagging information.
\item run LCFIPlus for each jet separately, use PandoraPFOs of each VLC jet separately as input for secondary vertexing and jet clustering, request two output jets. LCFIPlus creates for each VLC jet two refined subjets with b-tagging information.
\end{itemize}
The performance of the three settings is tested for the \zhsm events with $\PH\rightarrow\PQb\PAQb$. In all three settings, for the more massive jet one subjet with a high b-tag probability is found, but less frequently two refined subjets with high b-tags. The third procedure provides refined subjets with the largest b-tag values for $jet1$, and is thus used throughout the analysis. As expected, for $jet2$, far less of the refined subjets are compatible with originating from a b-quark.

\section{Monte Carlo simulation}
\label{sec:MCSimulation}

Both signal and backgrounds samples are produced by \whizard2.7.0, using luminosity spectra from \guineapig interfaced by \textsc{circe2}{} with initial state radiation enabled. Parton shower and hadronisation are handled by \pythia6. The \textsc{DD4hep}{} detector description toolkit has been used to implement the simulated model of CLICdet in \geant{}, version 10.02p02, via the \textsc{DDG4} package. \\
Backgrounds to all-hadronic \zhsm events originate from di-quark $\epem\rightarrow \qqbar$, four-quark $\epem\rightarrow \qqqq$ and six-quark $\epem\rightarrow \qqqqqq$ final states. In order to increase the statistics of the backgrounds at large effective \roots, additional di-jet and four-quark samples have been produced with phase-space selections on the di-quark mass $m_{\qqbar}>\SI{1}{TeV}$ and four quark mass $m_{\qqqq}>\SI{2}{TeV}$ respectively. It has been checked that all background events at large \roots are originating from that particular phase-space. Table \ref{Tab:MC_samples} lists the details of the produced samples for both negative and positive polarisation of 80\% of the electron beam. The weight of each event is calculated under the assumption of luminosity sharing of the ratio 4:1 between the negative and positive polarisation of the electron beam, thus $\mathrm{L}_{-80\%}=\SI{4}{\abinv}$ and $\mathrm{L}_{+80\%}=\SI{1}{\abinv}$ are used as values for the integrated luminosity. The polarisation has a moderate impact on the \zhsm signal, decreasing the cross section by about 28\% for positive compared to negative electron beam polarisation, a similar impact can be observed for the di-quark sample. The four-quark dataset cross section is largely reduced for positive polarisation by a factor of about 7.5. The six-quark dataset is split into 14 samples to cover all possible flavour combinations compatible with \ttbar. These datasets include the most relevant contributions from $\PW\PW\PZ$ tri-boson production as well. In the Table \ref{Tab:MC_samples} the six quark flavour combinations with the largest cross sections are shown. For the six quark dataset positive polarisation reduces the cross section considerably as well. Di-boson production from \PW\PW-fusion is expected to be negligible in the phase-space of this study.

\begin{table}[hbtp]
 \centering
 \caption{\label{Tab:MC_samples} Signal and background datasets with $y=\PQd,\PQs,\PQb$, $\mathrm{L}=\SI{4}{\abinv}$ for P(\Pem)=-80\%, $\mathrm{L}=\SI{1}{\abinv}$ for P(\Pem)=+80\%:}
 \begin{tabular}{|c|c|c|c|c|c|}
\hline
process & additional cuts & Events & $\sigma$[fb] & Polarisation & event weight \\
\hline
$\epem\rightarrow\zhsm$ & - & 114000 &3.83 & P(\Pem)=-80\% & 0.134 \\
$\epem\rightarrow\zhsm$ & - & 27840 & 2.76 & P(\Pem)=+80\% & 0.0959 \\
$\epem\rightarrow\qqbar$ & - & 1549464 & 1269 & P(\Pem)=-80\% & 3.28 \\
$\epem\rightarrow\qqbar$ & - & 388392 & 786 & P(\Pem)=+80\% & 2.02 \\
$\epem\rightarrow\qqbar$ & $m_{\qqbar}>\SI{1}{TeV}$ & 1519910 & 170.8 & P(\Pem)=-80\% & 0.445 \\
$\epem\rightarrow\qqbar$ & $m_{\qqbar}>\SI{1}{TeV}$ & 382464 & 73.5 & P(\Pem)=+80\% & 0.192 \\
$\epem\rightarrow\qqqq$ & - & 1915464 & 902 & P(\Pem)=-80\% & 1.88 \\
$\epem\rightarrow\qqqq$ & - & 479040 & 120 & P(\Pem)=+80\% & 0.251 \\
$\epem\rightarrow\qqqq$ & $m_{\qqqq}>\SI{2}{TeV}$ & 1522935 & 369.8 & P(\Pem)=-80\% & 0.971\\
$\epem\rightarrow\qqqq$ & $m_{\qqqq}>\SI{2}{TeV}$ & 380451 & 49.2 & P(\Pem)=+80\% & 0.129\\
$\epem\rightarrow\PQd\PQd\PQu  yy\PQu$ & - & 456336 & 14.5 & P(\Pem)=-80\% & 0.127\\
$\epem\rightarrow\PQd\PQd\PQu  yy\PQu$ & - &  121200 & 5.01 & P(\Pem)=+80\% & 0.0413 \\
$\epem\rightarrow yy\PQu\PQb\PQb\PQc$ & - & 428405  & 13.3 & P(\Pem)=-80\% & 0.124\\
$\epem\rightarrow yy\PQu\PQb\PQb\PQc$ & - & 123720 & 5.21 & P(\Pem)=+80\% & 0.0421\\
$\epem\rightarrow\PQs\PQs\PQc\PQb\PQb\PQc$ & - & 330096 & 12.5 & P(\Pem)=-80\% & 0.151\\
$\epem\rightarrow\PQs\PQs\PQc\PQb\PQb\PQc$ & - & 84240 & 4.89 & P(\Pem)=+80\% & 0.0581 \\
\hline
  \end{tabular}
 \end{table}

\section{Discriminating variables and signal selection}
\label{sec:variables}

\begin{figure}[htbp!]
\centering
\begin{minipage}[l]{0.45\textwidth}
\includegraphics[width=1.0\textwidth]{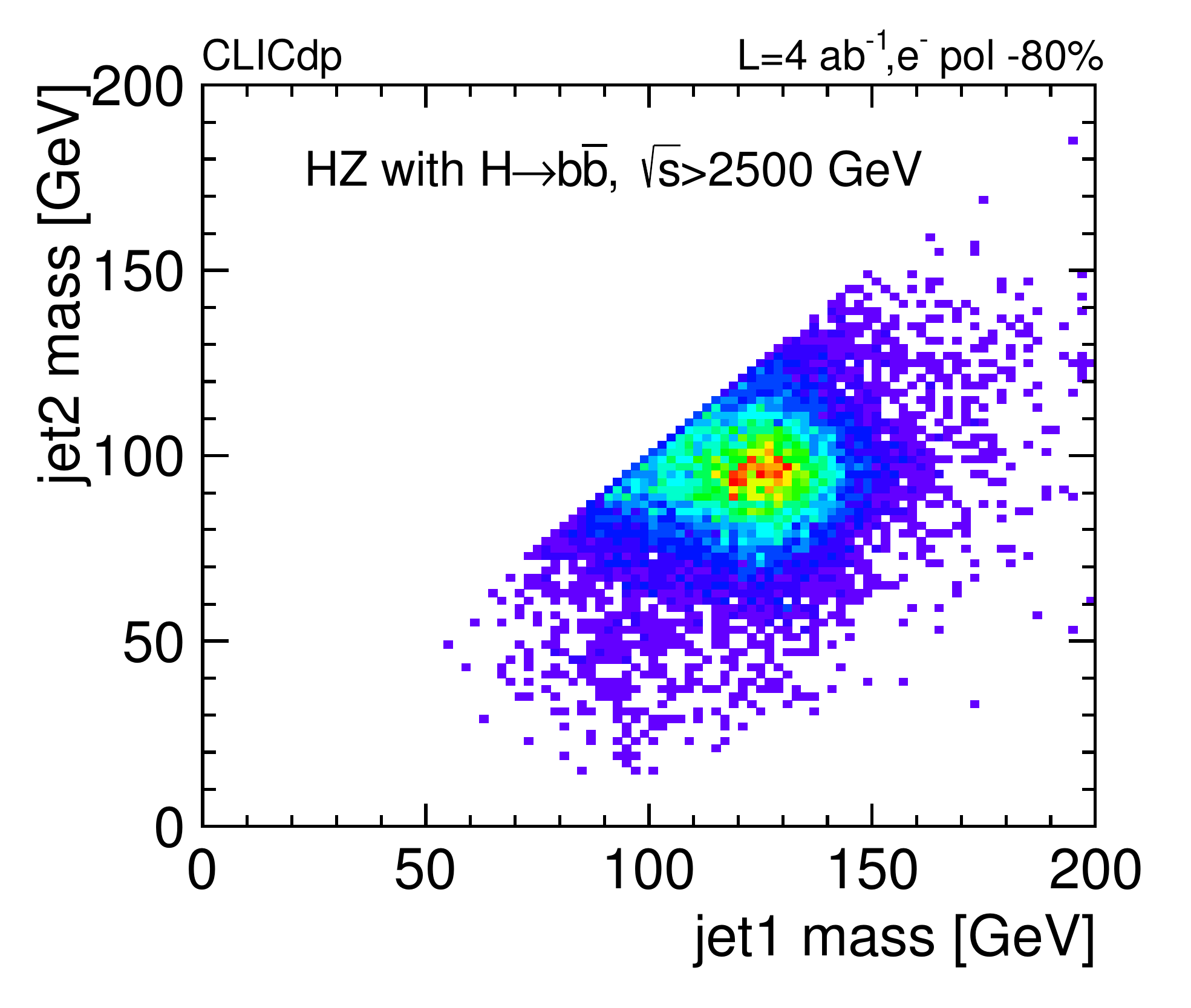}
\end{minipage}
\begin{minipage}[r]{0.45\textwidth}
\includegraphics[width=1.0\textwidth]{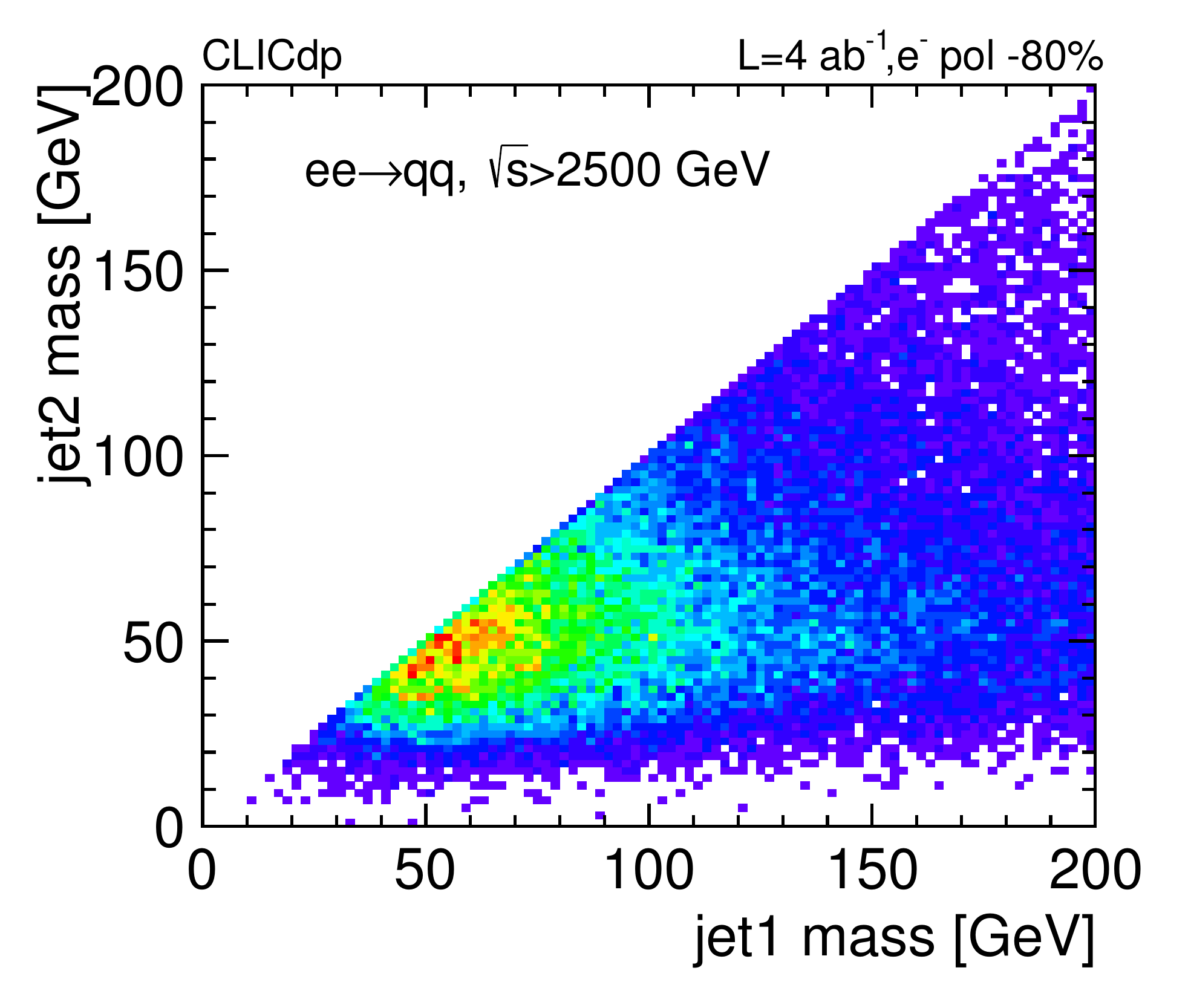}
\end{minipage}
\begin{minipage}[l]{0.45\textwidth}
\includegraphics[width=1.0\textwidth]{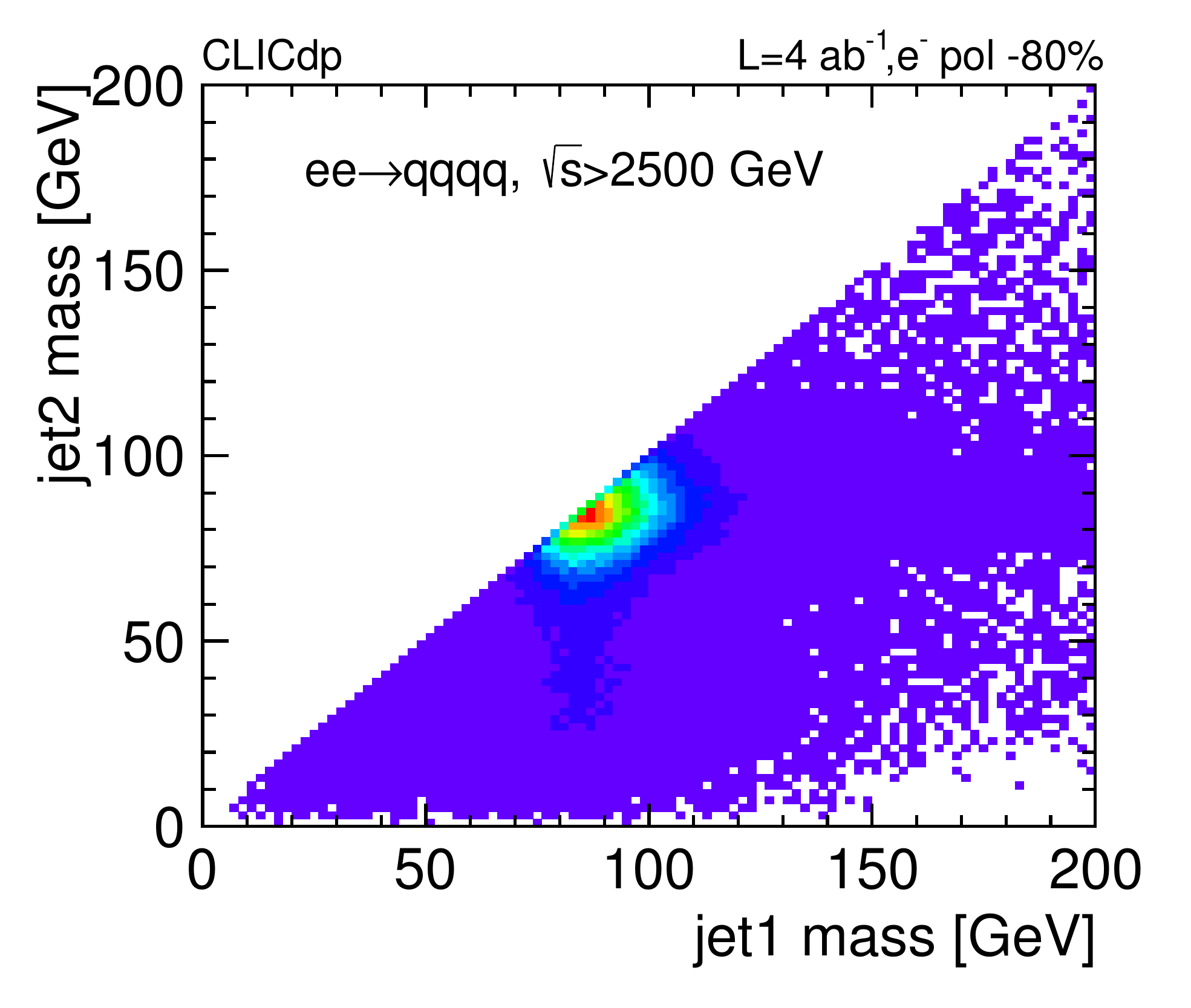}
\end{minipage}
\begin{minipage}[r]{0.45\textwidth}
\includegraphics[width=1.0\textwidth]{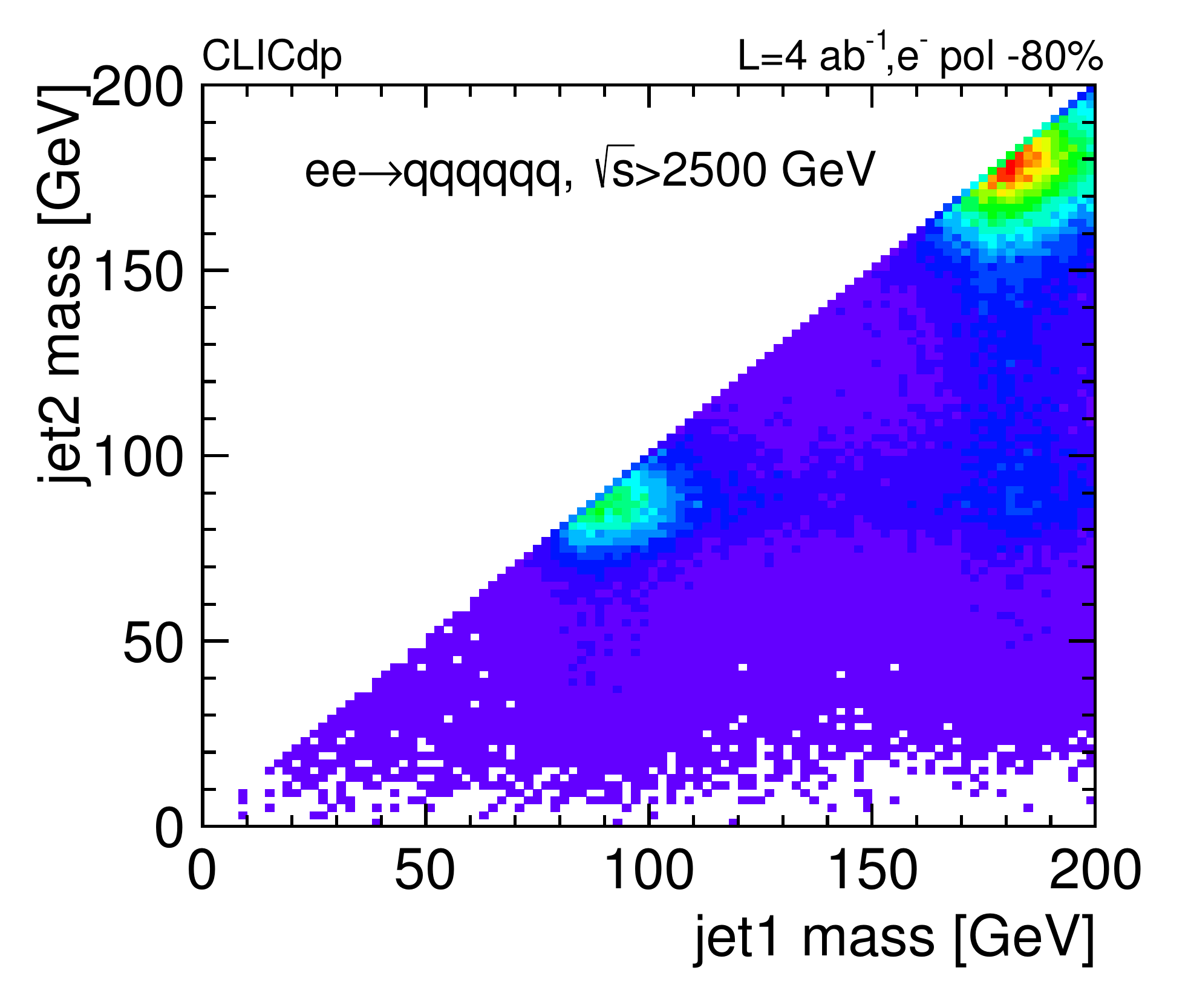}
\end{minipage}
\caption{The two-dimensional mass plane of the leading vs the second leading mass for \zhsm (top left), $\epem\rightarrow\qqbar$ (top right), $\epem\rightarrow\qqqq$ (bottom left), and $\epem\rightarrow \text{qqqqqq}$ events (bottom right) with negative electron beam polarisation.}
\label{Fig:2D_massplane}
\end{figure}

The jet mass distributions are most discriminating to differentiate between signal and background. The two dimensional distributions are shown in Fig.~\ref{Fig:2D_massplane} for signal and background events. In order to provide a more signal-like selection for further processing a preselection is applied on the two-dimensional plane of the jet masses, selecting events which fall within a radius of \SI{35}{GeV} around the \PZ-\PH mass point in the $jet2$-$jet1$ mass plane. This cut retains about 85\% of all \zhsm events, 89\% of \zhsm events with $\PH\rightarrow\bb$. For the di-quark dataset the jet masses peak at considerably lower values. For the six-quark dataset the 2D mass distribution has a maximum around the top-mass, as well as a sub-leading maximum around \PW and \PZ boson masses. For the four-quark dataset the mass peak is around the \PW and \PZ boson masses. The preselection cut on the 2D mass plane removes about 85\% of di-quark, over 90\% of six-quark, and about 62\% of four-quark events as presented in Table \ref{Tab:preselection_efficiency}. For events with large reconstructed centre-of-mass-energy, $\roots>\SI{2500}{GeV}$, the amount of \zhsm events is 0.38\% of all events for negative \Pem polarisation, for events with positive \Pem polarisation the signal purity is 1.5\% without any further selection. 

\begin{table}[hbtp]
 \centering
 \caption{Mass preselection efficiencies and event numbers\label{Tab:preselection_efficiency} for signal and background events, assuming an integrated luminosity of $\mathrm{L}=\SI{4}{\abinv}$ for runs with negative polarisation P(\Pem)=-80\%, and $\mathrm{L}=\SI{1}{\abinv}$ for runs with positive polarisation P(\Pem)=+80\%. All numbers are given for \mbox{$\roots>\SI{2500}{GeV}$}:}
\begin{tabular}{|c|c|c|c|c|c|c|}
\hline
process & Events & Evts after cut & Efficiency &  Events & Evts after cut & Efficiency  \\
  & neg. p.  & neg. p. & neg. p., in [\%] &  pos. p. & pos. p.  & pos. p., [in\%]\\
\hline
$\epem\rightarrow\zhsm$, $\PH\rightarrow\bb$ & 1740 & 1541 & 89 & 304 & 268 & 88 \\
$\epem\rightarrow\zhsm$, all \PH & 2600 & 2210 & 85 & 458 & 388 & 85 \\
$\epem\rightarrow\qqbar$ & 172 000 & 25 019 & 15 & 18 600 & 2680 & 14 \\
$\epem\rightarrow\qqqq$ & 248 000 & 91 100 & 37 & 8 450 & 3190 & 38 \\
$\epem\rightarrow\qqqqqq$ & 32 200 & 3190 & 9.9 & 3 090 & 145 & 4.7 \\
\hline
  \end{tabular}
 \end{table}

After applying the preselection the following classes of variables have been considered:
\begin{itemize}
\item polar angles $\theta$ of both leading jets: while the signal tends to be central, four- and six-quark events are peaked considerably more forward.
\item b-tagging information of the leading jet: For the signal a very high b-tagging probability is observed, most events having values larger than 0.9. The background peaks at lower values close to 0. No large discriminating power is observed using c-tagging or light flavour tagging probability on the leading jets, probabilities for the sub-leading jet are very similar as well.
\item event shape information, based on the three-jet resolution parameter $y_{23}$.
\item jet substructure information of both jets, notably the N-subjettiness ratio $\tau_{21}=\tau_{2}/\tau_{1}$~\cite{Thaler:2010tr}, and ratios between genereralised jet energy correlation functions. These are correlation functions based on energies and pair wise angles between particles within jets~\cite{Larkoski:2013eya}. Ratios between jet energy correlations lead to new jet substructure variables $C_{2}^{(\beta)}$, $C_{3}^{(\beta)}$, and $D_{2}^{(\beta)}$~\cite{Larkoski:2014gra}. In this analysis, a definition of jet energy correlations based on transverse momenta \pT and radial distances is used, with $\beta=1.0$ as weighting factor. 
\end{itemize}

The identification targets to discriminate between background events and signal events. Signal events are defined as \zhsm events with $\PH\rightarrow\bb$, since at \SI{3}{TeV} CLIC the coupling of \PH and \PQb will be known to best accuracy, and systematic uncertainties on the other \PH decays have a smaller impact. In addition to the distributions listed above, the acoplanarity $\Delta\phi(jet1,jet2)$, the four-jet resolution parameter $y_{34}$, the subjet merge clustering measures $d_{12}$, $d_{23}$, $d_{34}$, and energy correlation ratios $N_{2}^{(\beta)}$, $N_{3}^{(\beta)}$~\cite{Moult:2016cvt} have been studied. These are found to lead to no further improvement in discriminating power. 

\begin{figure}[htbp!]
\centering
\begin{minipage}[l]{0.49\textwidth}
\includegraphics[width=1.0\textwidth]{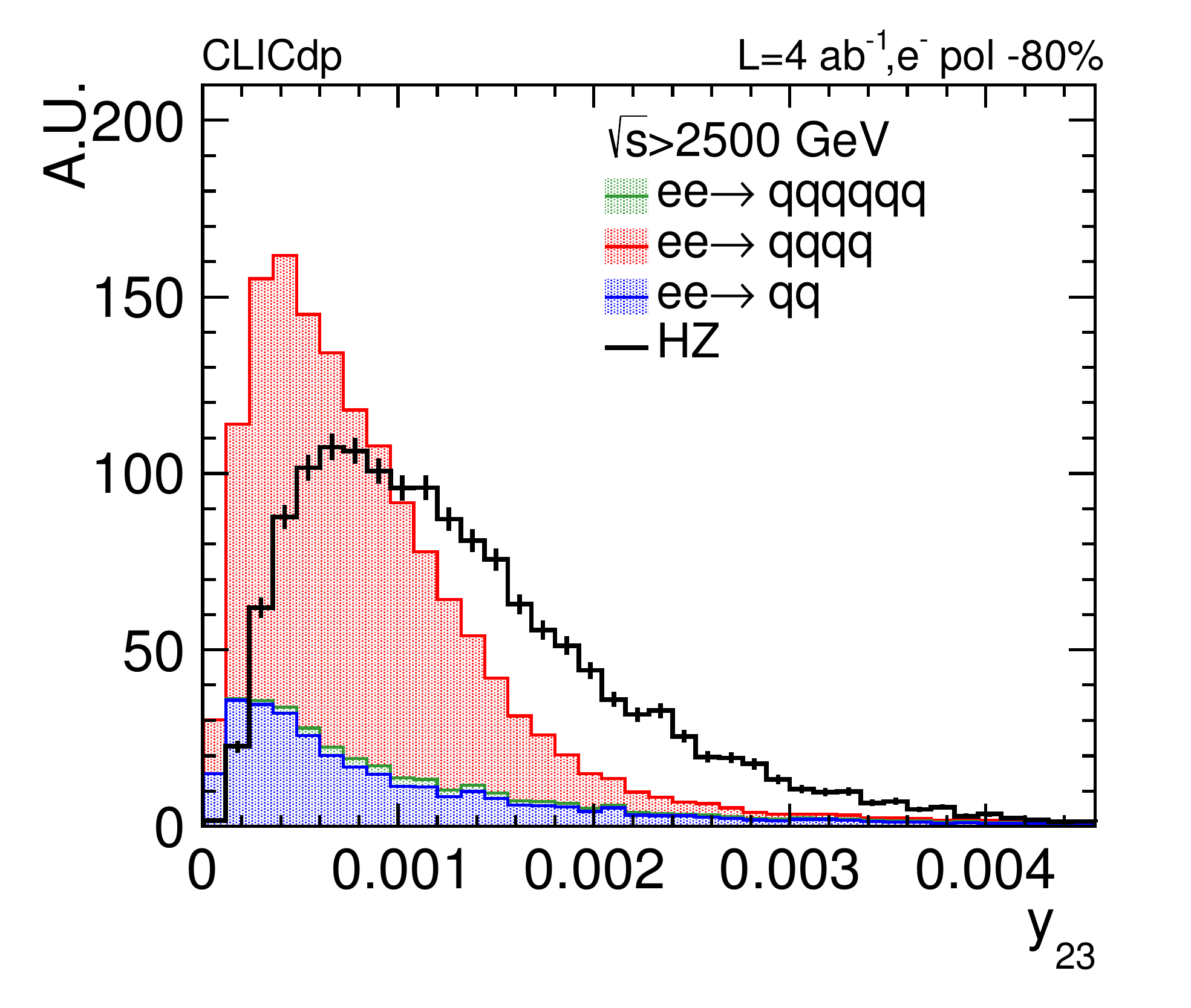}
\end{minipage}
\begin{minipage}[r]{0.49\textwidth}
\includegraphics[width=1.0\textwidth]{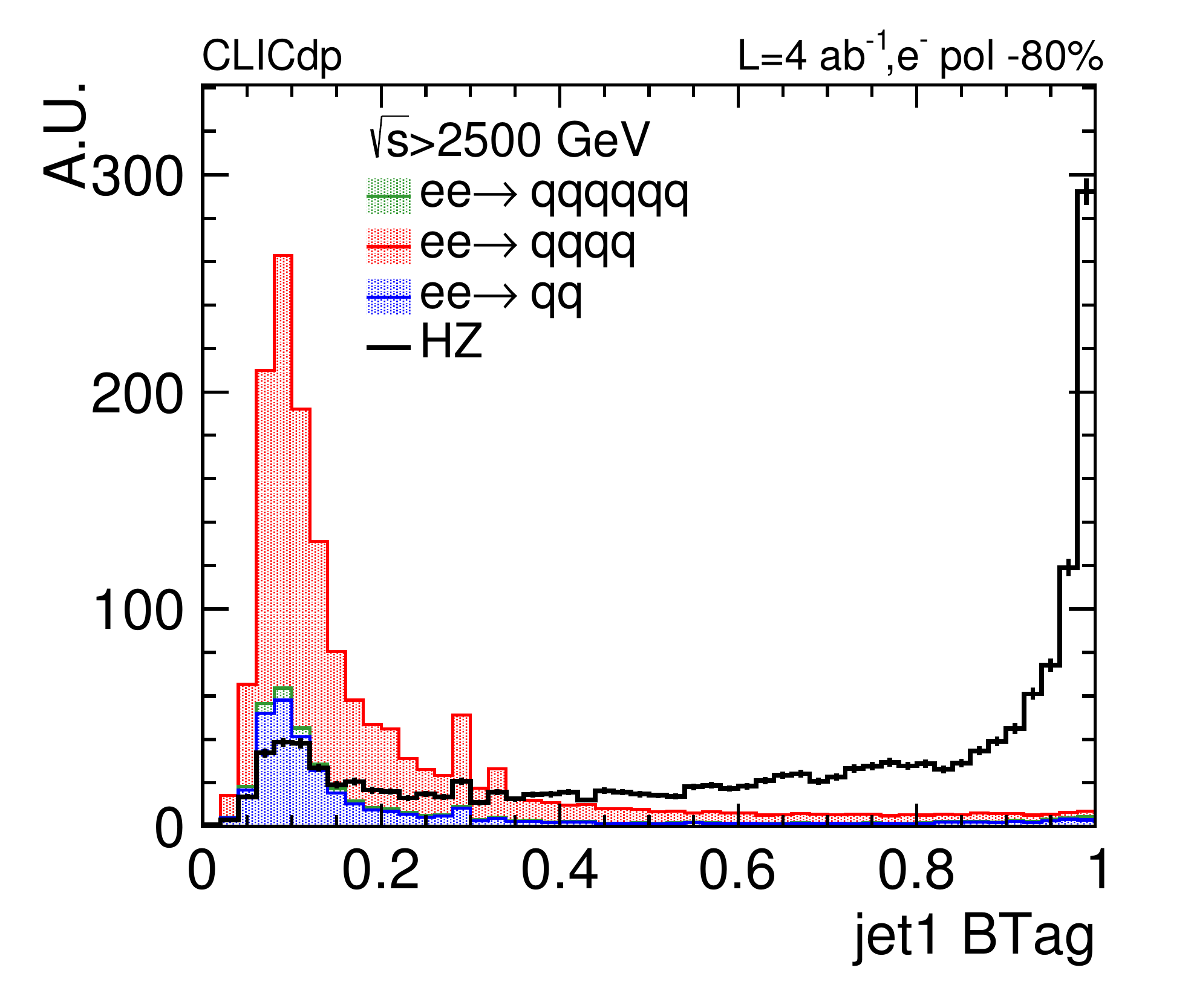}
\end{minipage}
\caption{The three-jet resolution parameter $y_{23}$ (left) and the b-tag distribution of the leading jet (left) for signal and background events with negative electron beam polarisation.}
\label{Fig:discrimination1}
\end{figure}

\begin{figure}[htbp!]
\centering
\begin{minipage}[l]{0.49\textwidth}
\includegraphics[width=1.0\textwidth]{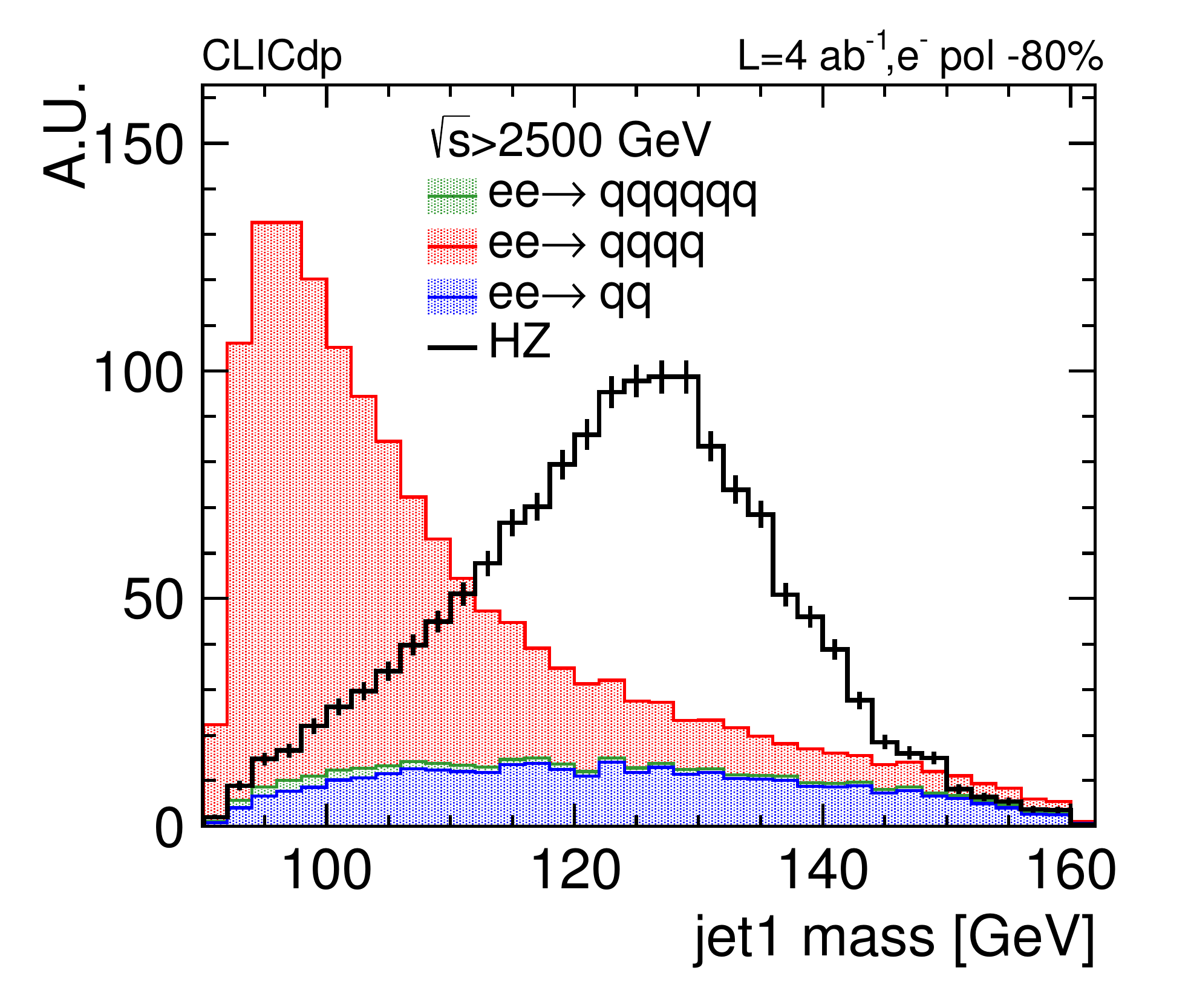}
\end{minipage}
\begin{minipage}[r]{0.49\textwidth}
\includegraphics[width=1.0\textwidth]{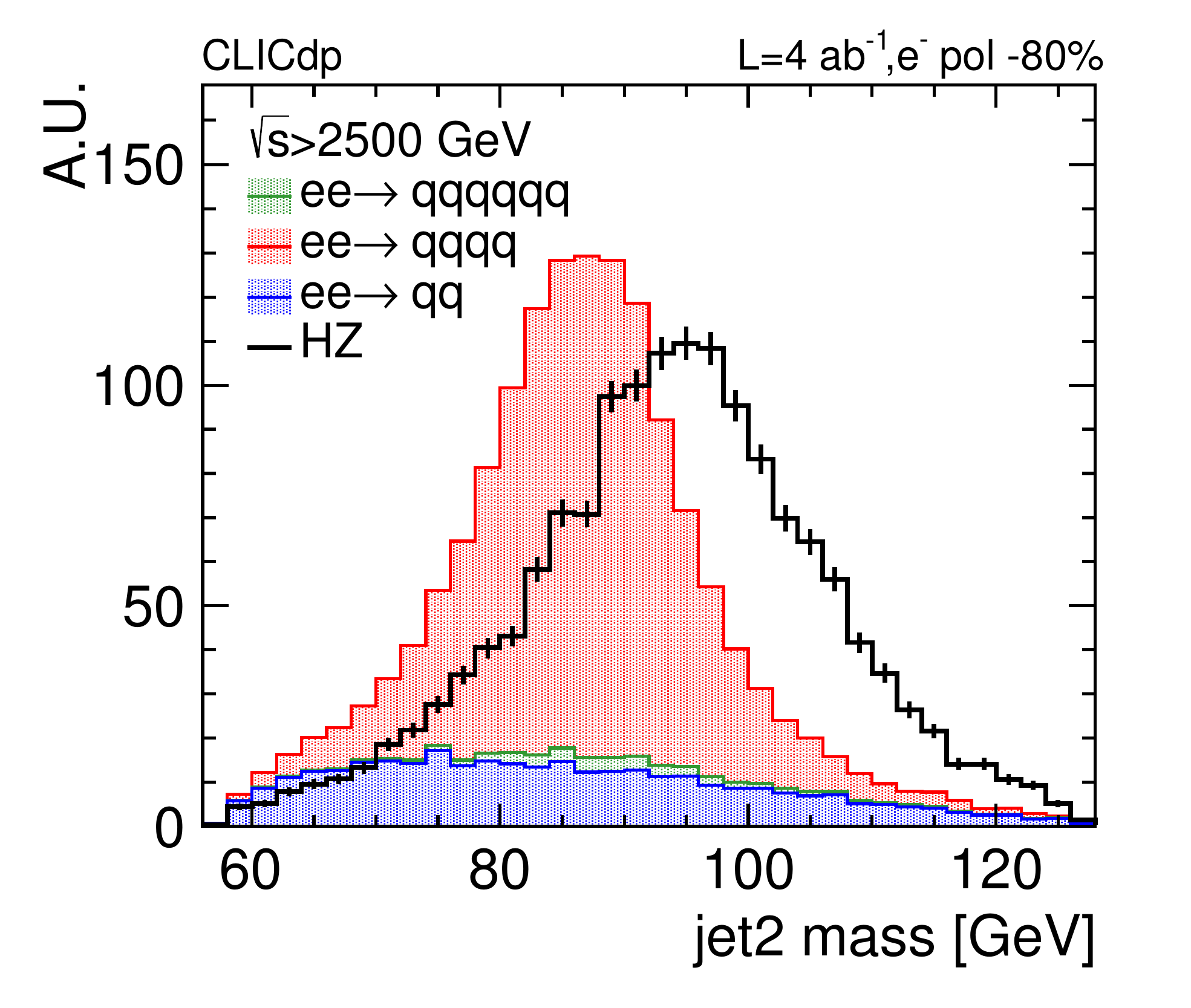}
\end{minipage}
\begin{minipage}[l]{0.49\textwidth}
\includegraphics[width=1.0\textwidth]{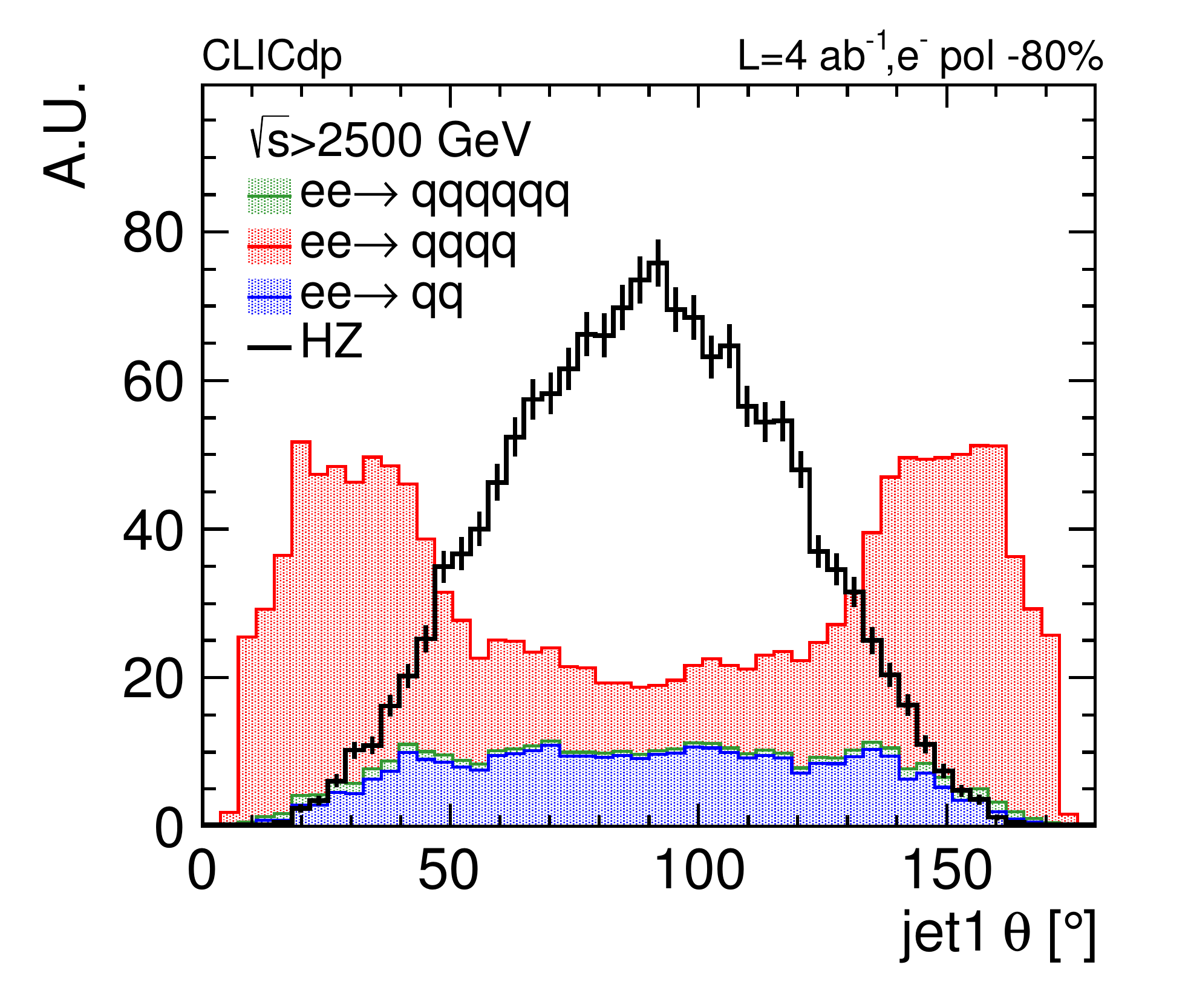}
\end{minipage}
\begin{minipage}[r]{0.49\textwidth}
\includegraphics[width=1.0\textwidth]{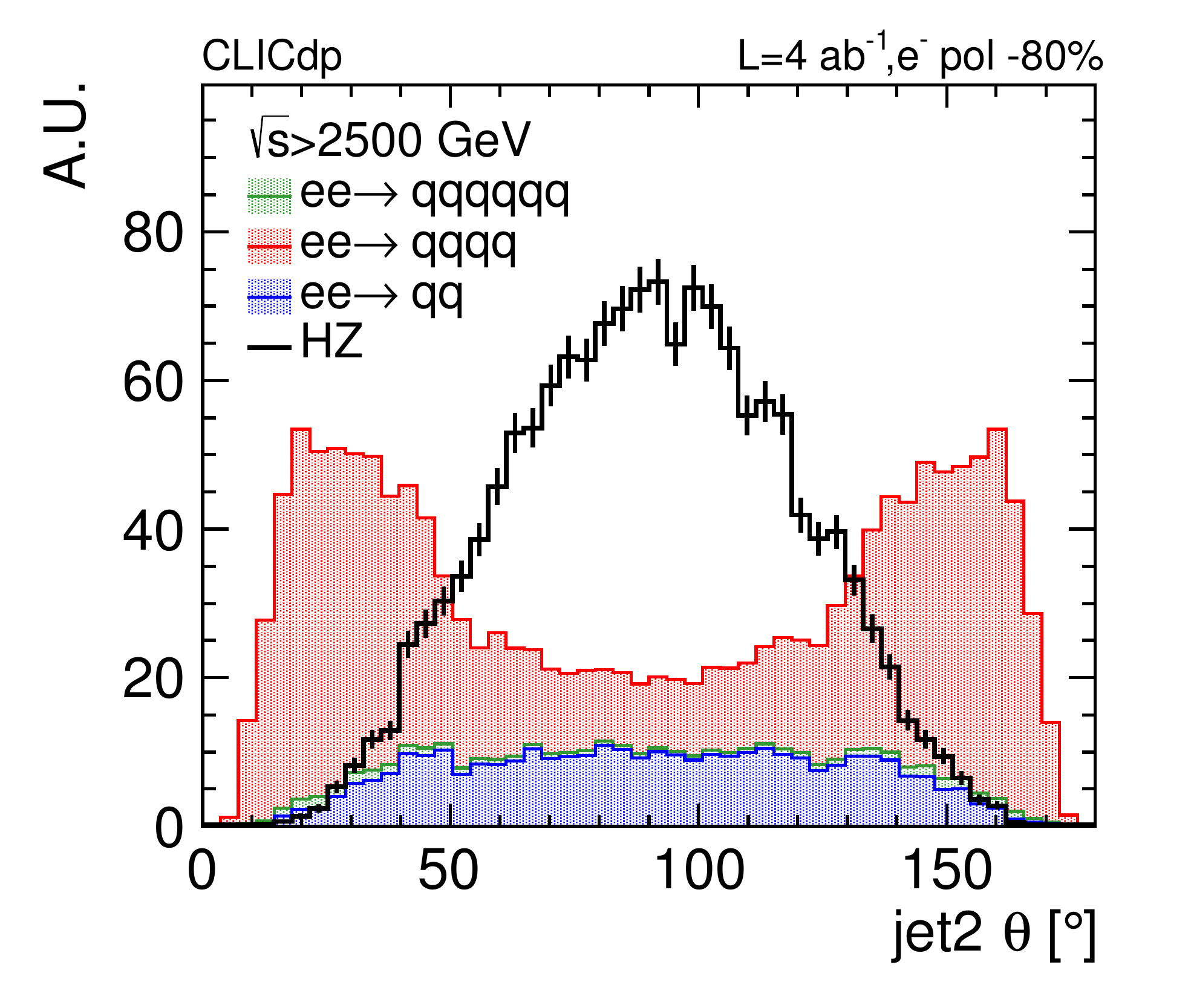}
\end{minipage}
\caption{The jet mass (top) and polar angle distributions (bottom) for the leading (left) and subleading (right) jet for signal and background events with negative electron beam polarisation.}
\label{Fig:discrimination2}
\end{figure}

\begin{figure}[htbp!]
\centering
\begin{minipage}[l]{0.49\textwidth}
\includegraphics[width=1.0\textwidth]{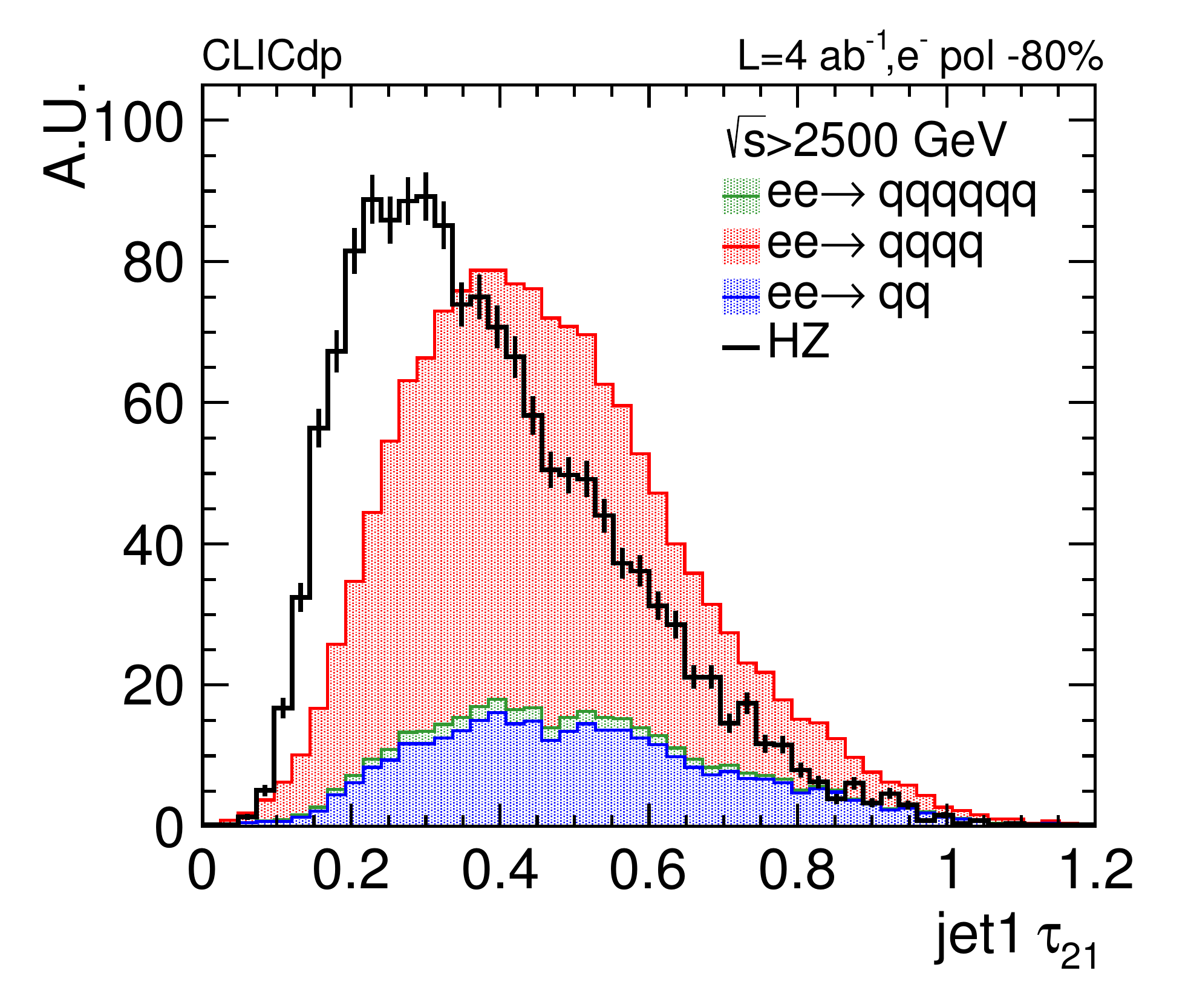}
\end{minipage}
\begin{minipage}[r]{0.49\textwidth}
\includegraphics[width=1.0\textwidth]{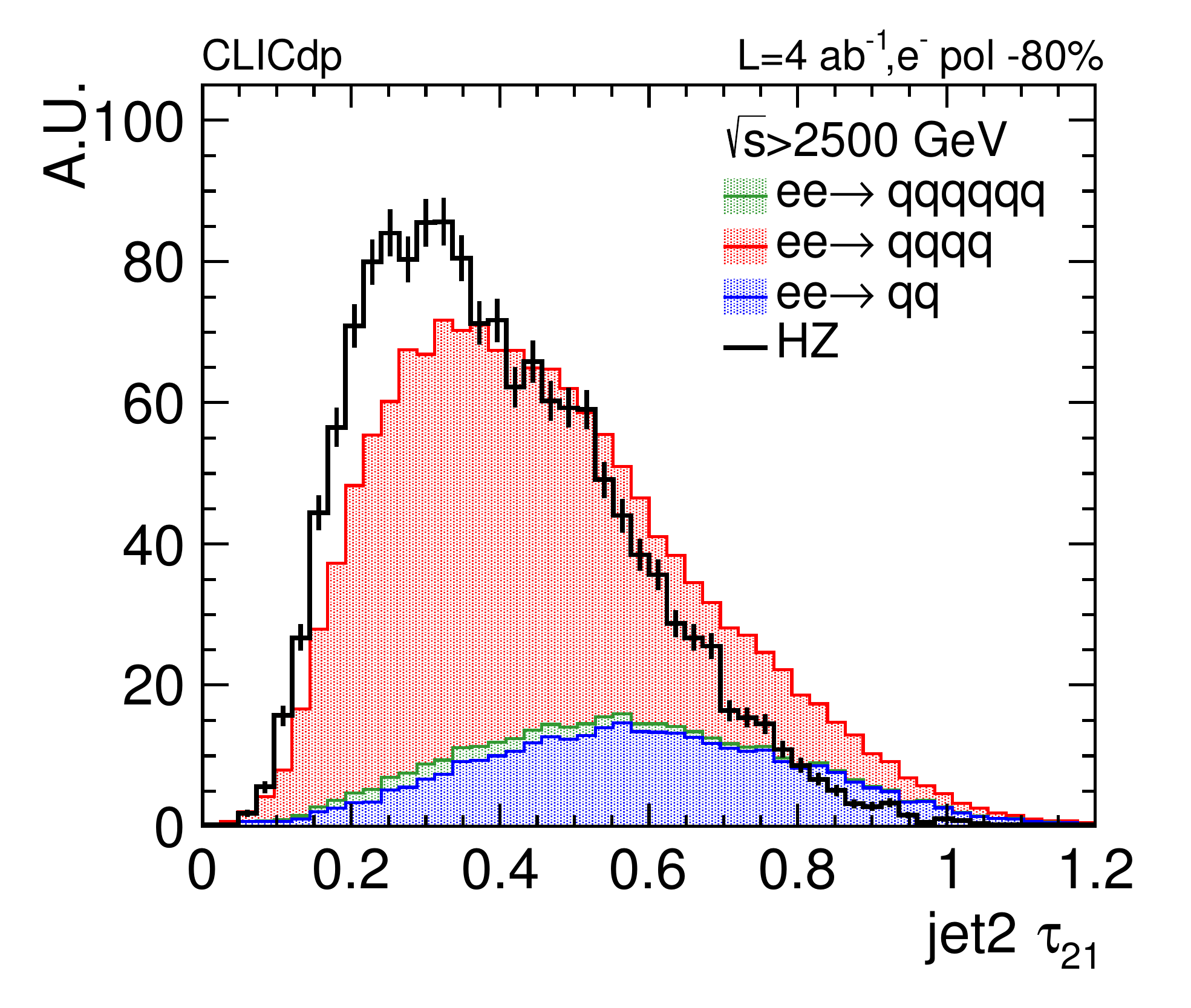}
\end{minipage}
\caption{The N-subjettiness ratio $\tau_{21}$ for the leading (left) and sub-leading jet (right) for signal and background events with negative electron beam polarisation.}
\label{Fig:discrimination3}
\end{figure}

\begin{figure}[htbp!]
\centering
\begin{minipage}[l]{0.49\textwidth}
\includegraphics[width=1.0\textwidth]{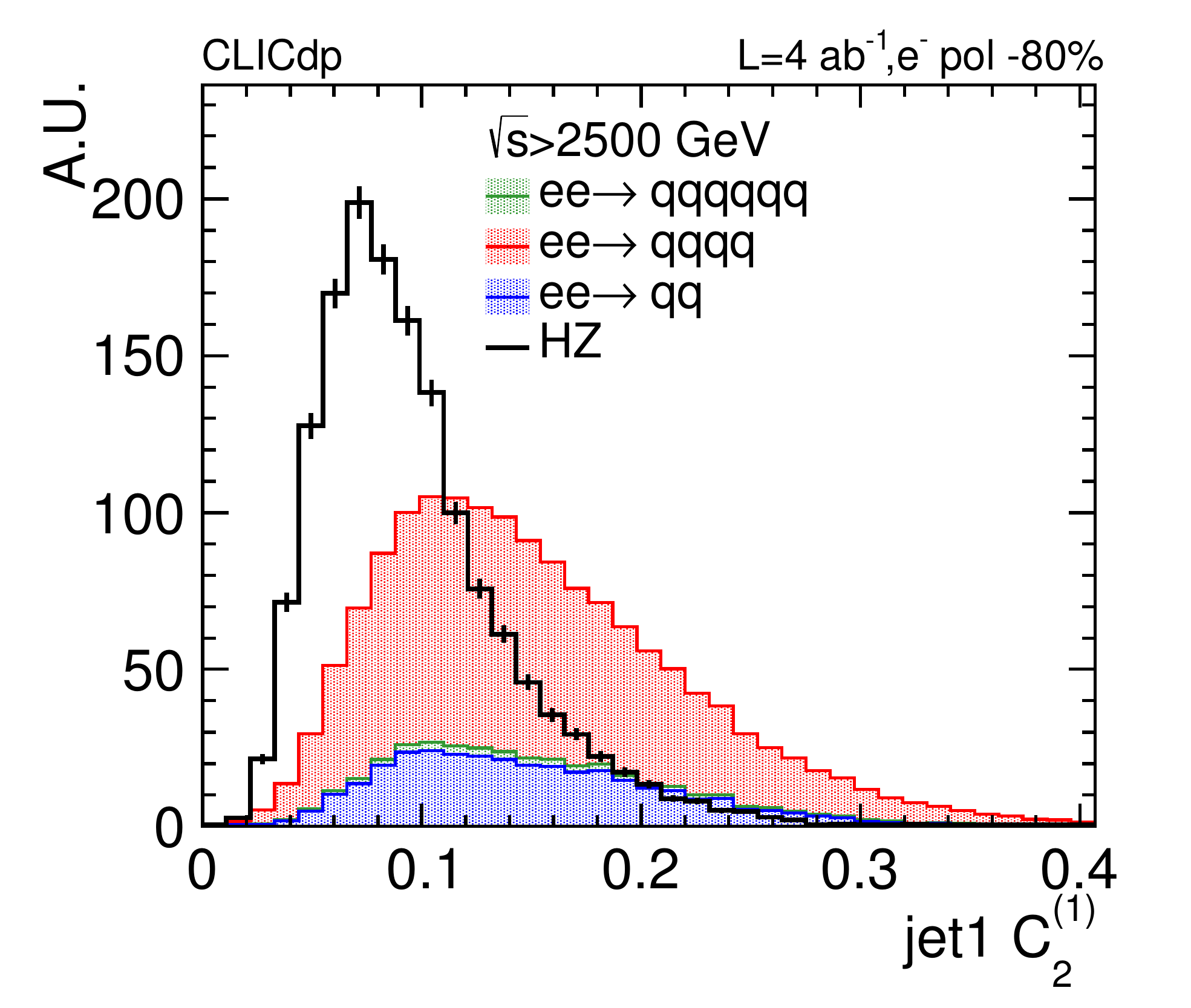}
\end{minipage}
\begin{minipage}[r]{0.49\textwidth}
\includegraphics[width=1.0\textwidth]{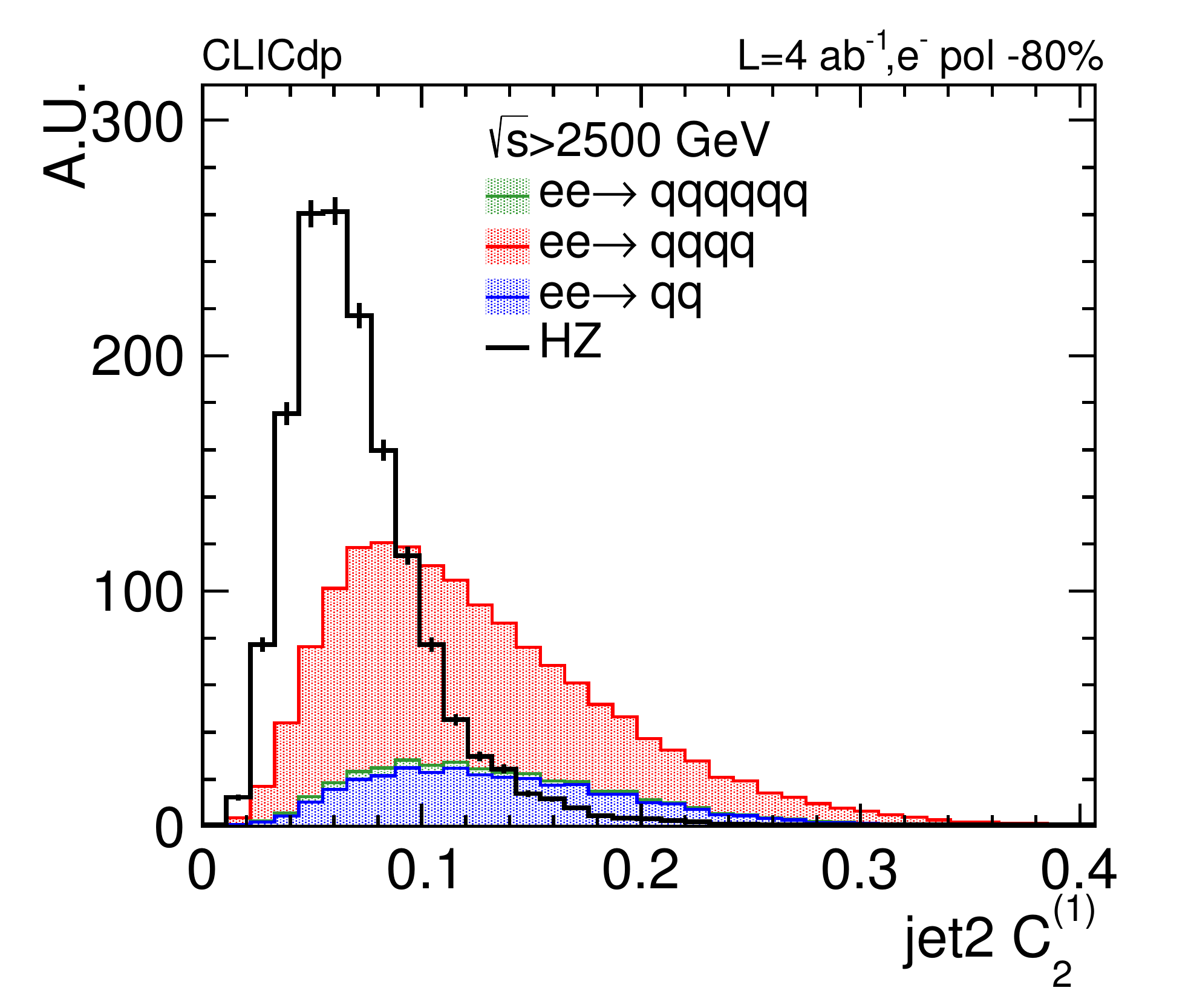}
\end{minipage}
\begin{minipage}[l]{0.49\textwidth}
\includegraphics[width=1.0\textwidth]{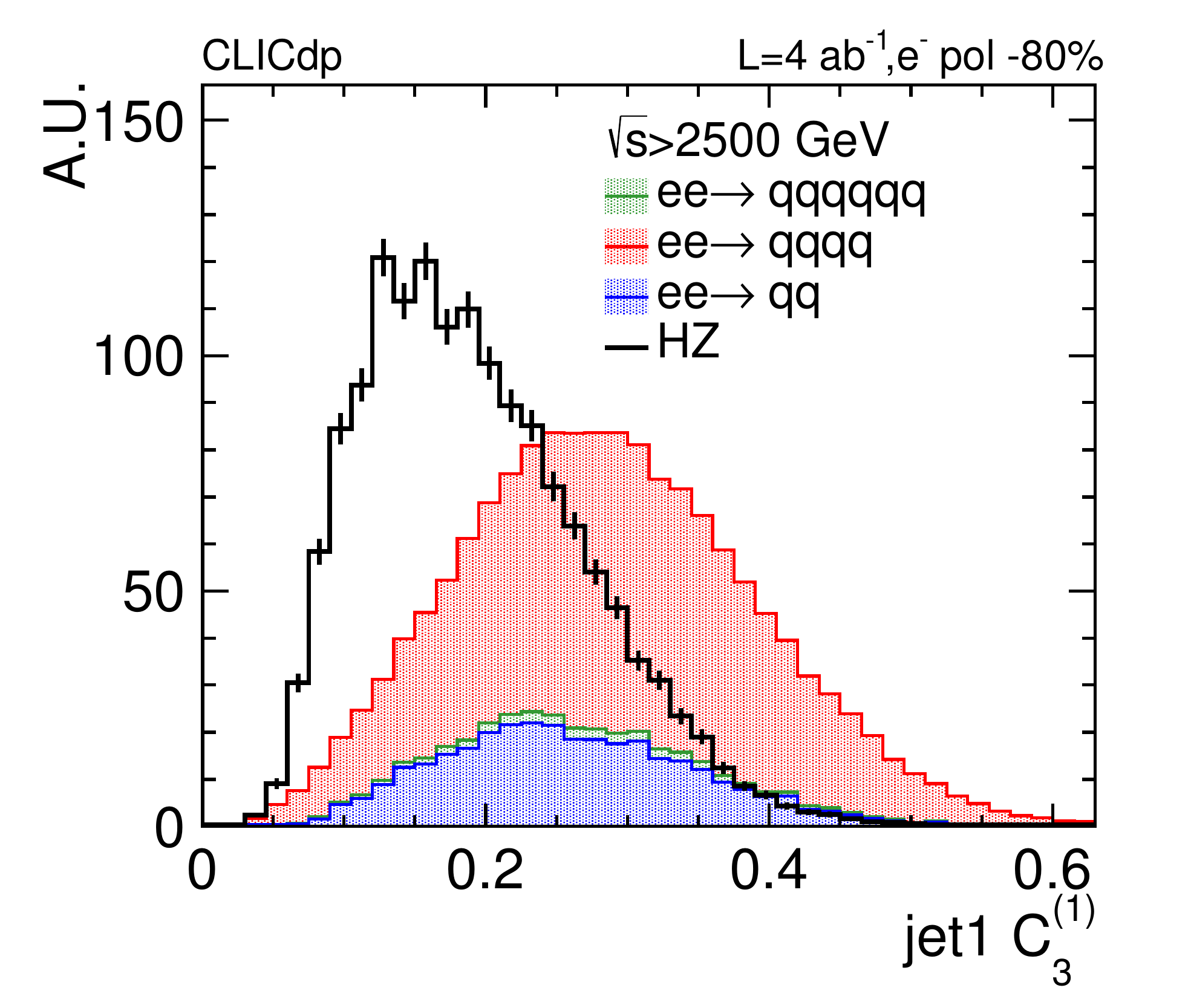}
\end{minipage}
\begin{minipage}[r]{0.49\textwidth}
\includegraphics[width=1.0\textwidth]{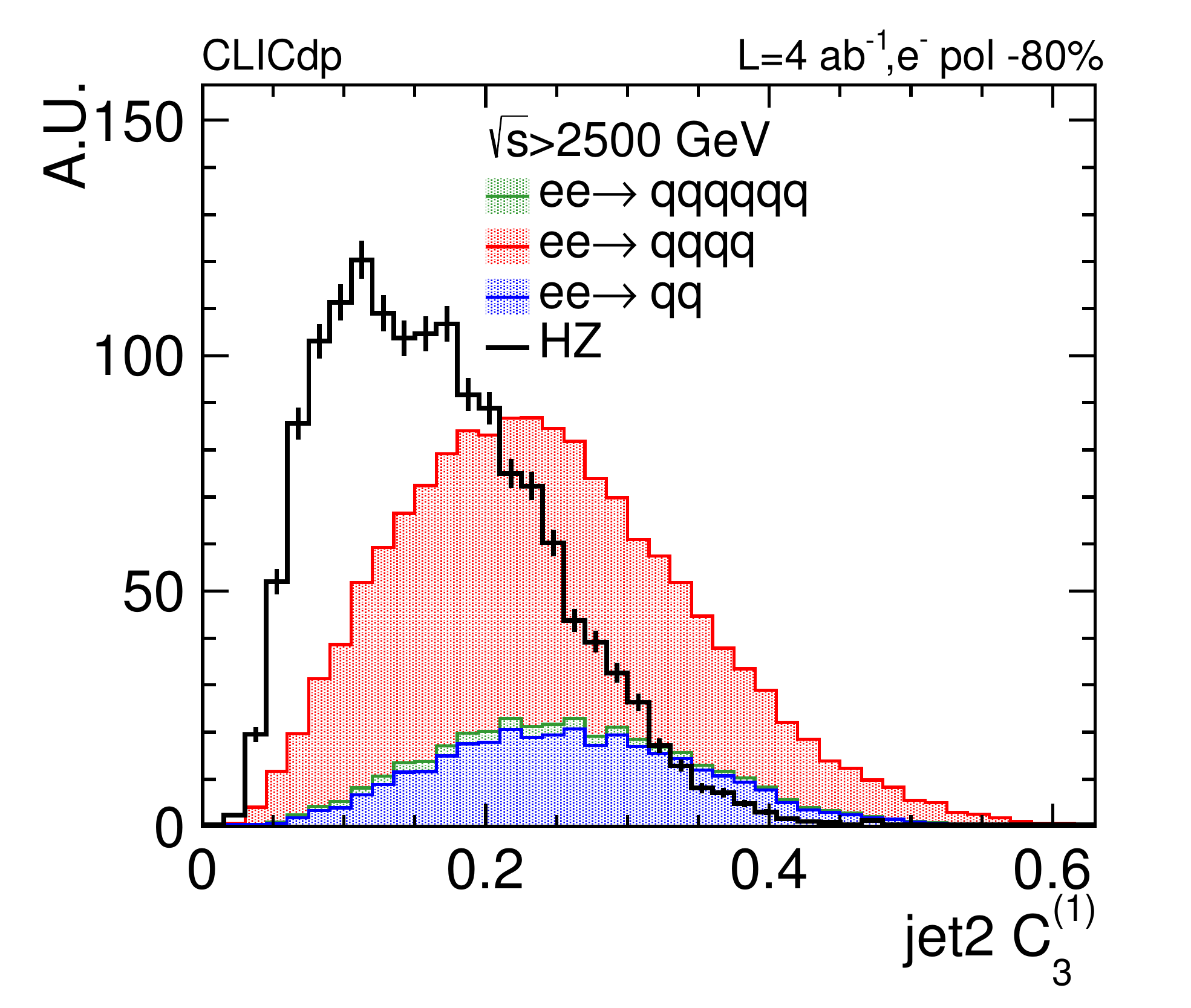}
\end{minipage}
\caption{The jet energy correlation ratio distribution $C_{2}^{(1)}$ (top) and $C_{3}^{(1)}$ (bottom) for the leading (left) and subleading (right) jet for signal and background events with negative electron beam polarisation.}
\label{Fig:discrimination4}
\end{figure}

\begin{figure}[htbp!]
\centering
\begin{minipage}[l]{0.49\textwidth}
\includegraphics[width=1.0\textwidth]{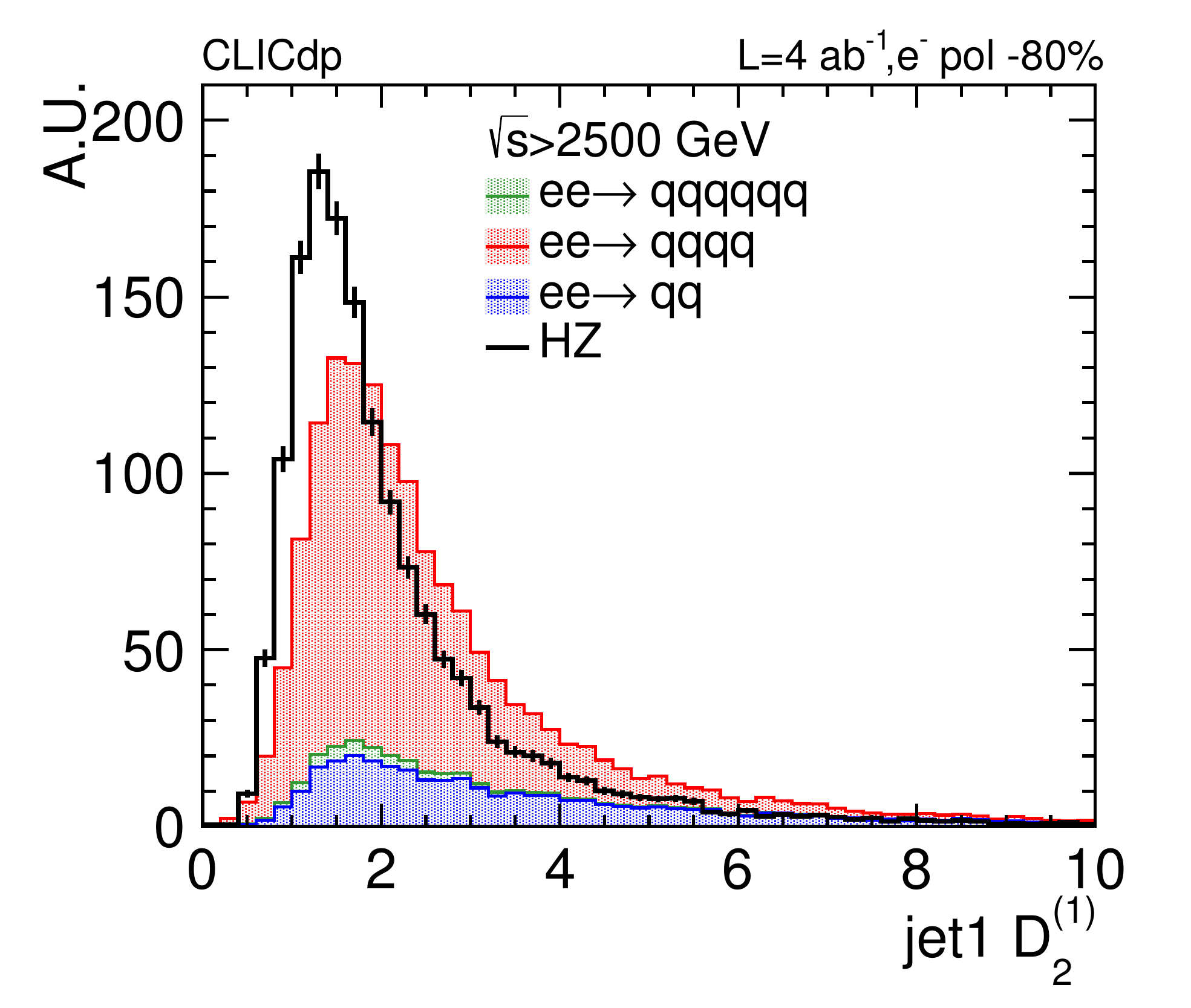}
\end{minipage}
\begin{minipage}[r]{0.49\textwidth}
\includegraphics[width=1.0\textwidth]{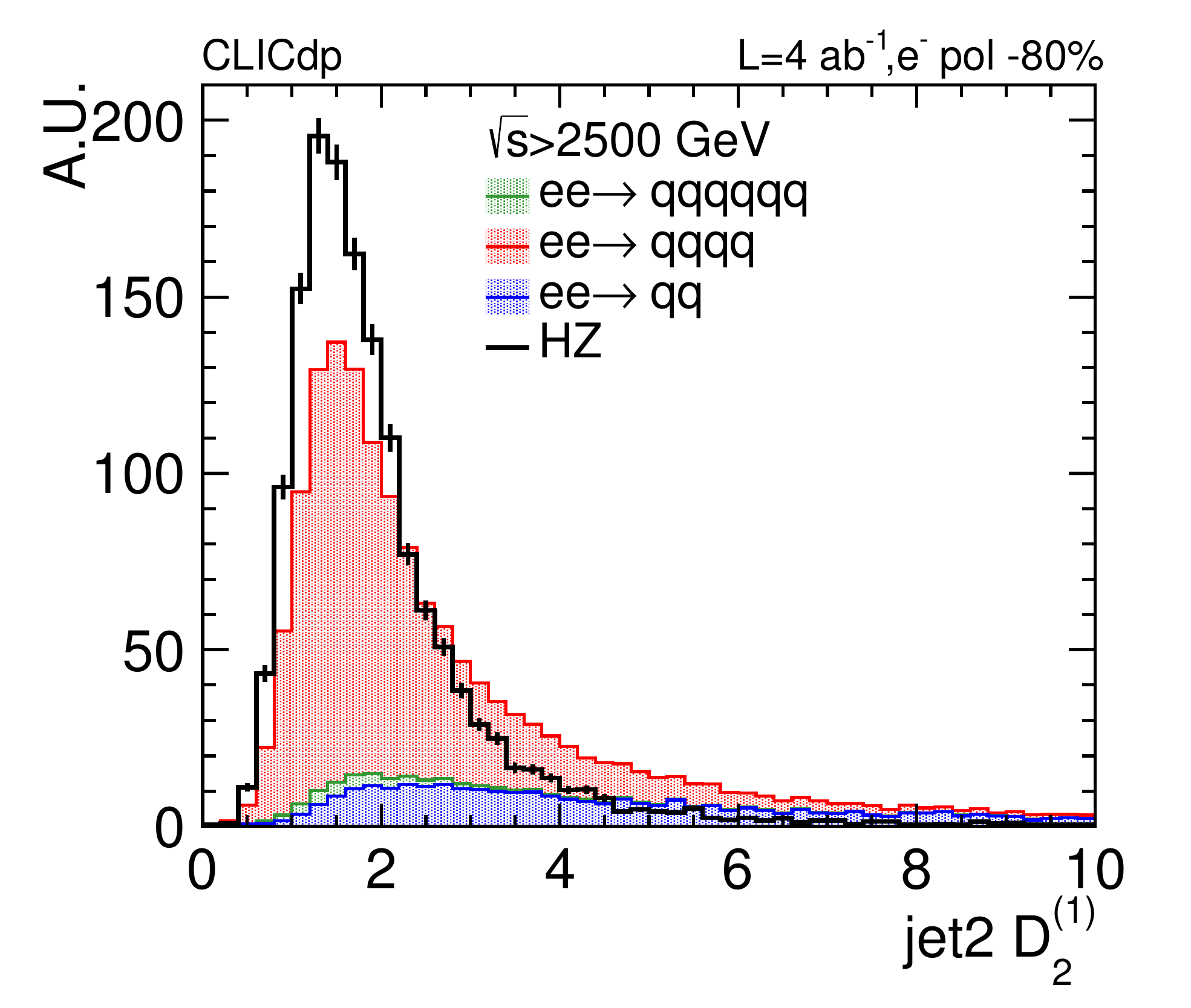}
\end{minipage}
\caption{The jet energy correlation ratio $D_{2}^{(1)}$ for the leading (left) and sub-leading jet (right) for signal and background events with negative electron beam polarisation.}
\label{Fig:discrimination5}
\end{figure}

Figures \ref{Fig:discrimination1}-\ref{Fig:discrimination5} display the most discriminating distributions after preselection for events with negative electron beam polarisation. Analogous plots for events with positive electron beam polarisation can be found in appendix~\ref{sec:polp80_plots}. The jet polar angle distributions of four quark and six-quark events after the preselection are peaked more forward, while for \zhsm both jet polar angle distributions are peaked in the central region of the detector around $\theta$ of \SI{90}{^{\circ}}. While the signal has a high content of b-jets, the backgrounds are largely dominated by light flavour jets. Substructure variables are particularly powerful for discriminating between the one-prong structure of jets from $\epem\rightarrow\qqbar$ and the two-prong structure of jets from $\epem\rightarrow\zhsm$. 

The background and signal discrimination is achieved using boosted decision trees as implemented in the Toolkit for MultiVariate data Analysis (TMVA)~\cite{Hocker:2007ht}, integrated into ROOT~\cite{Antcheva:2009zz}. Using adaptive boosting leads to slightly better results than using gradient boosting. In both cases the best results are achieved using the Gini-index as separation criteria. BDTs are trained separately for both polarisation states, restricting the signal to $\PH\rightarrow\bb$ decays. Training the BDT on all full-hadronic \PH final states does not lead to any significant increase in events compared to using the $\PH\rightarrow\bb$ only trained BDT. The angle $\theta_{1}$ depends on the inner structure of $jet2$, particularly the jet substructure variables based on jet energy correlations. Including these substructure variables in the BDT leads to a slight bias for the asymmetry distribution ${A}_{\theta_{1}}$ to larger negative values. Excluding these variables from the BDT training removes this bias with reconstructed ${A}_{\theta_{1}}$ value very close to the parton level prediction, but at the cost of a larger amount of background, particularly from the di-quark dataset. Including substructure from the second jet leads to very similar ${A}_{\theta_{1}}$ values for background and signal events. Removing these variables from the training leads to a considerable shift in ${A}_{\theta_{1}}$ to lower negative values for background events.

\begin{figure}[htbp!]
\centering
\begin{minipage}[l]{0.49\textwidth}
\includegraphics[width=1.0\textwidth]{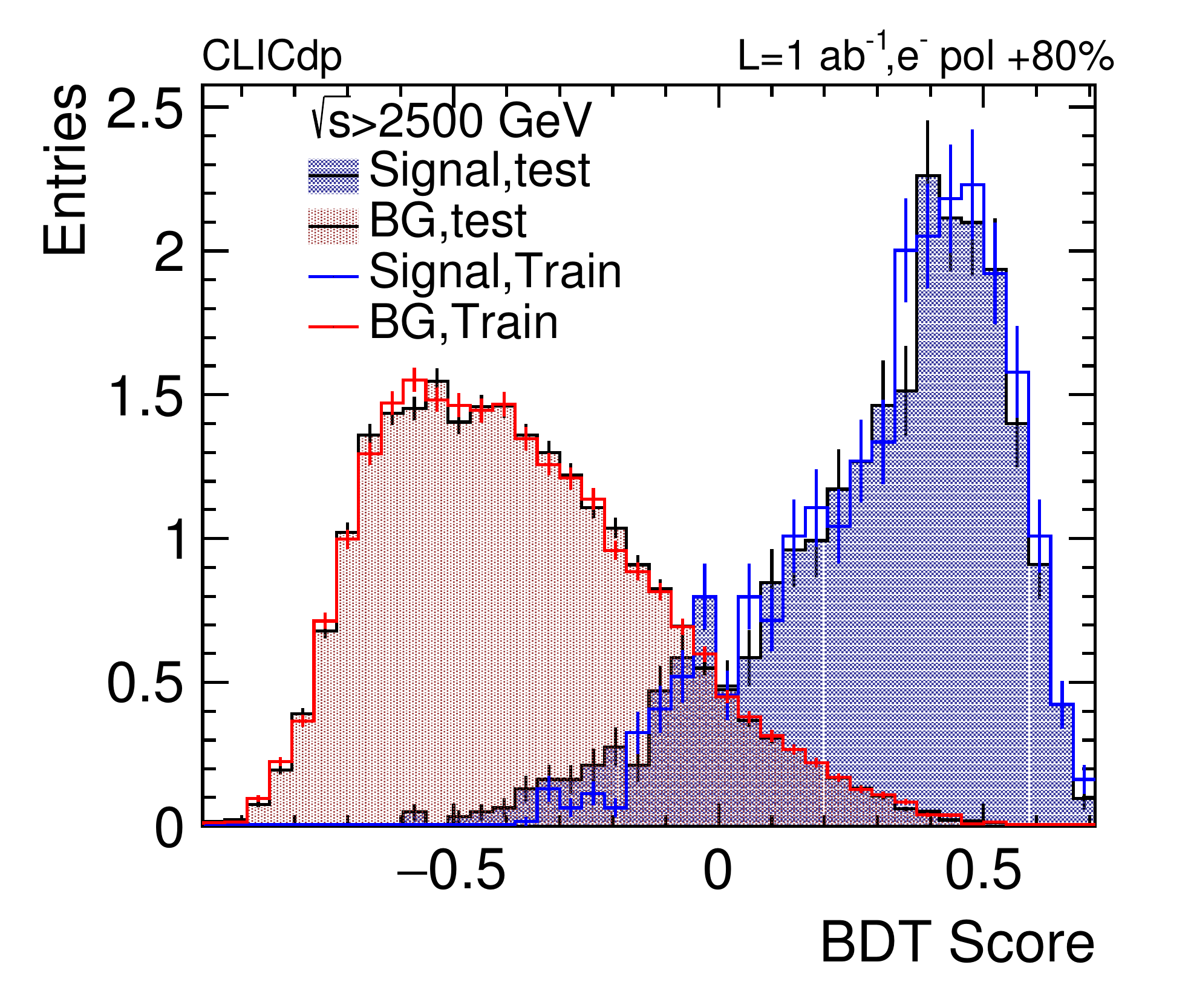}
\end{minipage}
\begin{minipage}[r]{0.49\textwidth}
\includegraphics[width=1.0\textwidth]{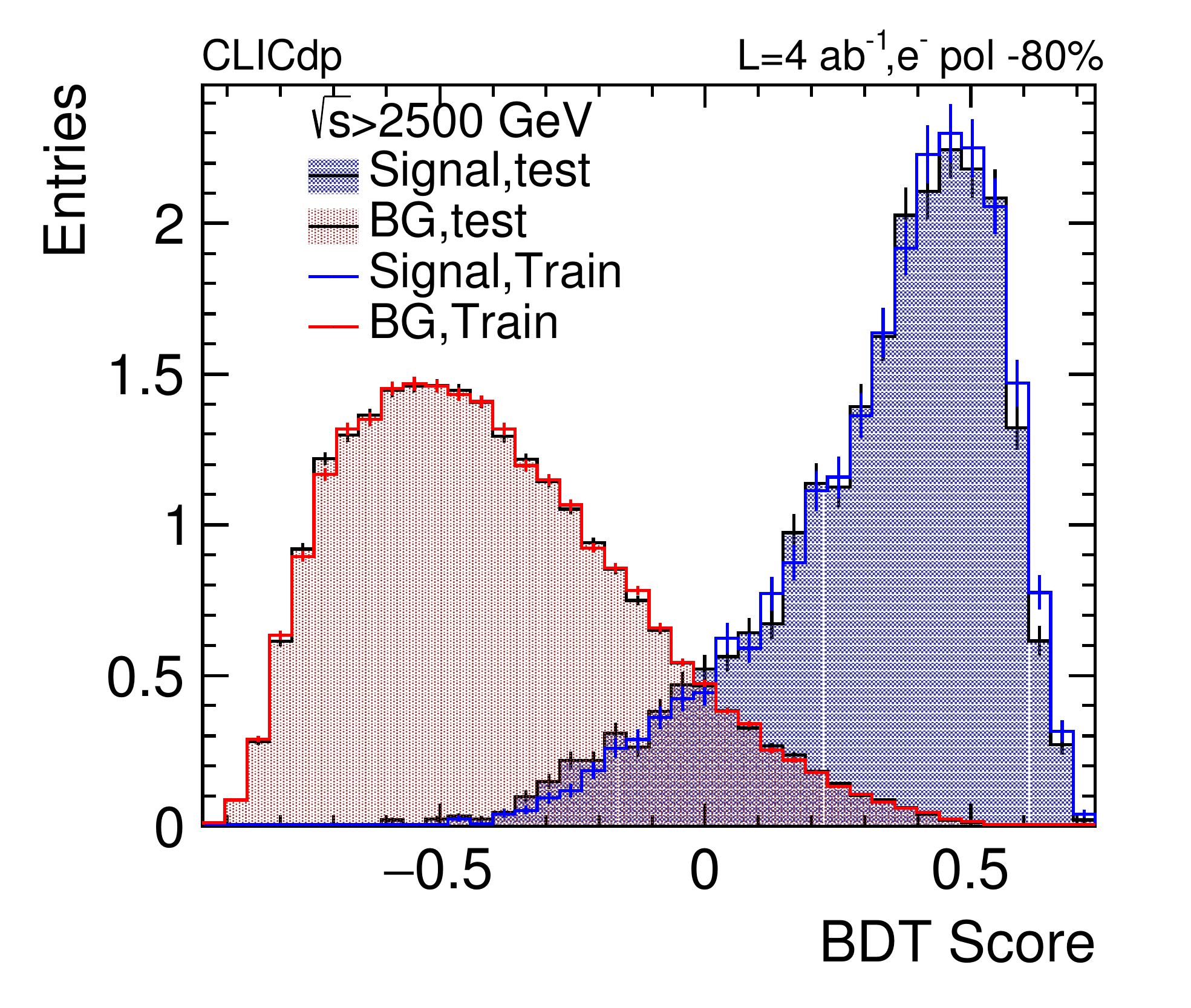}
\end{minipage}
\caption{The distribution of the BDT score for positive (left) and negative (right) electron beam polarisation for the training and testing samples.}
\label{Fig:BDT_overfittests}
\end{figure}

\begin{figure}[htbp!]
\centering
\begin{minipage}[l]{0.49\textwidth}
\includegraphics[width=1.0\textwidth]{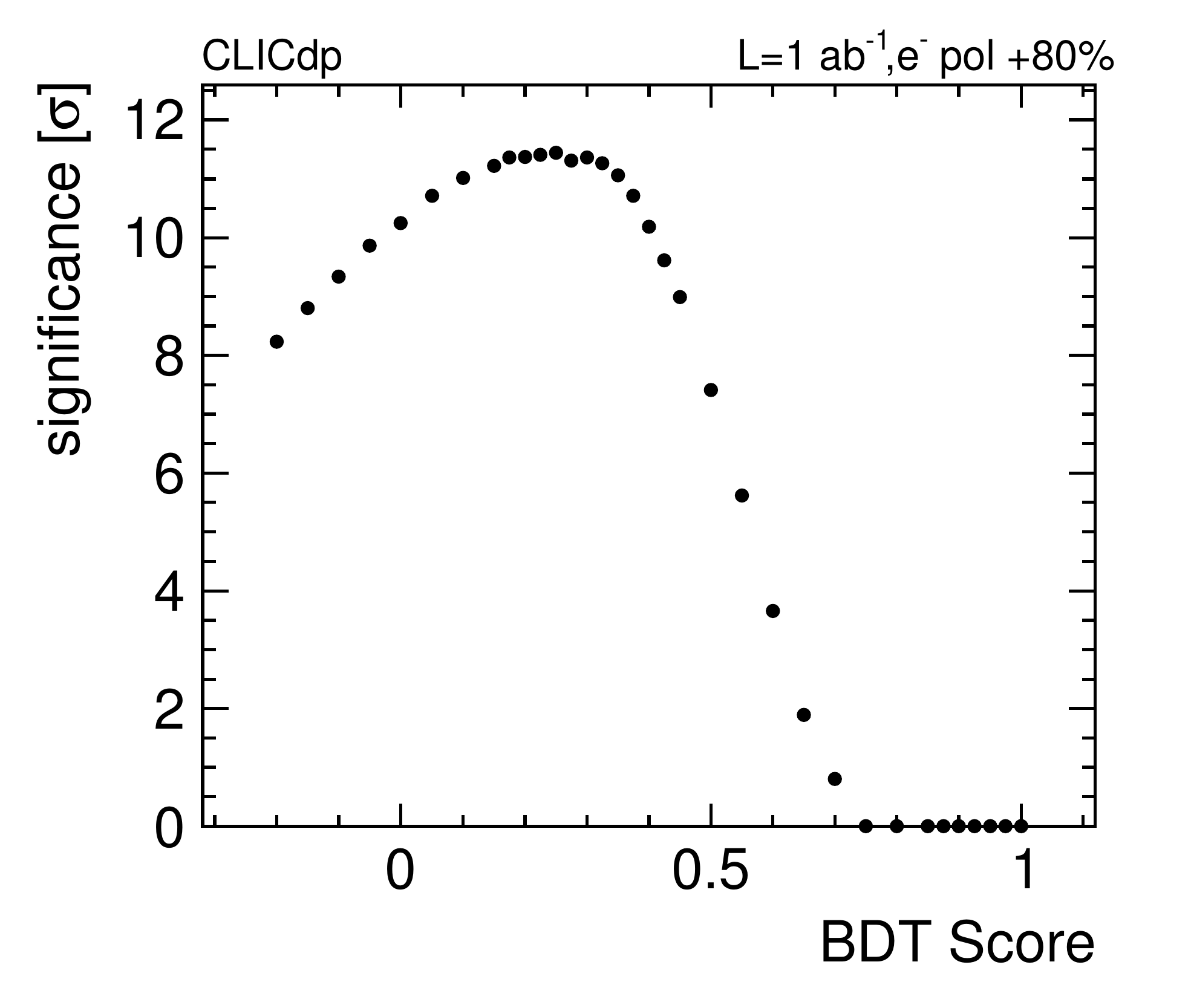}
\end{minipage}
\begin{minipage}[r]{0.49\textwidth}
\includegraphics[width=1.0\textwidth]{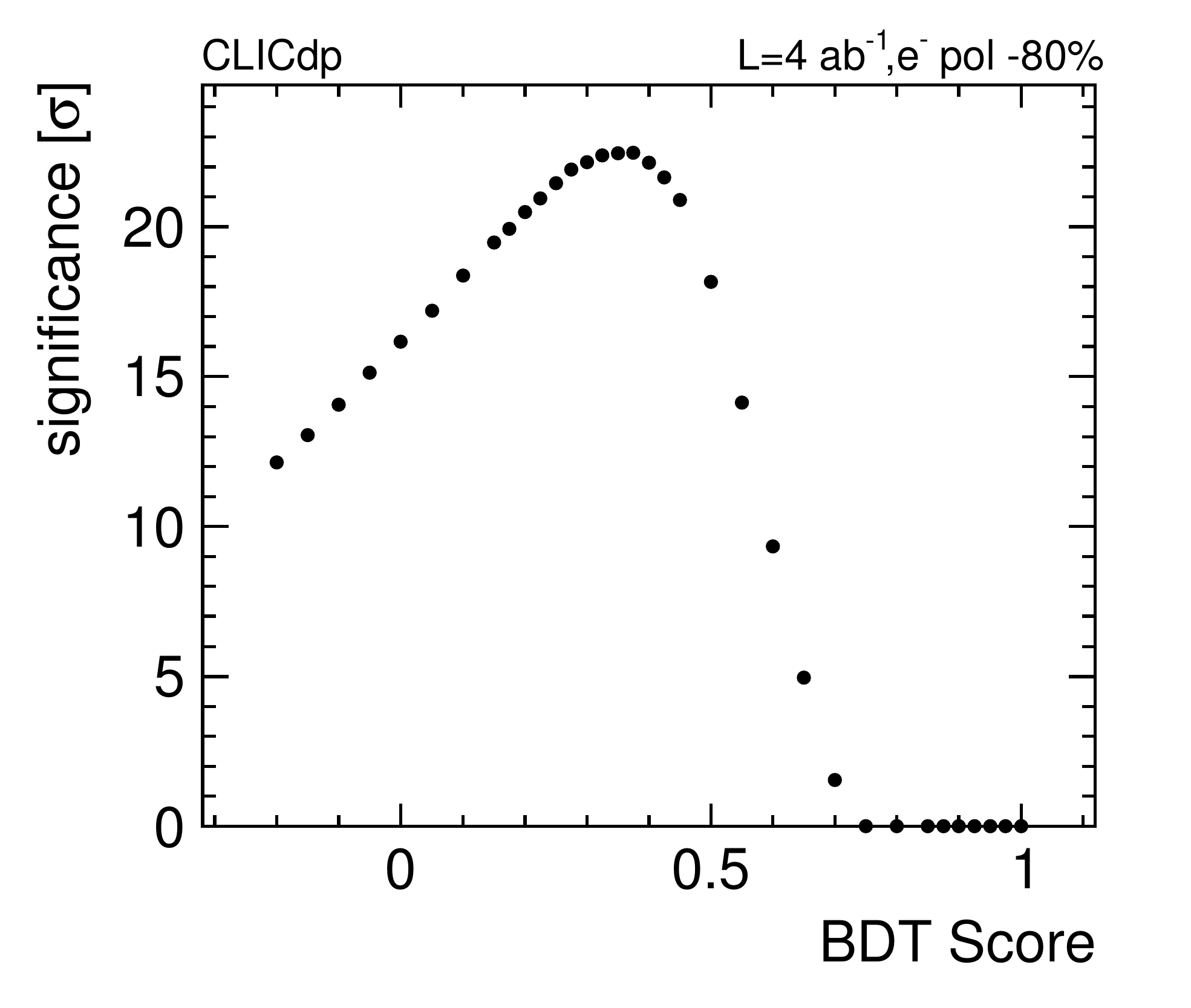}
\end{minipage}
\begin{minipage}[l]{0.49\textwidth}
\includegraphics[width=1.0\textwidth]{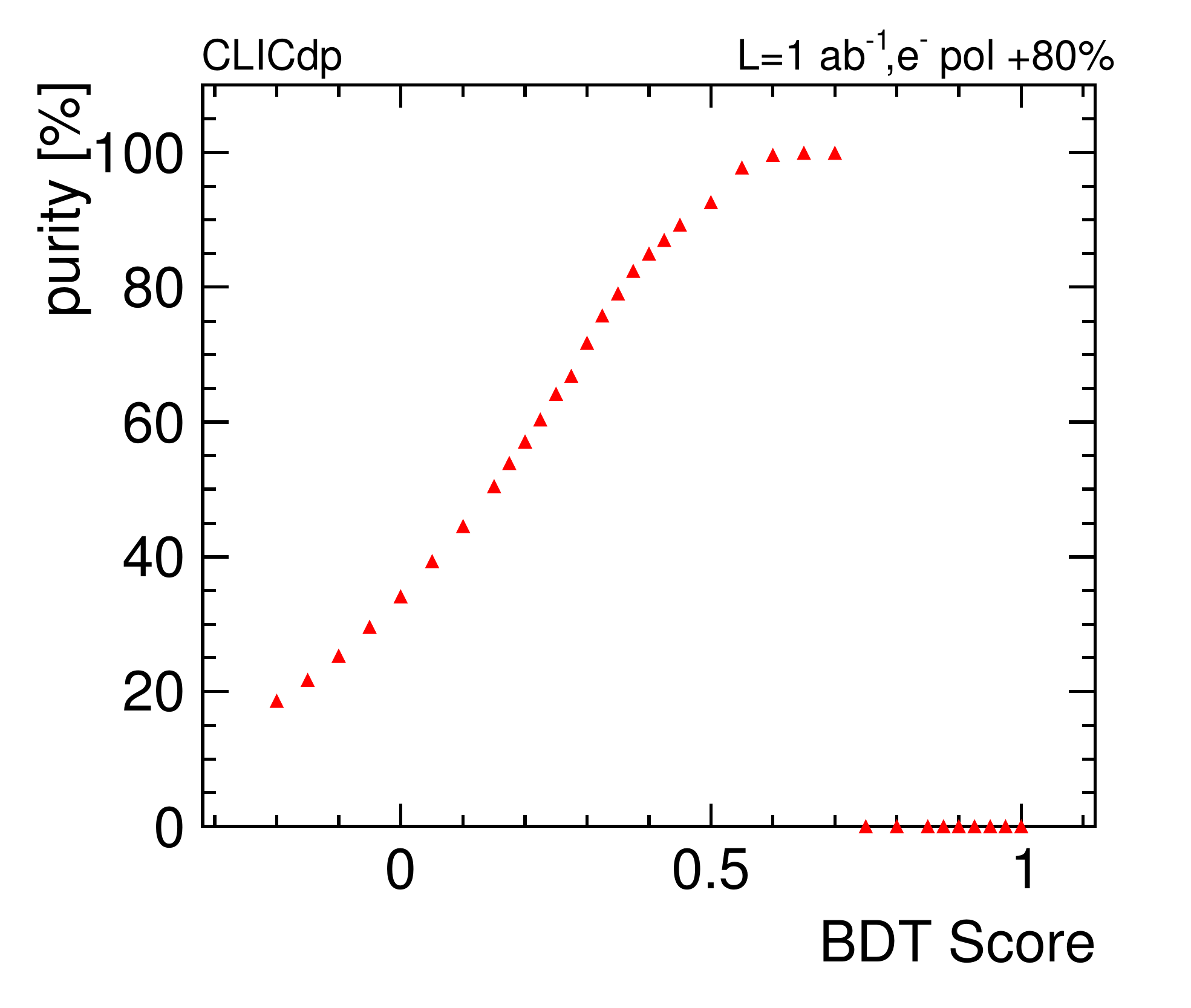}
\end{minipage}
\begin{minipage}[r]{0.49\textwidth}
\includegraphics[width=1.0\textwidth]{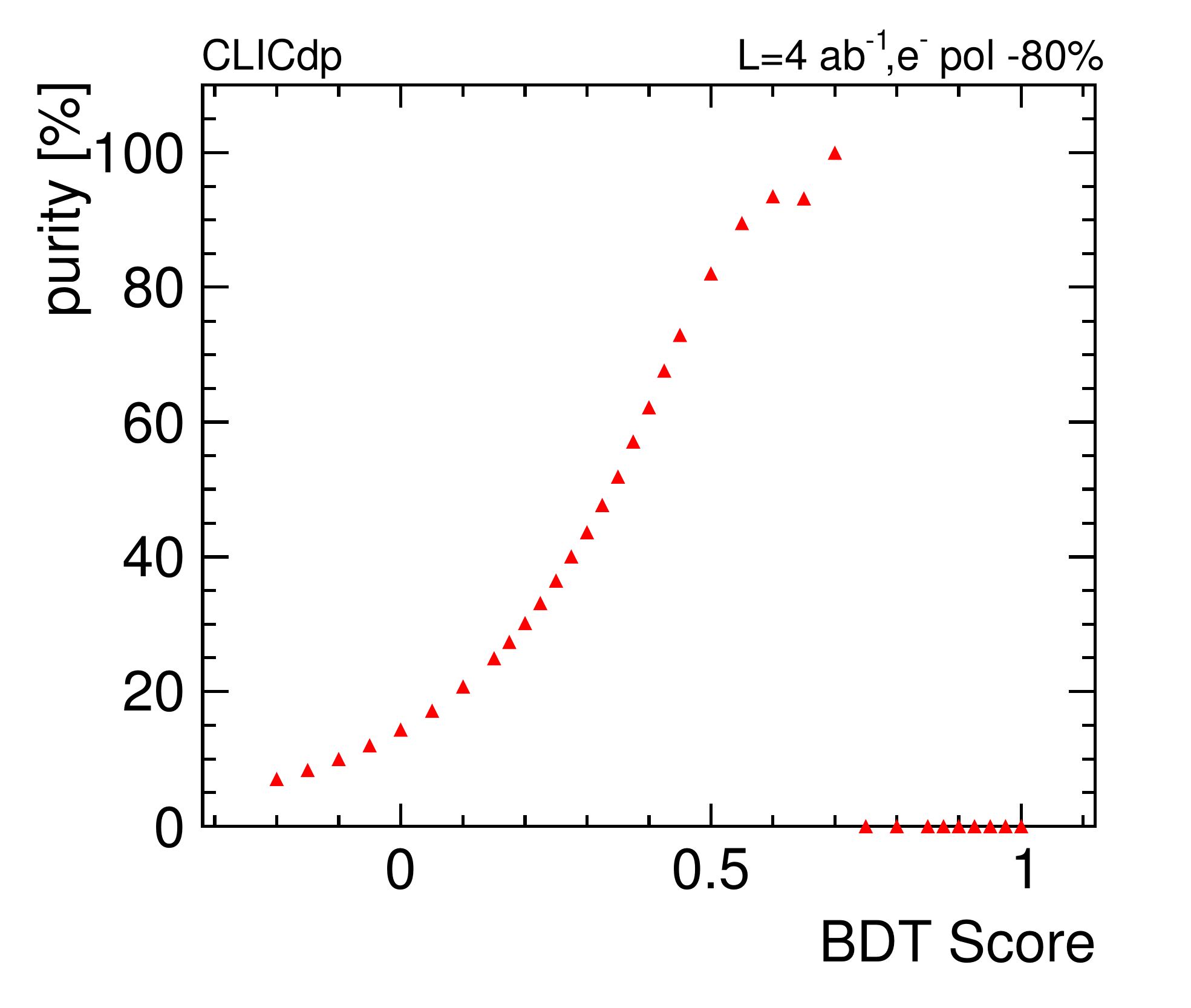}
\end{minipage}
\caption{The signal significance as function of the BDT score for positive (upper left) and negative (upper right) electron beam polarisation as well as the corresponding signal purities.}
\label{Fig:BDTScore}
\end{figure}

Figure~\ref{Fig:BDT_overfittests} shows the good agreement between the BDT score of events used to train the BDT and the other half of events, the testing dataset, for both polarisations. No sign of overtraining is observed. The significance of the \zhsm signal as well as the purity of the selection are shown as a function of the BDT score in Fig.~\ref{Fig:BDTScore} both for negative and positive polarisation final states. 

\section{Total cross section and angular distributions}
\label{sec:DifferentialDistributions}

The final event numbers are listed in Table~\ref{Tab:finalNumbers}. For events with negative electron beam polarisation a cut on the BDT score of $\mathrm{BDT}>0.375$ achieves for all \PH decays a significance of about 22.5 with a purity of 57\%. For events with positive beam polarisation data, a cut on the BDT score of $\mathrm{BDT}>0.30$ achieves a significance of about 11.4, with a slightly higher purity of 72\%. These numbers are equivalent to statistical precisions of about 4.4\%, and 8.8\% respectively. The higher purity is a result of the relatively lower background rates for positive electron beam polarisation, particularly from four-quark samples. Beyond $\zhsm\rightarrow\bb$, the selected other channels are \PH decays into \ww, \tptm, \gluglu, and to a lesser extend into \cc. The combined uncertainty of 4.0\% is in agreement with earlier estimates from 
a fast simulation study~\cite{Ellis:2017kfi}.

\begin{table}[hbtp]
 \centering
 \caption{Event numbers with purities and selection efficiencies after the final selection\label{Tab:finalNumbers} for signal and background events, assuming an integrated luminosity of $\mathrm{L}=\SI{4}{\abinv}$ for runs with negative polarisation P(\Pem)=-80\%, and $\mathrm{L}=\SI{1}{\abinv}$ for runs with positive polarisation P(\Pem)=+80\%. All numbers are given for \mbox{$\roots>\SI{2500}{GeV}$}:}
 \begin{tabular}{|c|c|c|c|c|c|c|}
\hline
process & Events & Purity & Efficiency &  Events & Purity & Efficiency  \\
  & neg. p.  & neg. p., in [\%] & neg. p., in [\%] &  pos. p. & pos. p., in [\%] & pos. p., in [\%]\\
\hline
$\epem\rightarrow\zhsm$, $\PH\rightarrow\bb$ & 811 & 52 & 47 & 162 & 64 & 53 \\
$\epem\rightarrow\zhsm$, all \PH & 884 & 57 & 34 & 180 & 72 & 39 \\
$\epem\rightarrow\qqbar$ & 256 & 17 & 0.15 & 33.7 & 13 & 0.18 \\
$\epem\rightarrow\qqqq$ & 335 & 22  & 0.12 & 30.8 & 12 & 0.36 \\
$\epem\rightarrow\qqqqqq$ & 71.1 & 4.6 & 0.22 & 6.28 & 2.5 & 0.20 \\
\hline
  \end{tabular}
 \end{table}

\begin{figure}[htbp!]
\centering
\begin{minipage}[l]{0.32\textwidth}
\includegraphics[width=1.0\textwidth]{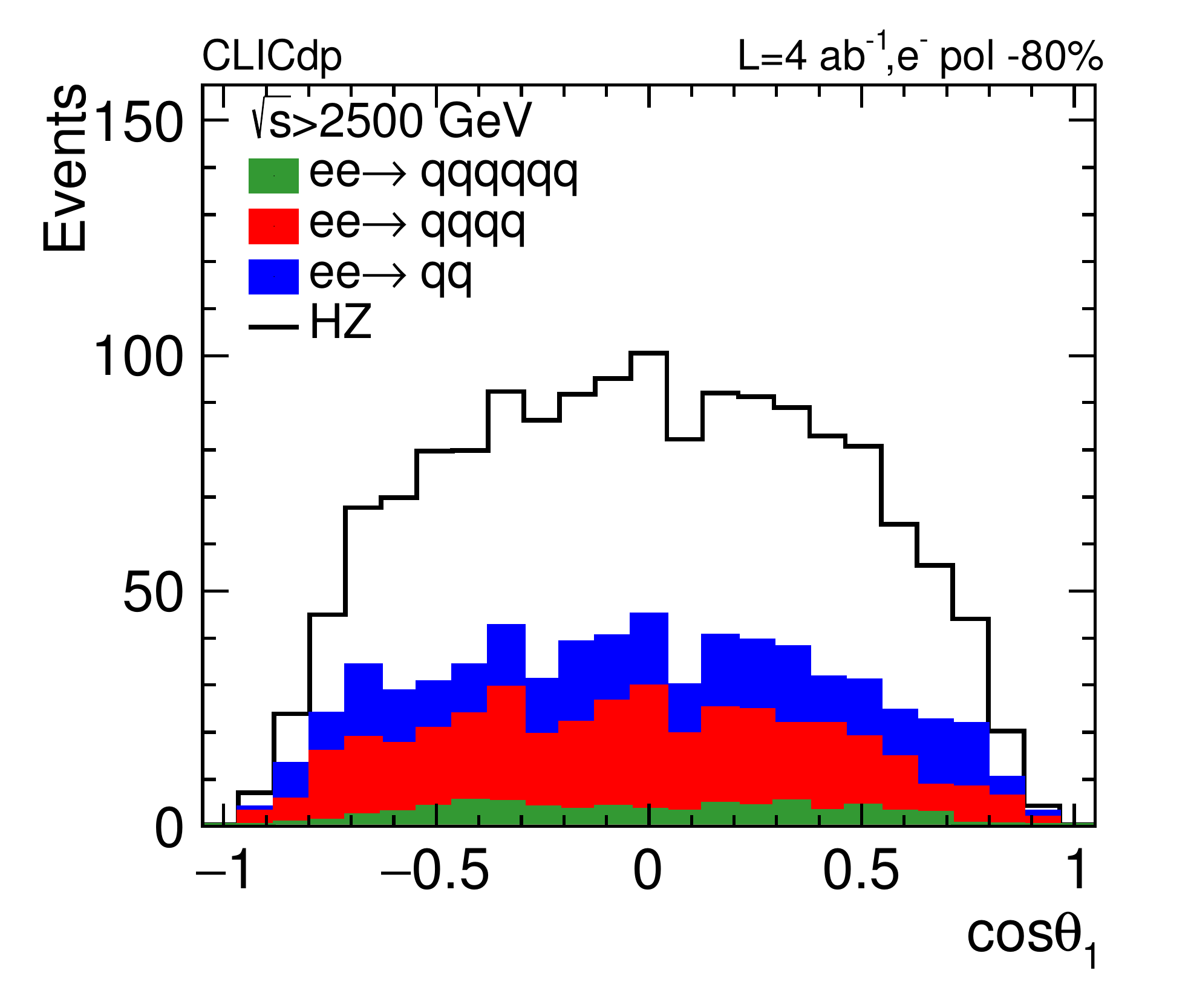}
\end{minipage}
\begin{minipage}[c]{0.32\textwidth}
\includegraphics[width=1.0\textwidth]{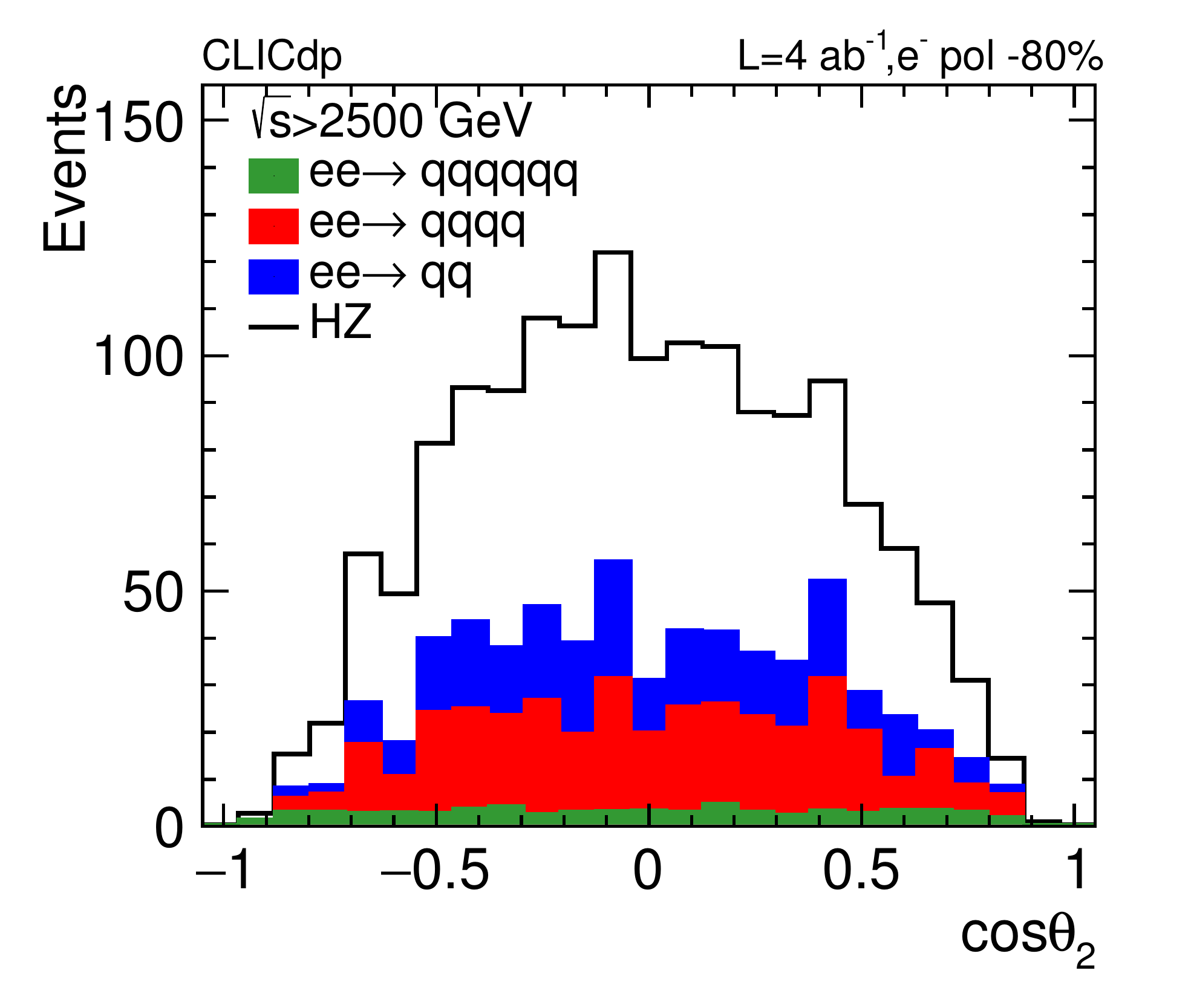}
\end{minipage}
\begin{minipage}[l]{0.32\textwidth}
\includegraphics[width=1.0\textwidth]{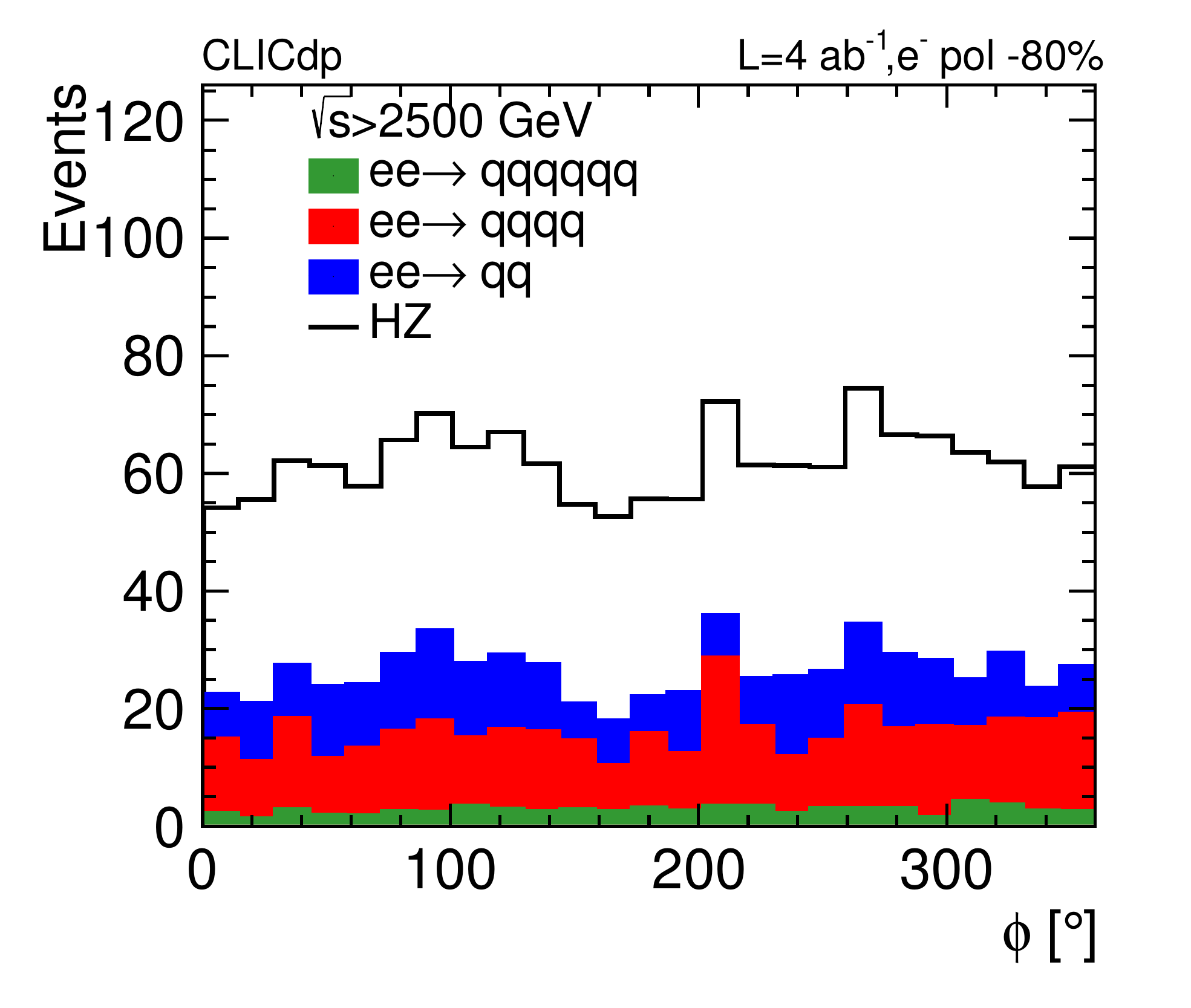}
\end{minipage}
\begin{minipage}[l]{0.32\textwidth}
\includegraphics[width=1.0\textwidth]{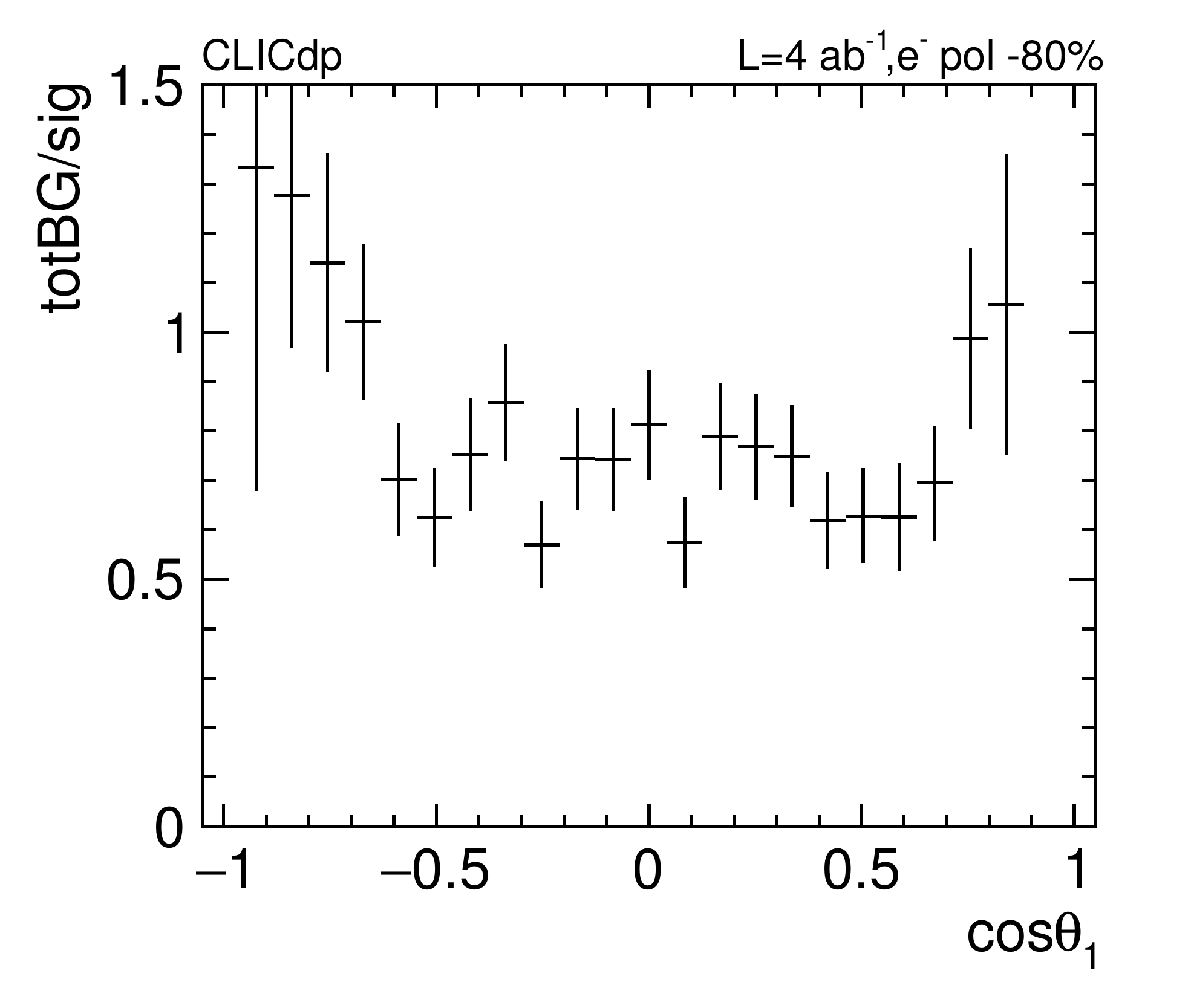}
\end{minipage}
\begin{minipage}[c]{0.32\textwidth}
\includegraphics[width=1.0\textwidth]{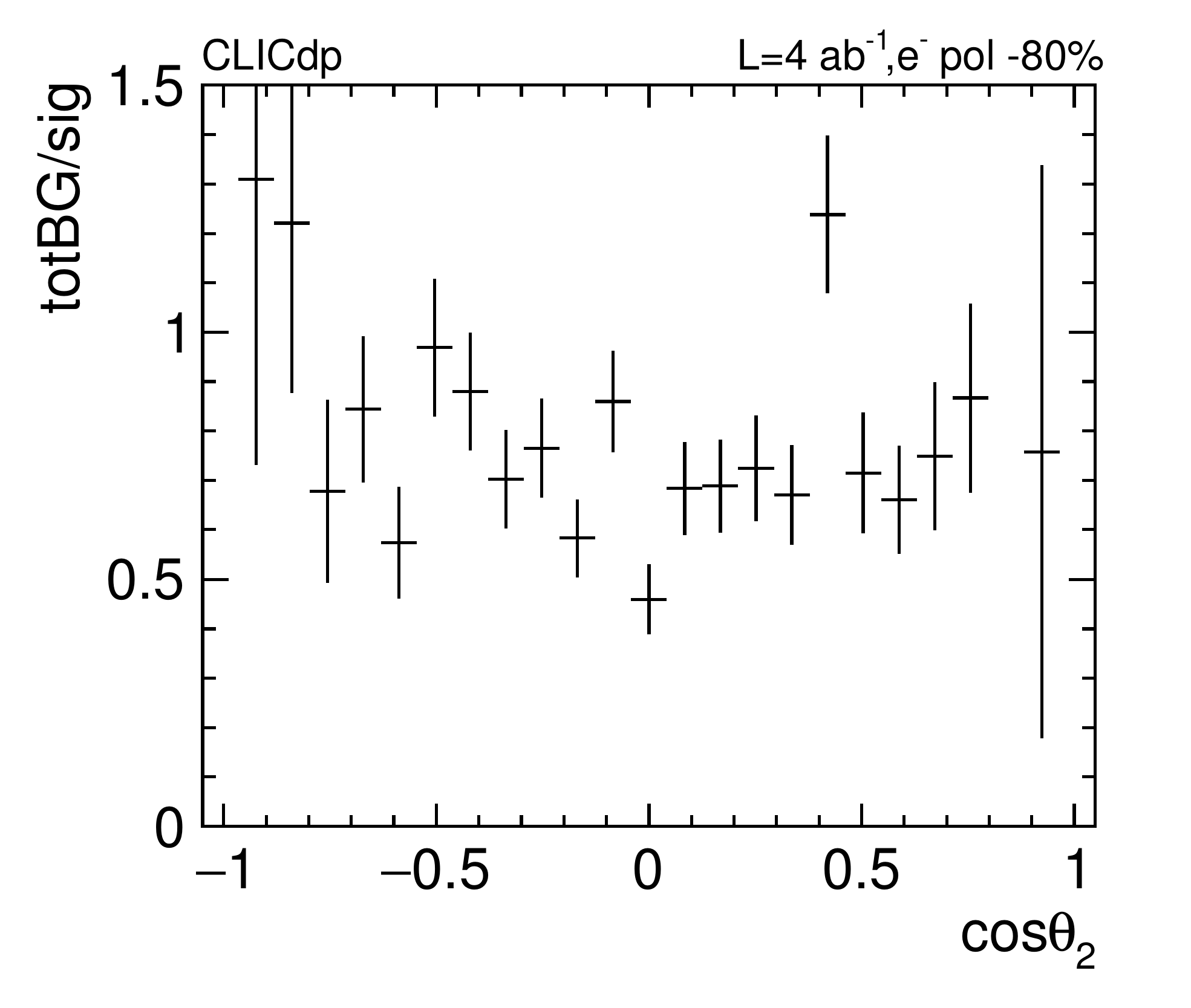}
\end{minipage}
\begin{minipage}[l]{0.32\textwidth}
\includegraphics[width=1.0\textwidth]{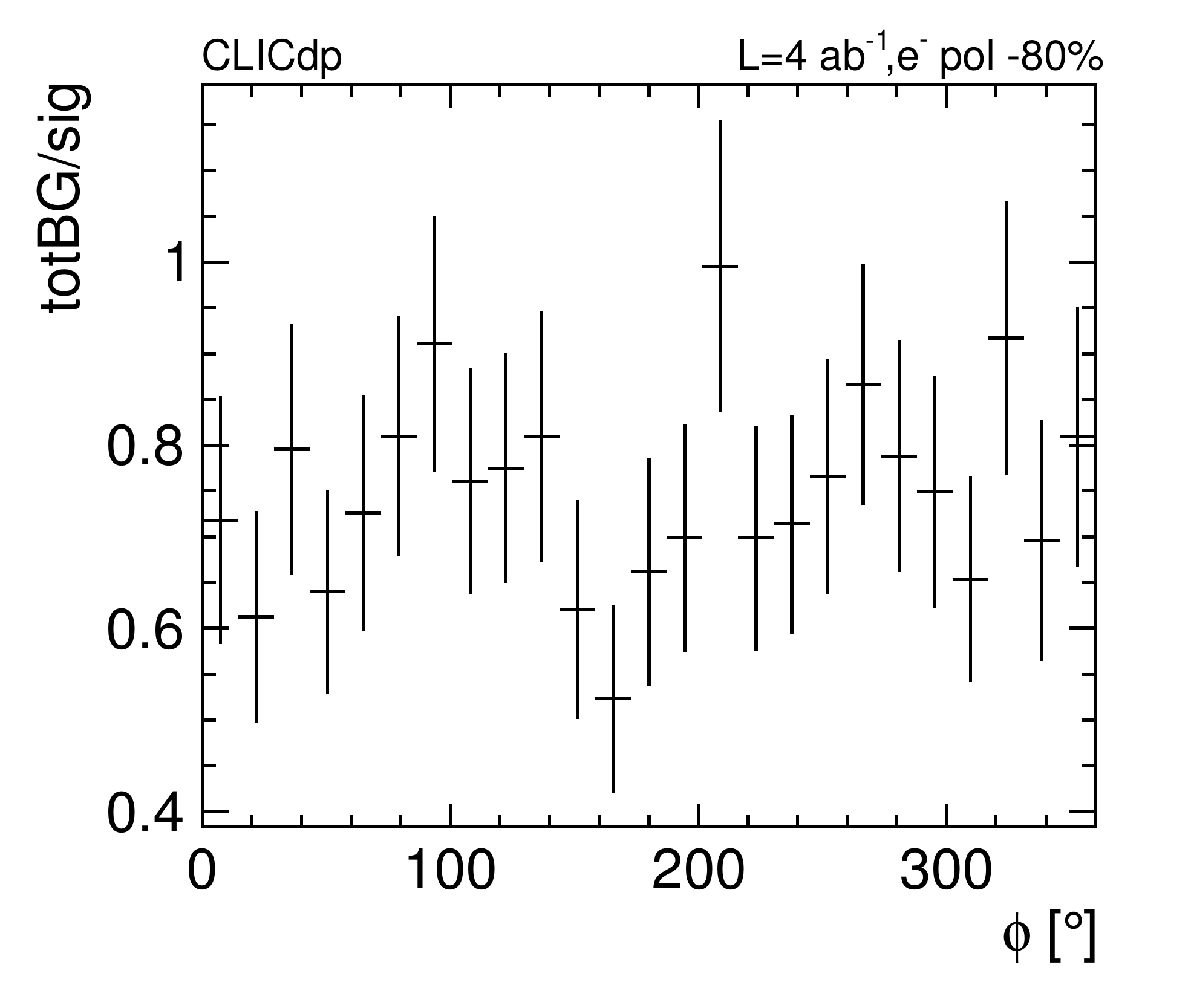}
\end{minipage}
\caption{The three reconstructed angular distributions of $\cos\theta_{1}$ (left), $\cos\theta_{2}$ (centre), and $\phi$ (right) for signal and background events with negative electron beam polarisation. The lower plots show the ratios between background and signal events.}
\label{Fig:angles_signal_background}
\end{figure}

The three reconstructed angular distributions of $\cos\theta_{1}$, $\cos\theta_{2}$, and $\phi$ are shown in Fig.~\ref{Fig:angles_signal_background} for \zhsm signal and background events. The lower plots display the ratio between background and signal events. Background events tend to be rather flat with respect to signal distributions after applying the signal selection. Thus it is expected that the background contamination has a mild effect on the extracted asymmetry values.

Tables \ref{Tab:asymmetries_neg} and \ref{Tab:asymmetries_pos} list the asymmetries with their statistical uncertainties separately for $\zhsm$ events with $\zhsm\rightarrow\bb$, $\zhsm$ with all \PH decays, the asymmetries for the sum of signal and background events, the asymmetries on reconstruction level after mass preselection, and the values for parton-level at large \roots. The asymmetries are considered separately for both polarisation schemes for the electron beam. Most asymmetry values are consistent with 0, the values for ${A}_{\theta_{1}}$ are shifted to more negative values of larger than -0.80, compared to partonic values of around -0.76. This slight bias is introduced by the cuts on the substructure observables of the second jet. The value of ${A}_{\theta_{1}}$ depends on \roots, e.g for the lowest energy stage values of about -0.6 are predicted. The statistical uncertainties of values extracted for events with positive electron beam polarisation is roughly twice as large as the uncertainties for event with negative electron beam polarisation. 

\begin{table}[hbtp]
 \centering
 \caption{\label{Tab:asymmetries_neg}Extracted values for asymmetry observables for signal and background events, assuming an integrated luminosity of $\mathrm{L}=\SI{4}{\abinv}$ for runs with negative polarisation P(\Pem)=-80\%. All numbers are given for \mbox{$\roots>\SI{2500}{GeV}$}:}
 \begin{tabular}{|c|c|c|c|c|c|}
\hline
Asymmetry & $\epem\rightarrow\zhsm$ & $\epem\rightarrow\zhsm$ & Backgrounds & $\epem\rightarrow\zhsm$ and BKG & Parton Level\\
 & $\PH\rightarrow\bb$ &all \PH &  & & \\
\hline
${A}_{c\theta_{1},c\theta_{2}}$ & 0.019$\pm$0.035 & 0.021$\pm$0.034  & 0.028$\pm$0.039 & 0.024$\pm$0.025 & -0.021$\pm$0.019 \\
${A}_{\theta_{1}}$ & -0.834$\pm$0.019 & -0.837$\pm$0.018  & -0.760$\pm$0.025 & -0.804$\pm$0.015 & -0.765$\pm$0.012 \\
${A}^{(1)}_{\phi}$ & -0.002$\pm$0.035 & -0.004$\pm$0.034  & -0.050$\pm$0.039 & -0.024$\pm$0.026 & -0.005$\pm$0.019 \\
${A}^{(2)}_{\phi}$ & -0.014$\pm$0.035 & -0.011$\pm$0.034  & -0.000$\pm$0.039 & -0.006$\pm$0.026 & -0.037$\pm$0.019 \\
${A}^{(3)}_{\phi}$ & -0.001$\pm$0.035 & -0.004$\pm$0.034  & 0.007$\pm$0.039 & 0.001$\pm$0.026 & 0.003$\pm$0.019 \\
${A}^{(4)}_{\phi}$ & -0.036$\pm$0.035 & -0.037$\pm$0.034  & -0.07`$\pm$0.039 & -0.049$\pm$0.026 & -0.015$\pm$0.019 \\
\hline
  \end{tabular}
 \end{table}

\begin{table}[hbtp]
 \centering 
 \caption{\label{Tab:asymmetries_pos}Extracted values for asymmetry observables for signal and background events, assuming an integrated luminosity of $\mathrm{L}=\SI{1}{\abinv}$ for runs with positive polarisation P(\Pem)=+80\%. All numbers are given for \mbox{$\roots>\SI{2500}{GeV}$}:}
 \begin{tabular}{|c|c|c|c|c|c|}
\hline
Asymmetry & $\epem\rightarrow\zhsm$ & $\epem\rightarrow\zhsm$ & Backgrounds & $\epem\rightarrow\zhsm$ and BKG & Parton Level\\
 & $\PH\rightarrow\bb$ &all \PH &  & & \\
\hline
${A}_{c\theta_{1},c\theta_{2}}$ & -0.015$\pm$0.079 & -0.019$\pm$0.075  & -0.035$\pm$0.119 & -0.027$\pm$0.063 & -0.021$\pm$0.046 \\
${A}_{\theta_{1}}$ & -0.829$\pm$0.044 & -0.835$\pm$0.041  & -0.713$\pm$0.083 & -0.801$\pm$0.038 & -0.761$\pm$0.030 \\
${A}^{(1)}_{\phi}$ & 0.015$\pm$0.079 & 0.009$\pm$0.075  & -0.102$\pm$0.118 & -0.022$\pm$0.063 & 0.021$\pm$0.046 \\
${A}^{(2)}_{\phi}$ & 0.011$\pm$0.079 & 0.007$\pm$0.075  & -0.011$\pm$0.119 & 0.002$\pm$0.063 & 0.051$\pm$0.046 \\
${A}^{(3)}_{\phi}$ & 0.059$\pm$0.079 & 0.058$\pm$0.074  & 0.014$\pm$0.119 & 0.046$\pm$0.063 & 0.041$\pm$0.046 \\
${A}^{(4)}_{\phi}$ & -0.053$\pm$0.079 & -0.039$\pm$0.074  & -0.015$\pm$0.119 & -0.037$\pm$0.063 & 0.004$\pm$0.046 \\
\hline
  \end{tabular}
 \end{table}

\section{Systematic uncertainties}
\label{sec:systematics}

Systematic uncertainties are hard to estimate without a detailed knowledge on the precise technical details of the detector. At this stage the impact of sources of potential systematic uncertainties of relevance for this final state is discussed. The total luminosity is expected to be measured to an accuracy of a few per mille with the luminometer of CLICdet, using Bhabha scattering events~\cite{Lukic:2013fw, Bozovic-Jelisavcic:2013aca}. The rate of events in the high-energy \roots region depends on the luminosity spectrum. The relative uncertainty of events with \roots close to the nominal centre-of-mass energy of \SI{3}{TeV} is about 0.2\%~\cite{Poss:2013oea}. The expected uncertainty on the beam polarisation is on the level of 0.2\%~\cite{Wilson:lcws2012}. These three sources lead to negligible effects compared to the statistical uncertainty. An uncertainty on the jet energy scale of 1\% leads to a systematic uncertainty of 0.08\% (0.2\%) for events with negative (positive) electron beam polarisation. The uncertainty on the b-tagging shape is estimated by reweighting b-tagging values with linear slope, which increases b-tagging values of 0 by 1\% and decreases b-tagging values of 1 by 1\% and vice versa. The reweighted shapes are normalized, keeping the number of events after the preselection on the jet masses constant. This leads to a systematic uncertainty of about 0.9\% on the cross section. Both the jet energy scale uncertainty of 1\% and the b-tagging shape uncertainty lead to negligible effects on the errors of the asymmetries and the value of ${A}_{\theta_{1}}$. 

\section{Summary and conclusions}
\label{sec:summary}

A study of Higgs production from \zhsm with hadronic Z decays has been presented at high centre-of-mass energies at the \SI{3}{TeV} energy stage of CLIC. At these high energies hadronically decaying boosted \PH and \PZ bosons are identified using boosted jets and jet substructure variables. This is the first analysis with the new detector model CLICdet and the new software. The analysis profits from the excellent jet energy and mass reconstruction of CLICdet based on high granularity calorimeters optimised for particle flow algorithms. The statistical uncertainty on the cross section of all-hadronic high-energy \zhsm production is about 4.4\% for events with negative electron beam polarisation, and 8.8\% for events with positive electron beam polarisation, and a combined uncertainty of 4.0\%. Angular asymmetries, which are particularly sensitive to deviations from the Standard Model in \zhsm production, have been extracted on reconstructed level. For most observables signal and background events are shape-wise similar. Backgrounds tend to produce more forward events for the $\cos{\theta_{1}}$ distribution. This counteracts the bias of the event selection on the signal, which is more efficient for central $\cos{\theta_{1}}$ values.

\section*{Acknowledgements}
This work benefited from services provided by the ILC Virtual Organisation, supported by the national resource providers of the EGI Federation. This research was done using resources provided by the Open Science Grid, which is supported by the National Science Foundation and the U.S. Department of Energy's Office of Science.
This project has received funding from the European Union's Horizon 2020 Research and Innovation programme under Grant Agreement no. 654168.


\printbibliography[title=References]

@article{CLICdet_note_2017,   
      author =       {Alipour Tehrani, N. and others},  
      title  =        {CLICdet: The post-CDR CLIC detector model},   
      year   =         {2017},   
      month  =        {mar},   
      note   =         {CLICdp-Note-2017-001},   
      url    =          {https://cds.cern.ch/record/2254048} 
}

@article{Hocker:2007ht,
	author 	=	"Hoecker, Andreas and Speckmayer, Peter and Stelzer, Joerg and Therhaag, Jan and von Toerne, Eckhard and Voss, Helge",
	title 	 = 	"{TMVA: Toolkit for Multivariate Data Analysis}",
	journal 	 = 	"PoS",
	volume 	 = 	"ACAT",
	year 	 = 	"2007",
	pages 	 = 	"040",
	eprint 	 = 	"physics/0703039",
	archivePrefix 	 = 	"arXiv",
	SLACcitation 	 = 	"%%CITATION = PHYSICS/0703039;%%"}

@article{Boronat:2016tgd,
      author         = "Boronat, M. and Fuster, J. and Garcia, I. and Roloff, Ph.
                        and Simoniello, R. and Vos, M.",
      title          = "{Jet reconstruction at high-energy electron-positron
                        colliders}",
      journal        = "Eur. Phys. J.",
      volume         = "C78",
      year           = "2018",
      number         = "2",
      pages          = "144",
      doi            = "10.1140/epjc/s10052-018-5594-6",
      eprint         = "1607.05039",
      archivePrefix  = "arXiv",
      primaryClass   = "hep-ex",
      reportNumber   = "CLICDP-PUB-2017-002",
      SLACcitation   = "%%CITATION = ARXIV:1607.05039;%%"
}

@article{Lukic:2013fw,
      author         = "Luki\'{c}, Strahinja and Bo\v{z}ovi\'{c}-Jelisav\v{c}i\'{c}, I. and
                        Pandurovi\'{c}, M. and Smiljani\'{c}, I.",
      title          = "{Correction of beam-beam effects in luminosity
                        measurement in the forward region at CLIC}",
      journal        = "JINST",
      volume         = "8",
      year           = "2013",
      pages          = "P05008",
      doi            = "10.1088/1748-0221/8/05/P05008",
      eprint         = "1301.1449",
      archivePrefix  = "arXiv",
      primaryClass   = "physics.acc-ph",
      reportNumber   = "LCD-NOTE-2012-008",
      SLACcitation   = "%%CITATION = ARXIV:1301.1449;%%"
}

@article{Bozovic-Jelisavcic:2013aca,
      author         = "Bo\v{z}ovi\'{c}-Jelisav\v{c}i\'{c}, I. and Luki\'{c}, S. and
                        Milutinovi\'{c} Dumbelovi\'{c}, G. and Pandurovi\'{c}, M. and
                        Smiljani\'{c}, I.",
      title          = "{Luminosity measurement at ILC}",
      journal        = "JINST",
      volume         = "8",
      year           = "2013",
      pages          = "P08012",
      doi            = "10.1088/1748-0221/8/08/P08012",
      eprint         = "1304.4082",
      archivePrefix  = "arXiv",
      primaryClass   = "physics.acc-ph",
      SLACcitation   = "%%CITATION = ARXIV:1304.4082;%%"
}

@unpublished{Wilson:lcws2012,   
      author = "Wilson, G.W.",   
      title = "{Beam Polarization Measurement Using Single Bosons with Missing Energy}",   
      year = "2012",   
      note = "International Workshop on Future Linear Colliders (LCWS12)" 
}

@article{Antcheva:2009zz,
      author         = "Antcheva, I. and others",
      title          = "{ROOT: A C++ framework for petabyte data storage,
                        statistical analysis and visualization}",
      journal        = "Comput. Phys. Commun.",
      volume         = "180",
      year           = "2009",
      pages          = "2499-2512",
      doi            = "10.1016/j.cpc.2009.08.005",
      eprint         = "1508.07749",
      archivePrefix  = "arXiv",
      primaryClass   = "physics.data-an",
      reportNumber   = "FERMILAB-PUB-09-661-CD",
      SLACcitation   = "%%CITATION = ARXIV:1508.07749;%%"
}

@article{Arominski:2018uuz,
      author         = {Arominski, Dominik and others},
      title          = {A detector for CLIC: main parameters and performance},
      collaboration  = {CLICdp Collaboration},
      year           = {2018},
      eprint         = {1812.07337},
      archivePrefix  = {arXiv},
      primaryClass   = {physics.ins-det},
      reportNumber   = {CLICdp-Note-2018-005},
      SLACcitation   = {%%CITATION = ARXIV:1812.07337;%%}
}

@article{Sirunyan:2017tyr,
      author         = "Sirunyan, Albert M and others",
      title          = "{Measurements of jet charge with dijet events in pp
                        collisions at $\sqrt{s}=8$ TeV}",
      collaboration  = "CMS Collaboration",
      journal        = "JHEP",
      volume         = "10",
      year           = "2017",
      pages          = "131",
      doi            = "10.1007/JHEP10(2017)131",
      eprint         = "1706.05868",
      archivePrefix  = "arXiv",
      primaryClass   = "hep-ex",
      reportNumber   = "CMS-SMP-15-003, CERN-EP-2017-085",
      SLACcitation   = "%%CITATION = ARXIV:1706.05868;%%"
}

@article{Aad:2015cua,
      author         = "Aad, Georges and others",
      title          = "{Measurement of jet charge in dijet events from
                        $\sqrt{s}$=8  TeV pp collisions with the ATLAS detector}",
      collaboration  = "ATLAS Collaboration",
      journal        = "Phys. Rev.",
      volume         = "D93",
      year           = "2016",
      number         = "5",
      pages          = "052003",
      doi            = "10.1103/PhysRevD.93.052003",
      eprint         = "1509.05190",
      archivePrefix  = "arXiv",
      primaryClass   = "hep-ex",
      reportNumber   = "CERN-PH-EP-2015-207",
      SLACcitation   = "%%CITATION = ARXIV:1509.05190;%%"
}

@article{Larkoski:2013eya,
      author         = "Larkoski, Andrew J. and Salam, Gavin P. and Thaler,
                        Jesse",
      title          = "{Energy Correlation Functions for Jet Substructure}",
      journal        = "JHEP",
      volume         = "06",
      year           = "2013",
      pages          = "108",
      doi            = "10.1007/JHEP06(2013)108",
      eprint         = "1305.0007",
      archivePrefix  = "arXiv",
      primaryClass   = "hep-ph",
      reportNumber   = "MIT-CTP-4446, CERN-PH-TH-2013-066, LPN13-026",
      SLACcitation   = "%%CITATION = ARXIV:1305.0007;%%"
}

@article{Larkoski:2014gra,
      author         = "Larkoski, Andrew J. and Moult, Ian and Neill, Duff",
      title          = "{Power Counting to Better Jet Observables}",
      journal        = "JHEP",
      volume         = "12",
      year           = "2014",
      pages          = "009",
      doi            = "10.1007/JHEP12(2014)009",
      eprint         = "1409.6298",
      archivePrefix  = "arXiv",
      primaryClass   = "hep-ph",
      reportNumber   = "MIT--CTP-4588",
      SLACcitation   = "%%CITATION = ARXIV:1409.6298;%%"
}

@ARTICLE{Tran:2017tgrSoftwareCompensation,
  author = {Tran, H.L. and Kr{\"u}ger, K. and Sefkow, F. and Green, S. and Marshall,
	J.S and Thomson,M.A. and Simon, F.},
  title = {Software compensation in Particle Flow reconstruction},
  journal = {Eur. Phys. J.},
  year = {2017},
  volume = {C77},
  pages = {698},
  number = {10},
  archiveprefix = {arXiv},
  doi = {10.1140/epjc/s10052-017-5298-3},
  eprint = {1705.10363},
  primaryclass = {physics.ins-det},
  reportnumber = {DESY-17-083, MPP-2017-98},
  slaccitation = {%%CITATION = ARXIV:1705.10363;%%}
}

@article{deBlas:2018mhx,
      author         = "de Blas, J. and others",
      title          = "{The CLIC Potential for New Physics}",
      year           = "2018",
      doi            = "10.23731/CYRM-2018-003",
      eprint         = "1812.02093",
      archivePrefix  = "arXiv",
      primaryClass   = "hep-ph",
      reportNumber   = "CERN-TH-2018-267, CERN-2018-009-M",
      SLACcitation   = "%%CITATION = ARXIV:1812.02093;%%"
}

@article{deBlas:2019wgy,
      author         = "De Blas, Jorge and Durieux, Gauthier and Grojean,
                        Christophe and Gu, Jiayin and Paul, Ayan",
      title          = "{On the future of Higgs, electroweak and diboson
                        measurements at lepton colliders}",
      year           = "2019",
      eprint         = "1907.04311",
      archivePrefix  = "arXiv",
      primaryClass   = "hep-ph",
      reportNumber   = "DESY-19-077, MITP/19-028",
      SLACcitation   = "%%CITATION = ARXIV:1907.04311;%%"
}

@article{deBlas:2019rxi,
      author         = "de Blas, J. and others",
      title          = "{Higgs Boson Studies at Future Particle Colliders}",
      year           = "2019",
      eprint         = "1905.03764",
      archivePrefix  = "arXiv",
      primaryClass   = "hep-ph",
      SLACcitation   = "%%CITATION = ARXIV:1905.03764;%%"
}

@article{Beneke:2014sba,
      author         = "Beneke, Martin and Boito, Diogo and Wang, Yu-Ming",
      title          = "{Anomalous Higgs couplings in angular asymmetries of $H
                        \to Z\ell^{+} \ell^{-}$ and e$^{+}$ e$^{-} \to HZ$}",
      journal        = "JHEP",
      volume         = "11",
      year           = "2014",
      pages          = "028",
      doi            = "10.1007/JHEP11(2014)028",
      eprint         = "1406.1361",
      archivePrefix  = "arXiv",
      primaryClass   = "hep-ph",
      reportNumber   = "TUM-HEP-849-14, TTK-14-11, SFB-CPP-14-28, TUM-HEP-949-14",
      SLACcitation   = "%%CITATION = ARXIV:1406.1361;%%"
}

@article{Aicheler:2019dhf,
      author         = "Aicheler, M. and Burrows, P. N. and Catalan Lasheras, N.
                        and Corsini, R. and Draper, M. and Osborne, J. and
                        Schulte, D. and Stapnes, S. and Stuart, M. J.",
      title          = "{The Compact Linear Collider (CLIC) - Project
                        Implementation Plan}",
      collaboration  = "CLIC Collaboration",
      doi            = "10.23731/CYRM-2018-004",
      year           = "2019",
      eprint         = "1903.08655",
      archivePrefix  = "arXiv",
      primaryClass   = "physics.acc-ph",
      reportNumber   = "CERN-2018-010-M",
      SLACcitation   = "%%CITATION = ARXIV:1903.08655;%%"
}

@article{Abramowicz:2016zbo,
      author         = "Abramowicz, H. and others",
      title          = {Higgs Physics at the CLIC electron-positron linear collider},
      journal        = {Eur. Phys. J.},
      volume         = {C77},
      pages          = {475},
      doi            = {10.1140/epjc/s10052-017-4968-5},
      year           = {2017},
      eprint         = "1608.07538",
      archivePrefix  = "arXiv",
      primaryClass   = "hep-ex",
      reportNumber   = "CLICDP-PUB-2016-001",
      SLACcitation   = |%%CITATION = ARXIV:1608.07538;%%"
}
%@article{Abramowicz:2016zbo,
%      author         = "Abramowicz, H. and others",
%      title          = "{Higgs physics at the CLIC electron–positron linear
%                        collider}",
%      journal        = "Eur. Phys. J.",
%      volume         = "C77",
%      year           = "2017",
%      number         = "7",
%      pages          = "475",
%      doi            = "10.1140/epjc/s10052-017-4968-5",
%      eprint         = "1608.07538",
%      archivePrefix  = "arXiv",
%      primaryClass   = "hep-ex",
%      reportNumber   = "CLICDP-PUB-2016-001",
%      SLACcitation   = {%%CITATION = ARXIV:1608.07538;%%}
%}

@article{Ellis:2017kfi,
      author         = "Ellis, John and Roloff, Philipp and Sanz, Veronica and
                        You, Tevong",
      title          = "{Dimension-6 Operator Analysis of the CLIC Sensitivity to
                        New Physics}",
      journal        = "JHEP",
      volume         = "05",
      year           = "2017",
      pages          = "096",
      doi            = "10.1007/JHEP05(2017)096",
      eprint         = "1701.04804",
      archivePrefix  = "arXiv",
      primaryClass   = "hep-ph",
      reportNumber   = "KCL-PH-TH-2017-04, CERN-TH-2017-009, CAVENDISH-HEP-17-01,
                        CERN-PH-TH-2017-009, DAMTP-2017-01",
      SLACcitation   = "%%CITATION = ARXIV:1701.04804;%%"
}

@article{Craig:2015wwr,
      author         = "Craig, Nathaniel and Gu, Jiayin and Liu, Zhen and Wang,
                        Kechen",
      title          = "{Beyond Higgs Couplings: Couplings: Probing the Higgs with Angular Observables at Future e$^{+}$e$^{-}$ Colliders}",
%      title          = "{Beyond Higgs Couplings: Probing the Higgs with Angular
%                        Observables at Future e+e− Colliders}",
      journal        = "JHEP",
      volume         = "03",
      year           = "2016",
      pages          = "050",
      doi            = "10.1007/JHEP03(2016)050",
      eprint         = "1512.06877",
      archivePrefix  = "arXiv",
      primaryClass   = "hep-ph",
      reportNumber   = "FERMILAB-PUB-15-569-T",
      SLACcitation   = "%%CITATION = ARXIV:1512.06877;%%"
}

@article{Durieux:2017rsg,
      author         = "Durieux, Gauthier and Grojean, Christophe and Gu, Jiayin
                        and Wang, Kechen",
      title          = "{The leptonic future of the Higgs}",
      journal        = "JHEP",
      volume         = "09",
      year           = "2017",
      pages          = "014",
      doi            = "10.1007/JHEP09(2017)014",
      eprint         = "1704.02333",
      archivePrefix  = "arXiv",
      primaryClass   = "hep-ph",
      reportNumber   = "DESY-17-018",
      SLACcitation   = "%%CITATION = ARXIV:1704.02333;%%"
}

@article{CLIC:2016zwp,
      author         = "Boland, M J and others",
      editor         = "Lebrun, P and Linssen, L and Schulte, D and Sicking, E
                        and Stapnes, S and Thomson, M A and Burrows, P N",
      title          = "{Updated baseline for a staged Compact Linear Collider}",
      collaboration  = "CLIC and CLICdp Collaborations",
      doi            = "10.5170/CERN-2016-004",
      year           = "2016",
      eprint         = "1608.07537",
      archivePrefix  = "arXiv",
      primaryClass   = "physics.acc-ph",
      reportNumber   = "CERN-2016-004",
      SLACcitation   = "%%CITATION = ARXIV:1608.07537;%%"
}

@article{Thaler:2010tr,
      author         = "Thaler, Jesse and Van Tilburg, Ken",
      title          = "{Identifying Boosted Objects with N-subjettiness}",
      journal        = "JHEP",
      volume         = "03",
      year           = "2011",
      pages          = "015",
      doi            = "10.1007/JHEP03(2011)015",
      eprint         = "1011.2268",
      archivePrefix  = "arXiv",
      primaryClass   = "hep-ph",
      reportNumber   = "MIT-CTP-4191",
      SLACcitation   = "%%CITATION = ARXIV:1011.2268;%%"
}

@article{Suehara:2015ura,
      author         = "Suehara, Taikan and Tanabe, Tomohiko",
      title          = "{LCFIPlus: A Framework for Jet Analysis in Linear
                        Collider Studies}",
      journal        = "Nucl. Instrum. Meth.",
      volume         = "A808",
      year           = "2016",
      pages          = "109-116",
      doi            = "10.1016/j.nima.2015.11.054",
      eprint         = "1506.08371",
      archivePrefix  = "arXiv",
      primaryClass   = "physics.ins-det",
      SLACcitation   = "%%CITATION = ARXIV:1506.08371;%%"
}

@article{Poss:2013oea,
      author         = "Poss, S. and Sailer, A.",
      title          = "{Luminosity Spectrum Reconstruction at Linear Colliders}",
      journal        = "Eur. Phys. J.",
      volume         = "C74",
      year           = "2014",
      number         = "4",
      pages          = "2833",
      doi            = "10.1140/epjc/s10052-014-2833-3",
      eprint         = "1309.0372",
      archivePrefix  = "arXiv",
      primaryClass   = "physics.ins-det",
      reportNumber   = "LCD-NOTE-2013-008",
      SLACcitation   = "%%CITATION = ARXIV:1309.0372;%%"
}

@article{Brondolin:2019awm,
      author         = "Brondolin, E. and Gaede, F. and Hynds, D. and
                        Leogrande, E. and Petri{\v{c}}, M. and Sailer, A.
                        and Simoniello, R.",
      title          = "{Conformal Tracking for all-silicon trackers at future
                        electron-positron colliders}",
      journal        = {submitted to Nucl. Instrum. Meth. A},
      year           = "2019",
      eprint         = "1908.00256",
      archivePrefix  = "arXiv",
      primaryClass   = "physics.ins-det",
      SLACcitation   = "%%CITATION = ARXIV:1908.00256;%%"
}

@article{Frank:2015ivo,
      author         = "Frank, M. and Gaede, F. and Nikiforou, N. and Petric, M.
                        and Sailer, A.",
      title          = "{DDG4 A Simulation Framework based on the DD4hep Detector
                        Description Toolkit}",
      booktitle      = "{Proceedings, 21st International Conference on Computing
                        in High Energy and Nuclear Physics (CHEP 2015): Okinawa,
                        Japan, April 13-17, 2015}",
      journal        = "J. Phys. Conf. Ser.",
      volume         = "664",
      year           = "2015",
      number         = "7",
      pages          = "072017",
      doi            = "10.1088/1742-6596/664/7/072017",
      SLACcitation   = "%%CITATION = 00462,664,072017;%%"
}

@article{Sailer:2017rnh,
      author         = "Sailer, A. and Frank, M. and Gaede, F. and Hynds, D. and
                        Lu, S. and Nikiforou, N. and Petric, M. and Simoniello, R.
                        and Voutsinas, G.",
      title          = "{DD4Hep based event reconstruction}",
      booktitle      = "{Proceedings, 22nd International Conference on Computing
                        in High Energy and Nuclear Physics (CHEP2016): San
                        Francisco, CA, October 14-16, 2016}",
      collaboration  = "CLICdp and ILD Collaborations",
      journal        = "J. Phys. Conf. Ser.",
      volume         = "898",
      year           = "2017",
      number         = "4",
      pages          = "042017",
      doi            = "10.1088/1742-6596/898/4/042017",
      reportNumber   = "CLICdp-Conf-2017-002",
      SLACcitation   = "%%CITATION = 00462,898,042017;%%"
}

@article{Agostinelli:2002hh,
      author         = "Agostinelli, S. and others",
      title          = "{GEANT4: A Simulation toolkit}",
      collaboration  = "GEANT4 Collaboration",
      journal        = "Nucl. Instrum. Meth.",
      volume         = "A506",
      year           = "2003",
      pages          = "250-303",
      doi            = "10.1016/S0168-9002(03)01368-8",
      reportNumber   = "SLAC-PUB-9350, FERMILAB-PUB-03-339",
      SLACcitation   = "%%CITATION = NUIMA,A506,250;%%"
}

@article{Cacciari:2011ma,
      author         = {Cacciari, Matteo and Salam, Gavin P. and Soyez, Gregory},
      title          = {FastJet User Manual},
      journal        = {Eur. Phys. J.},
      volume         = {C72},
      year           = {2012},
      pages          = {1896},
      doi            = {10.1140/epjc/s10052-012-1896-2},
      eprint         = {1111.6097},
      archivePrefix  = {arXiv},
      primaryClass   = {hep-ph},
      reportNumber   = {CERN-PH-TH-2011-297},
      SLACcitation   = {%%CITATION = ARXIV:1111.6097;%%}
}

@collection{cdrvol2,
      editor         = {Linssen, Lucie and Miyamoto, Akiya and Stanitzki, Marcel
                        and Weerts, Harry},
      title          = {CLIC Conceptual Design Report: Physics and Detectors at CLIC},
      location       = {CERN},
      year           = {2012},
      eprint         = {1202.5940},
      archivePrefix  = {arXiv},
      primaryClass   = {physics.ins-det},
      series         = {CERN-2012-003},
      SLACcitation   = {%%CITATION = ARXIV:1202.5940;%%}
}

@article{Schulte:382453,
      author        = "Schulte, Daniel",
      title         = "{Beam-Beam Simulations with GUINEA-PIG}",
      month         = "Mar",
      year          = "1999",
      reportNumber  = "CERN-PS-99-014-LP",
      url           = "https://cds.cern.ch/record/382453",
}

@article{Robson:2018zje,
      author         = "Robson, Aidan and Roloff, Philipp",
      title          = "{Updated CLIC luminosity staging baseline and Higgs
                        coupling prospects}",
      year           = "2018",
      eprint         = "1812.01644",
      archivePrefix  = "arXiv",
      primaryClass   = "hep-ex",
      reportNumber   = "CLICdp-Note-2018-002",
      SLACcitation   = "%%CITATION = ARXIV:1812.01644;%%"
}

@article{Marshall:2015rfa,
      author         = {Marshall, J. S. and Thomson, M. A.},
      title          = {The Pandora Software Development Kit for Pattern
                        Recognition},
      journal        = {Eur. Phys. J.},
      volume         = {C75},
      year           = {2015},
      number         = {9},
      pages          = {439},
      doi            = {10.1140/epjc/s10052-015-3659-3},
      eprint         = {1506.05348},
      archivePrefix  = {arXiv},
      primaryClass   = {physics.data-an},
      SLACcitation   = {%%CITATION = ARXIV:1506.05348;%%}
}

@article{Marshall:2012ry,
      author         = {Marshall, J. S. and M{\"u}nnich, A. and Thomson, M. A.},
      journal        = {Nucl. Instrum. Meth.},
      title = {Performance of Particle Flow Calorimetry at CLIC},
      volume         = {A700},
      year           = {2013},
      pages          = {153-162},
      doi            = {10.1016/j.nima.2012.10.038},
      eprint         = {1209.4039},
      archivePrefix  = {arXiv},
      primaryClass   = {physics.ins-det},
      reportNumber   = {CU-HEP-12-12, AIDA-PUB-2013-002},
      SLACcitation   = {%%CITATION = ARXIV:1209.4039;%%}
}

@article{Moult:2016cvt,
      author         = {Moult, Ian and Necib, Lina and Thaler, Jesse},
      title          = {New Angles on Energy Correlation Functions},
      journal        = {JHEP},
      volume         = {12},
      year           = {2016},
      pages          = {153},
      doi            = {10.1007/JHEP12(2016)153},
      eprint         = {1609.07483},
      archivePrefix  = {arXiv},
      primaryClass   = {hep-ph},
      reportNumber   = {MIT-CTP-4825},
      SLACcitation   = {%%CITATION = ARXIV:1609.07483;%%}
}

\newpage
\appendix
\section{Distribution of events with positive electron beam polarisation}
\label{sec:polp80_plots}

\begin{figure}[htbp!]
\centering
\begin{minipage}[l]{0.45\textwidth}
\includegraphics[width=1.0\textwidth]{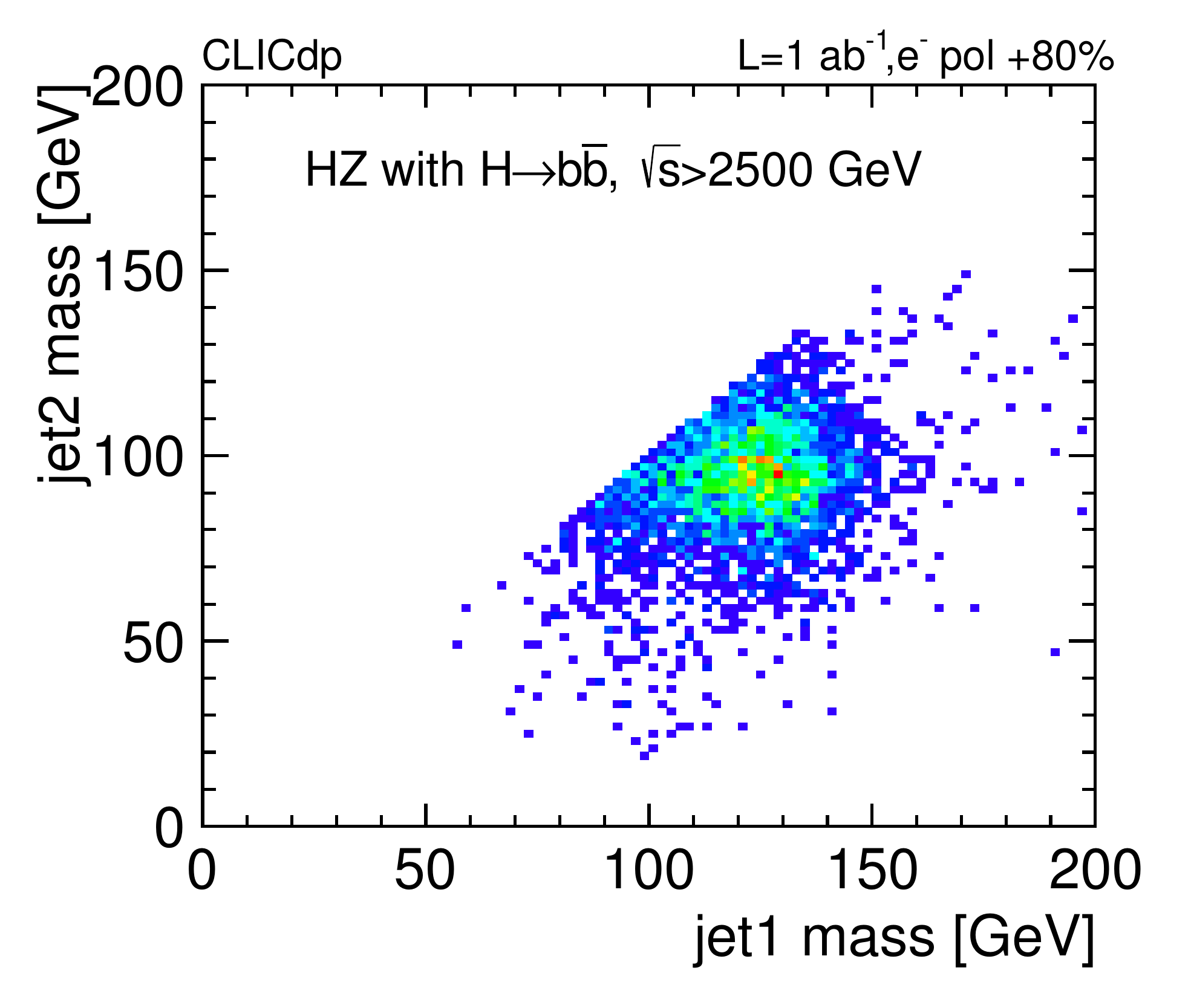}
\end{minipage}
\begin{minipage}[r]{0.45\textwidth}
\includegraphics[width=1.0\textwidth]{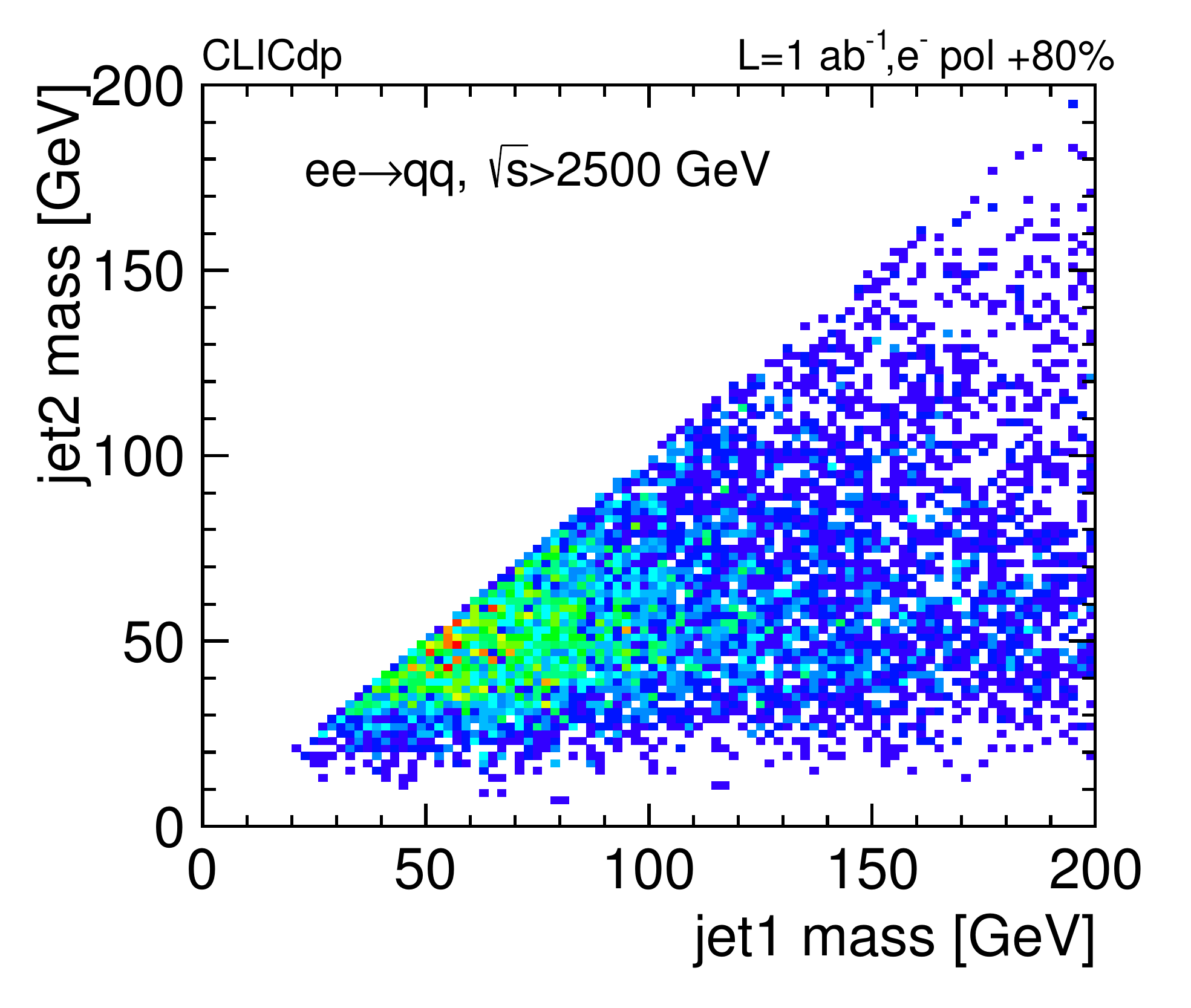}
\end{minipage}
\begin{minipage}[l]{0.45\textwidth}
\includegraphics[width=1.0\textwidth]{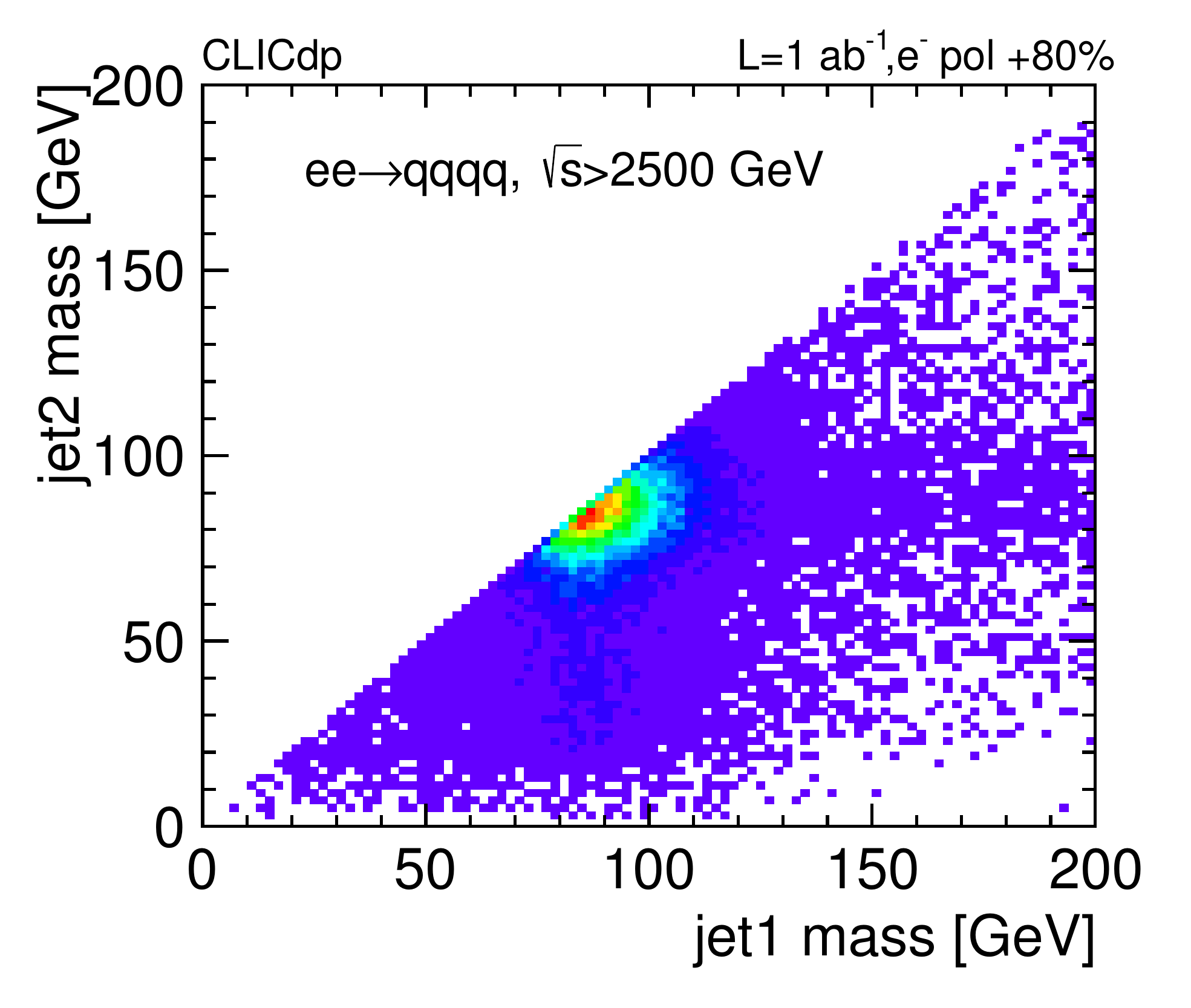}
\end{minipage}
\begin{minipage}[r]{0.45\textwidth}
\includegraphics[width=1.0\textwidth]{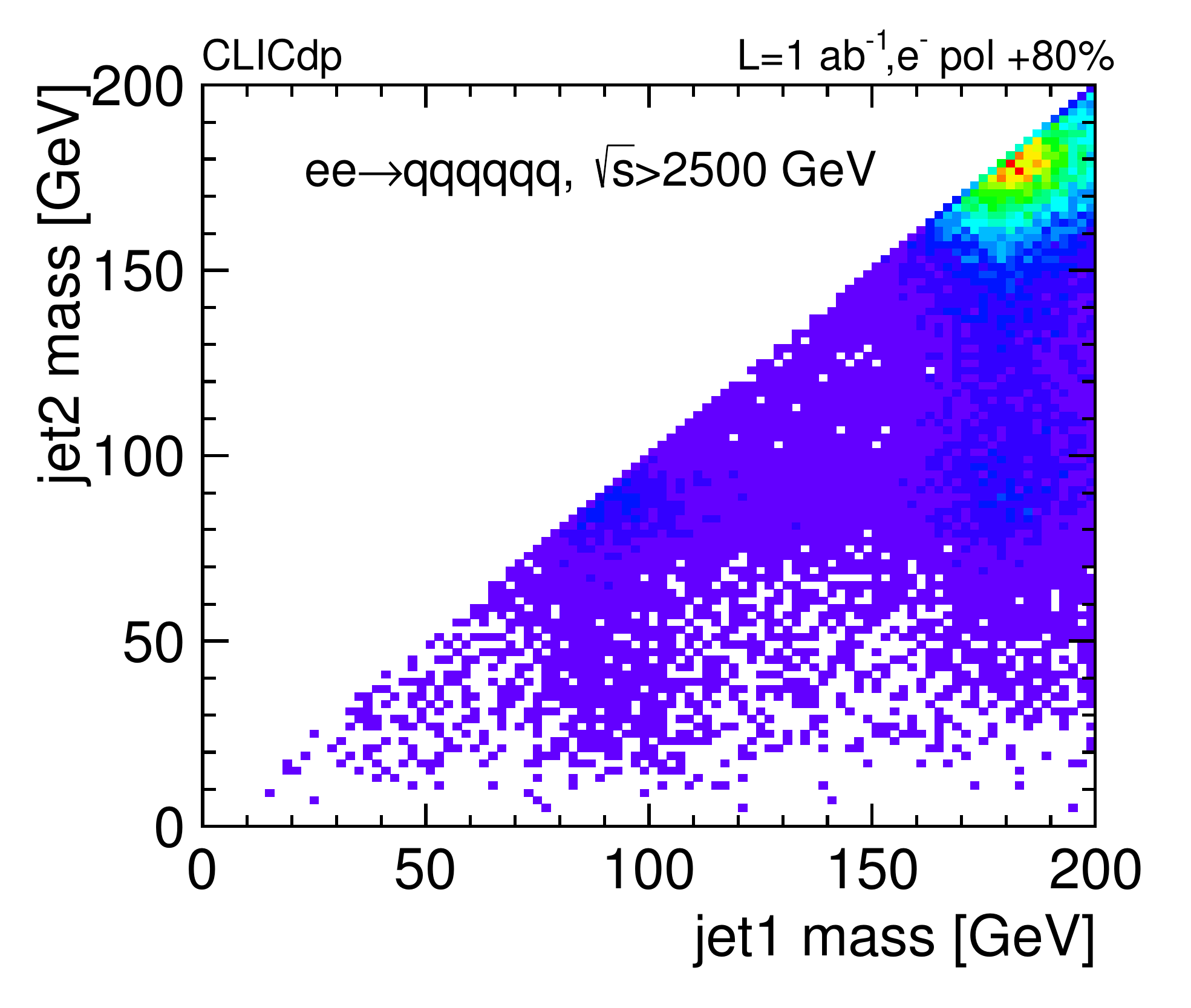}
\end{minipage}
\caption{The two-dimensional mass plane of the leading vs the second leading mass for \zhsm (top left), $\epem\rightarrow\qqbar$ (top right), $\epem\rightarrow\qqqq$ (bottom left), and $\epem\rightarrow \text{qqqqqq}$ events (bottom right) with positive electron beam polarisation.}
\label{Fig:2D_massplane_polp80}
\end{figure}

\begin{figure}[htbp!]
\centering
\begin{minipage}[l]{0.48\textwidth}
\includegraphics[width=1.0\textwidth]{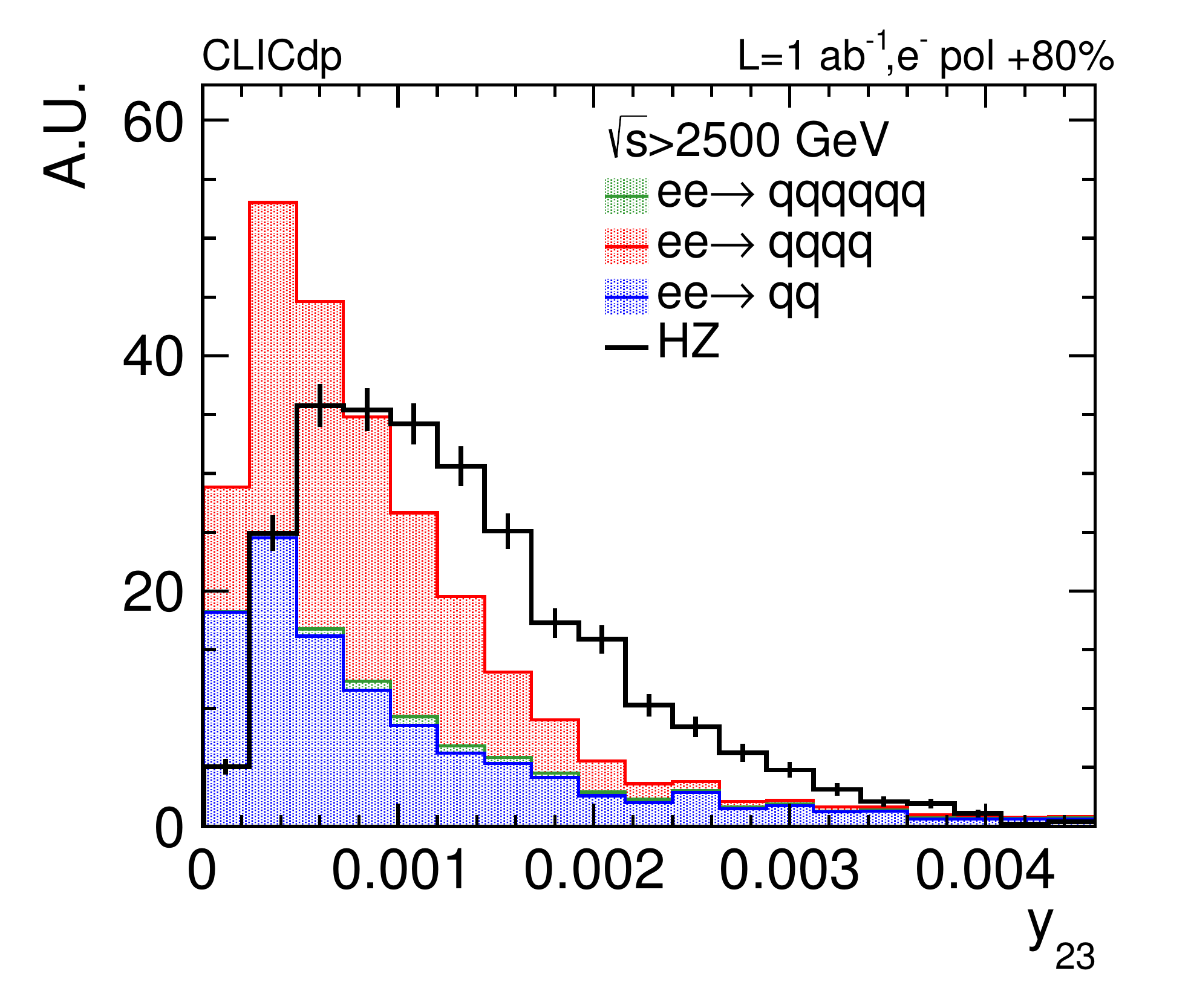}
\end{minipage}
\begin{minipage}[r]{0.48\textwidth}
\includegraphics[width=1.0\textwidth]{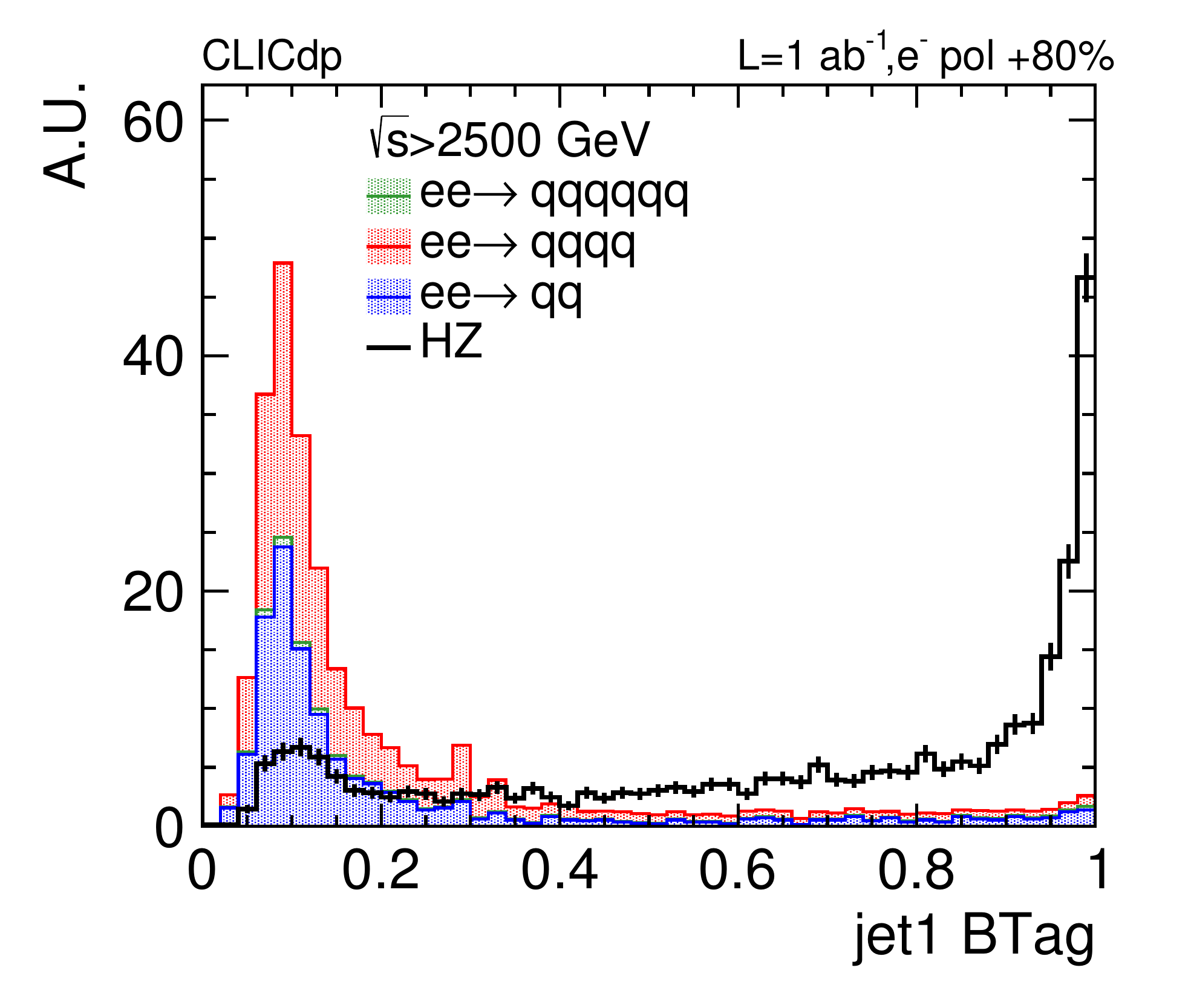}
\end{minipage}
\caption{The three-jet resolution parameter $y_{23}$ (left) and the b-tag distribution of the leading jet (right) for signal and background events with positive electron beam polarisation after the preselection on jet masses.}
\label{Fig:discrimination1_polp80}
\end{figure}

\begin{figure}[htbp!]
\centering
\begin{minipage}[l]{0.48\textwidth}
\includegraphics[width=1.0\textwidth]{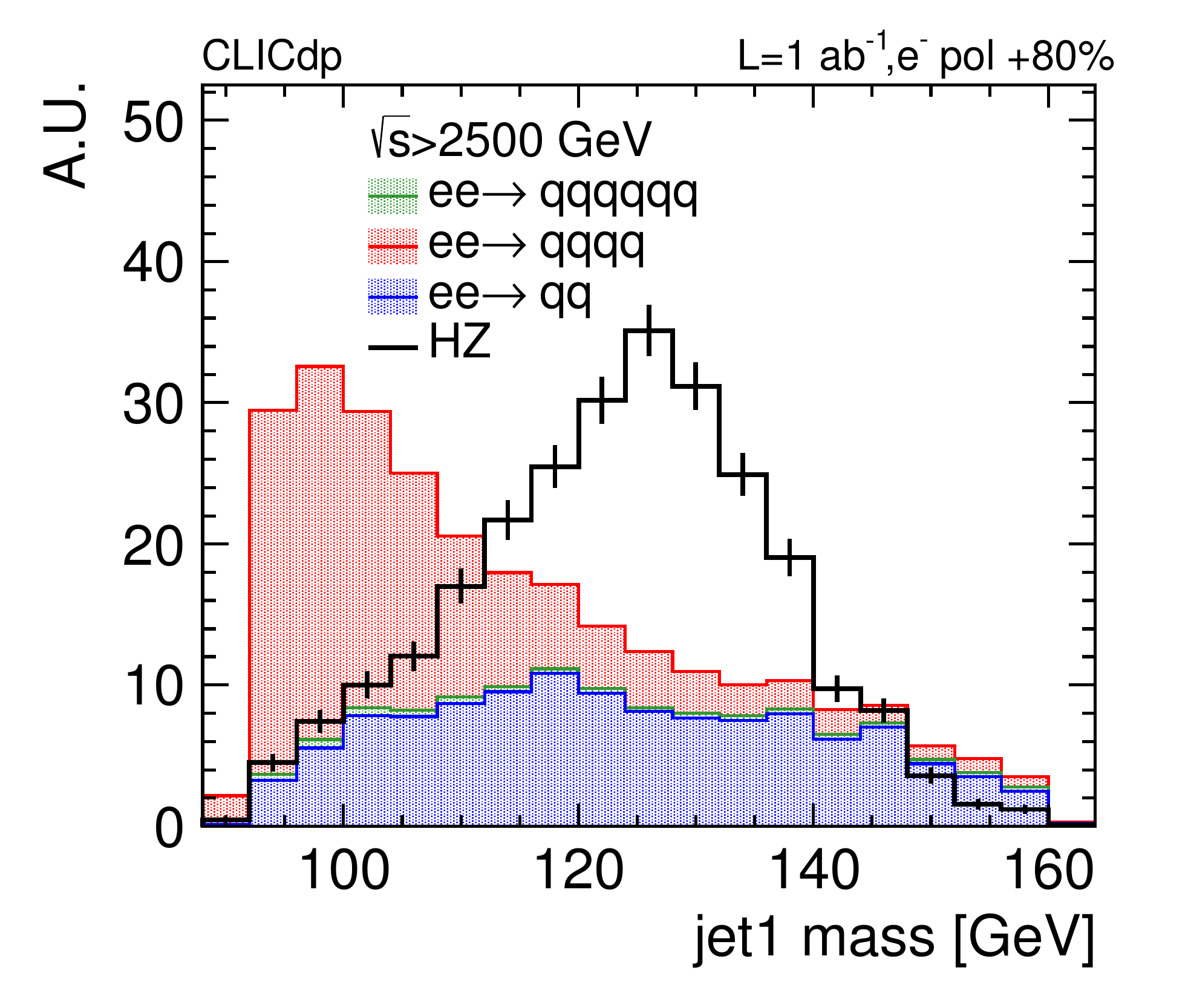}
\end{minipage}
\begin{minipage}[r]{0.48\textwidth}
\includegraphics[width=1.0\textwidth]{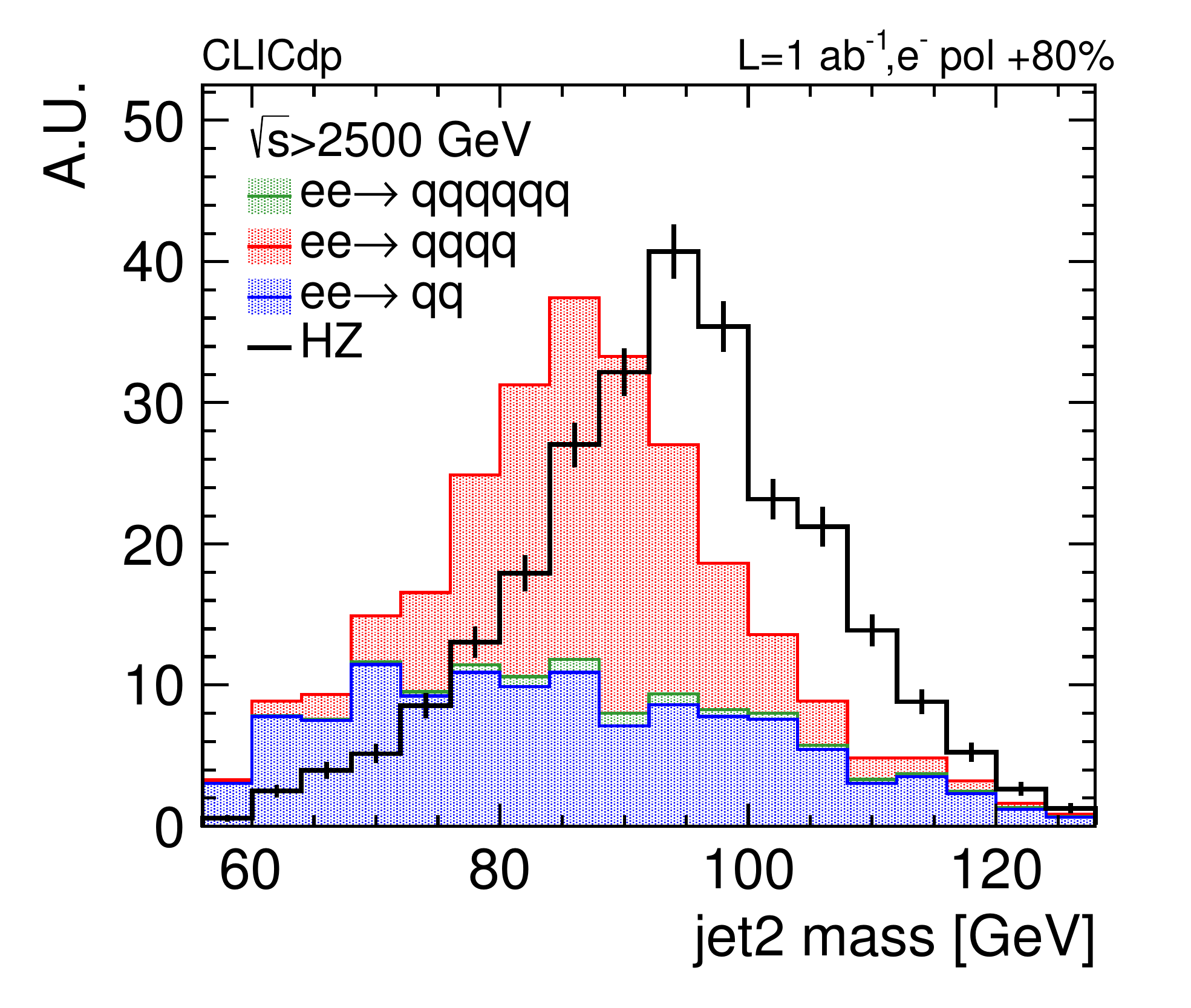}
\end{minipage}
\begin{minipage}[l]{0.48\textwidth}
\includegraphics[width=1.0\textwidth]{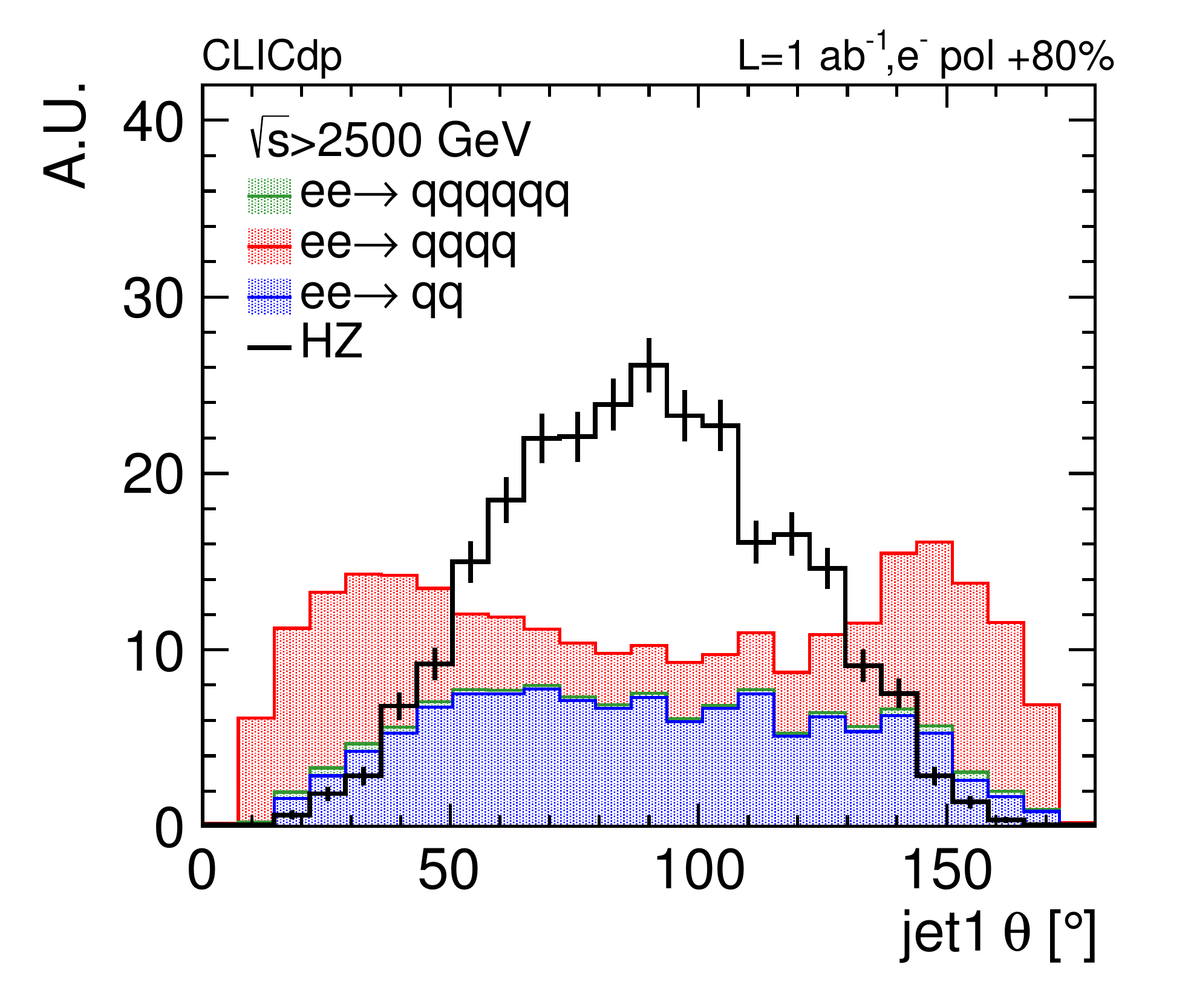}
\end{minipage}
\begin{minipage}[r]{0.48\textwidth}
\includegraphics[width=1.0\textwidth]{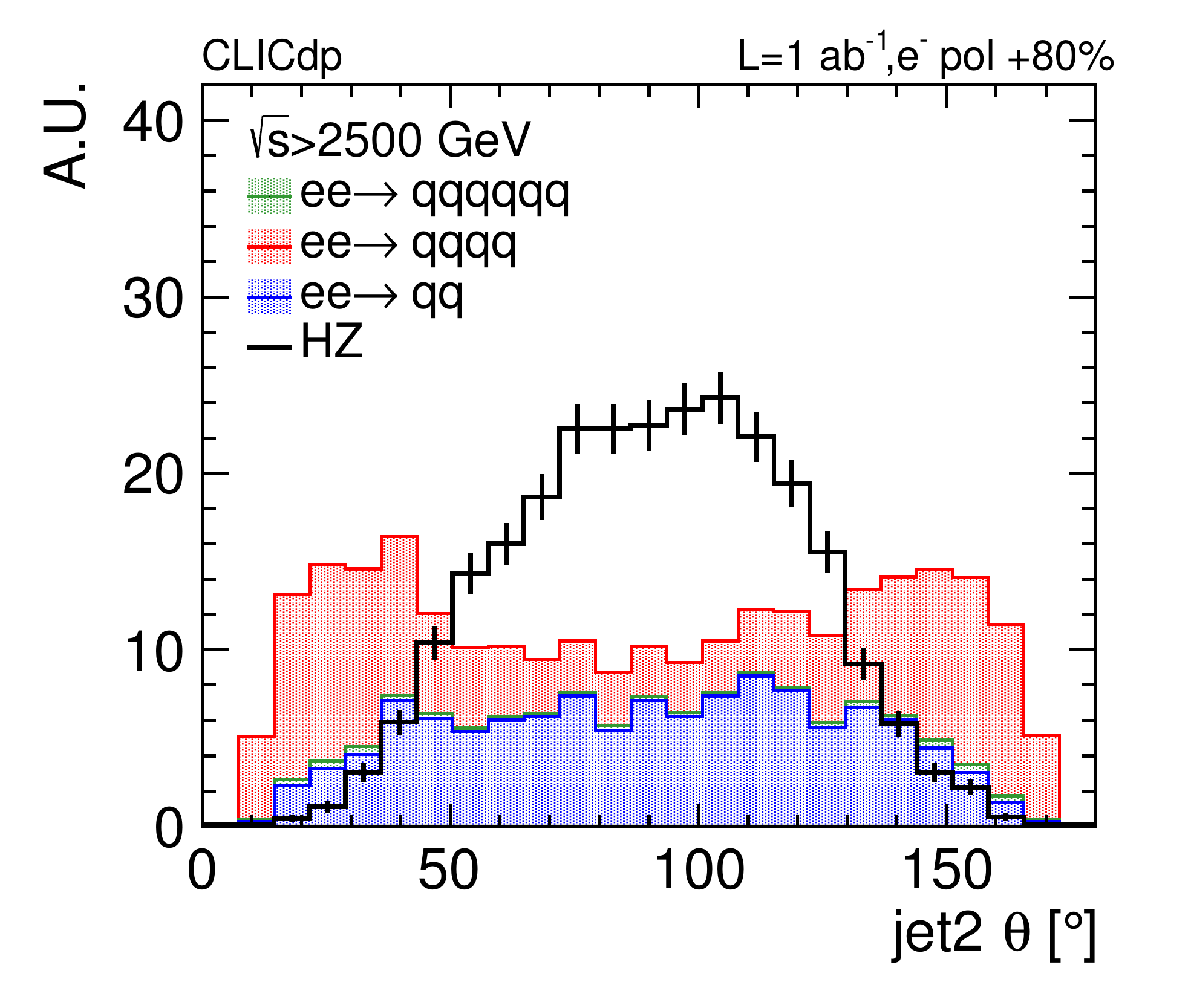}
\end{minipage}
\caption{The jet mass (top) and polar angle distributions (bottom) for the leading (left) and subleading (right) jet for signal and background events with positive electron beam polarisation after the preselection on jet masses.}
\label{Fig:discrimination2_polp80}
\end{figure}

\begin{figure}[htbp!]
\centering
\begin{minipage}[l]{0.48\textwidth}
\includegraphics[width=1.0\textwidth]{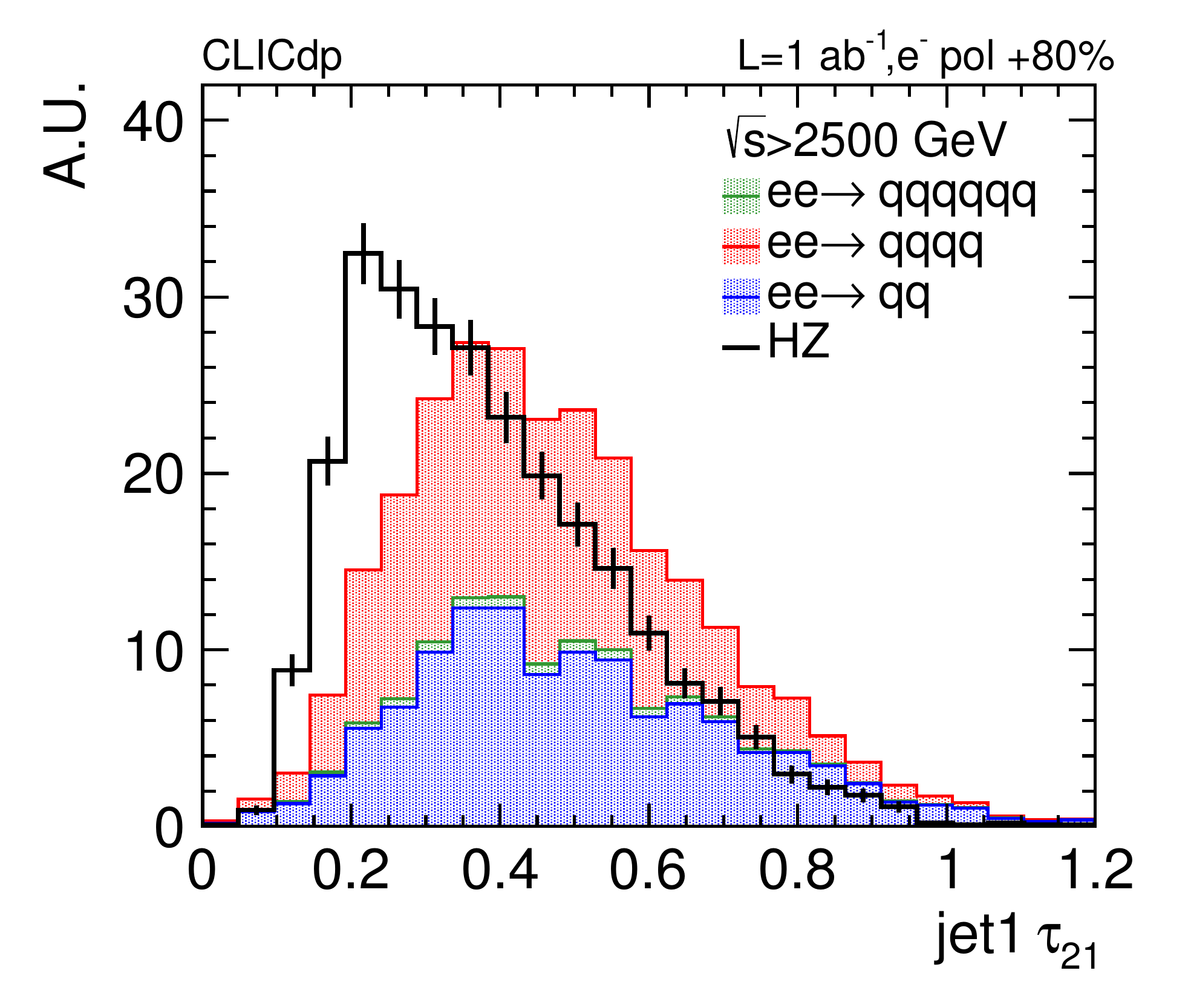}
\end{minipage}
\begin{minipage}[r]{0.48\textwidth}
\includegraphics[width=1.0\textwidth]{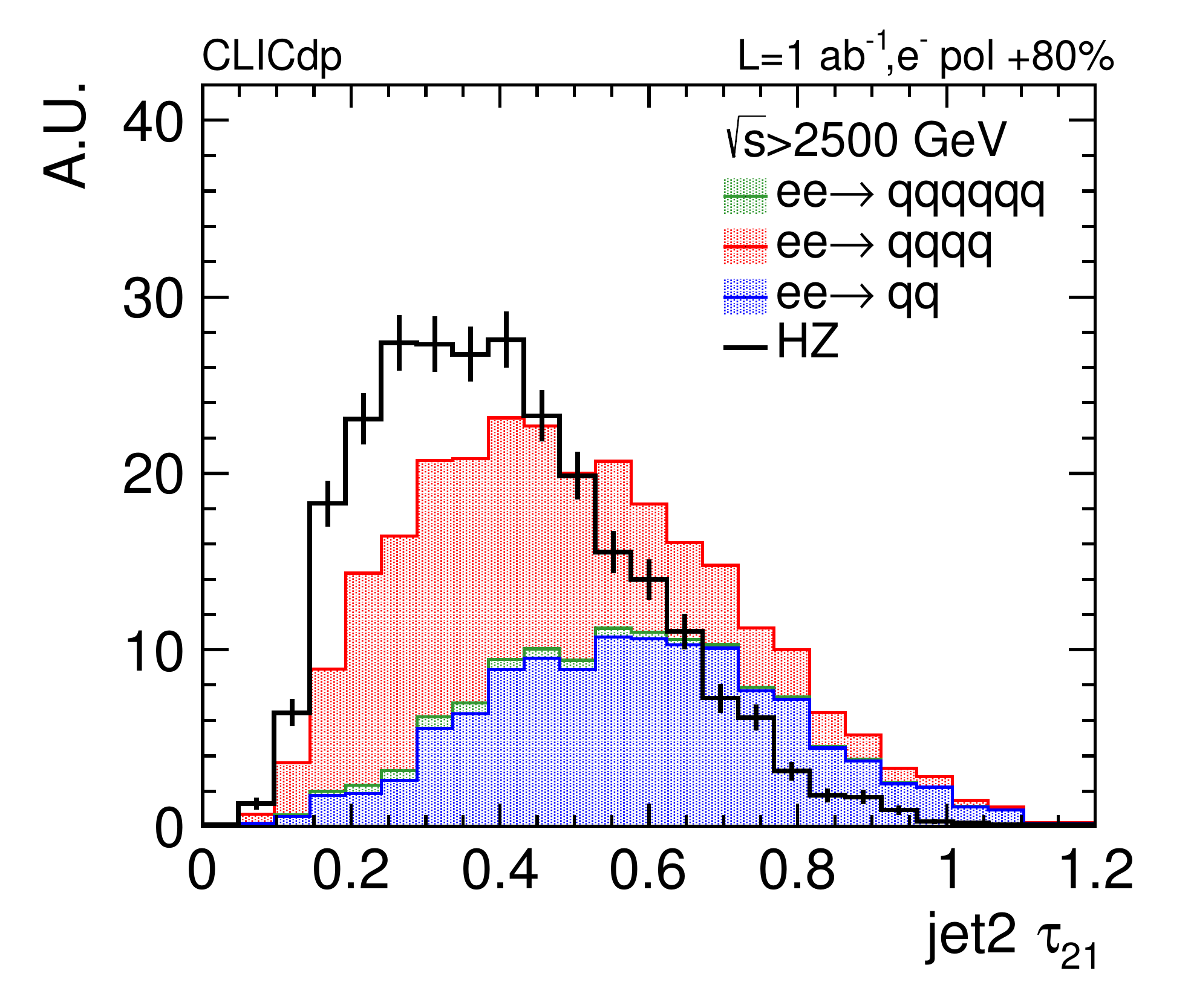}
\end{minipage}
\caption{The N-subjettiness ratio $\tau_{21}$ for the leading (left) and sub-leading jet (right) for signal and background events with positive electron beam polarisation after the preselection on jet masses.}
\label{Fig:discrimination3_polp80}
\end{figure}

\begin{figure}[htbp!]
\centering
\begin{minipage}[l]{0.48\textwidth}
\includegraphics[width=1.0\textwidth]{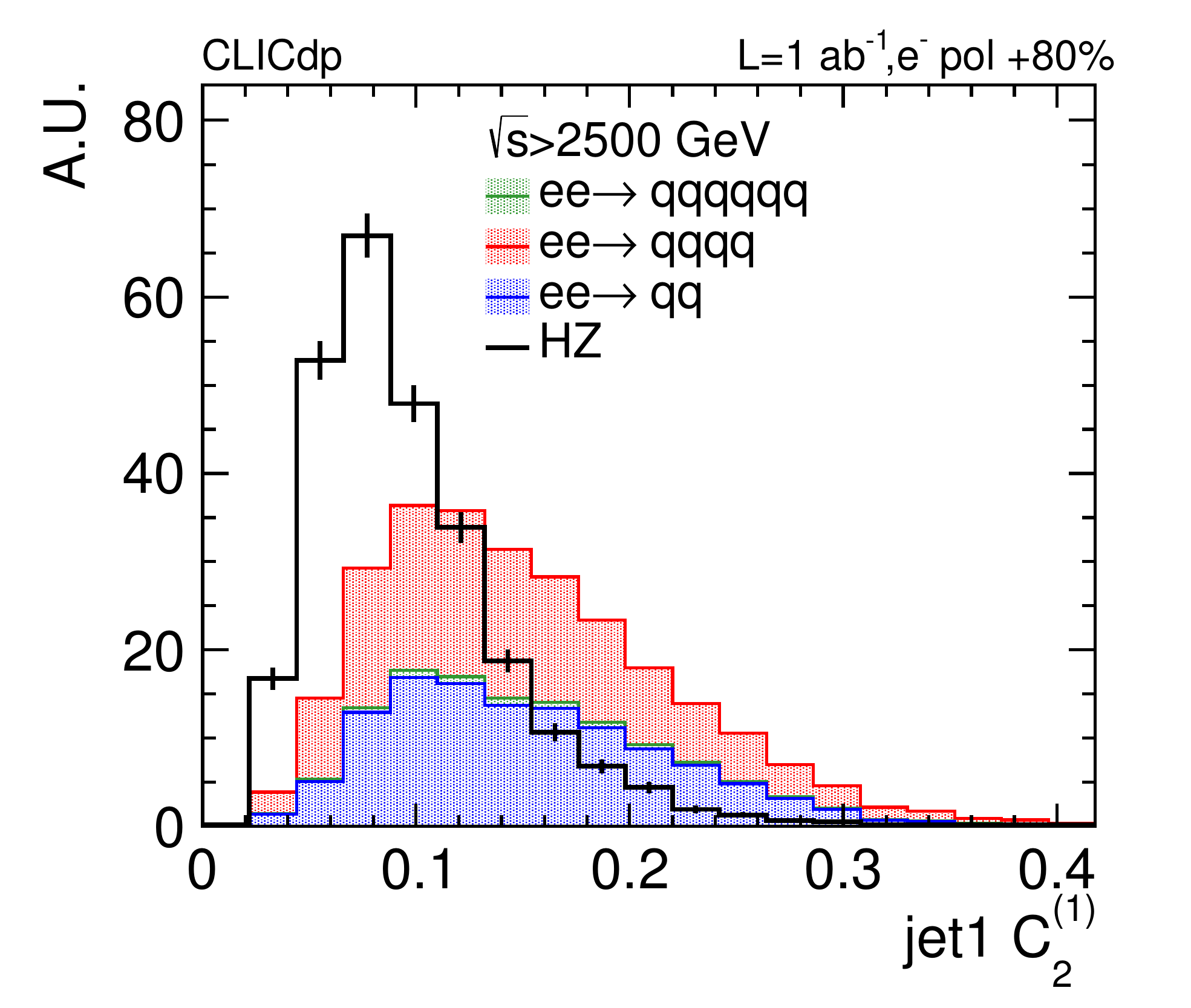}
\end{minipage}
\begin{minipage}[r]{0.48\textwidth}
\includegraphics[width=1.0\textwidth]{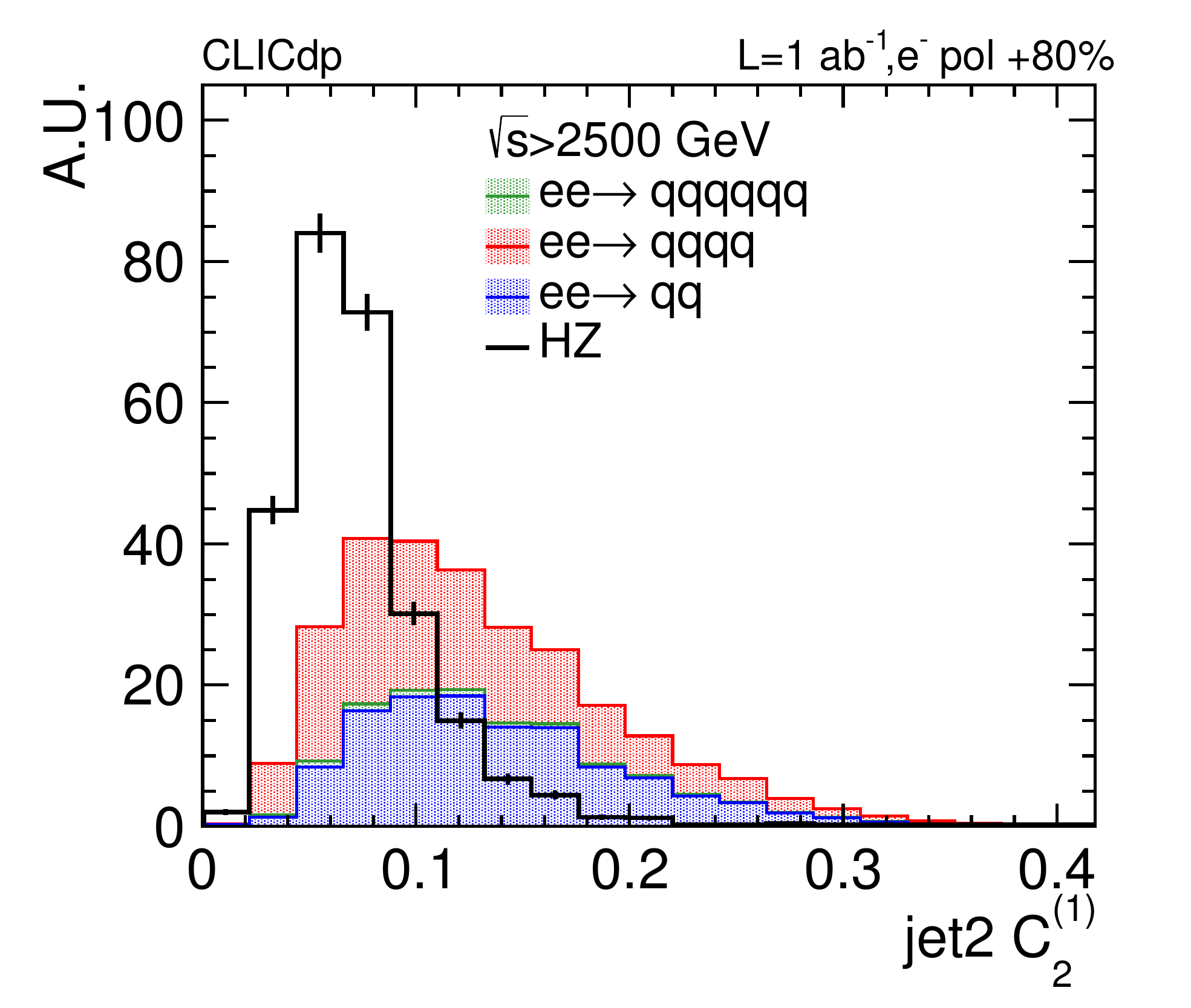}
\end{minipage}
\begin{minipage}[l]{0.48\textwidth}
\includegraphics[width=1.0\textwidth]{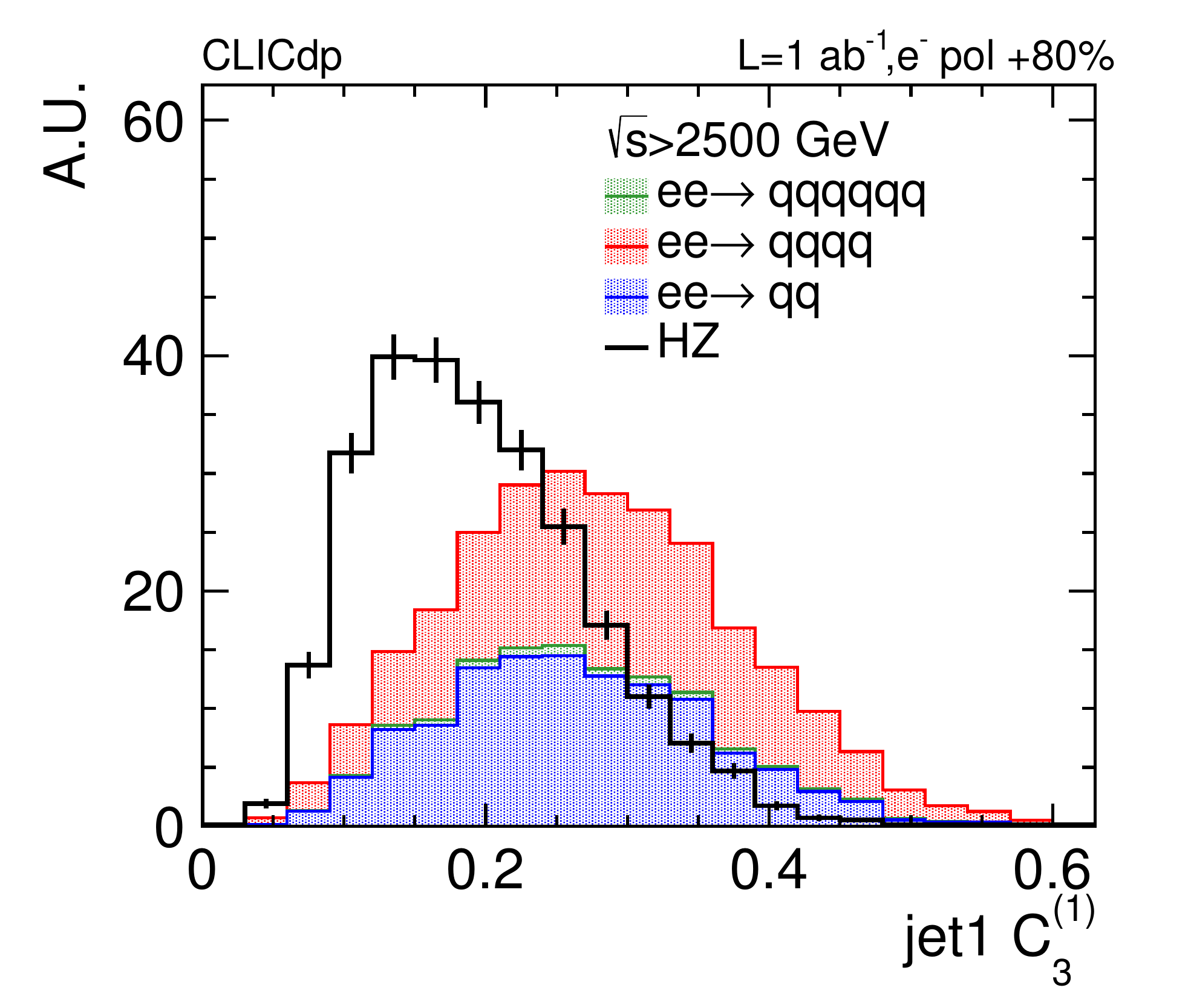}
\end{minipage}
\begin{minipage}[r]{0.48\textwidth}
\includegraphics[width=1.0\textwidth]{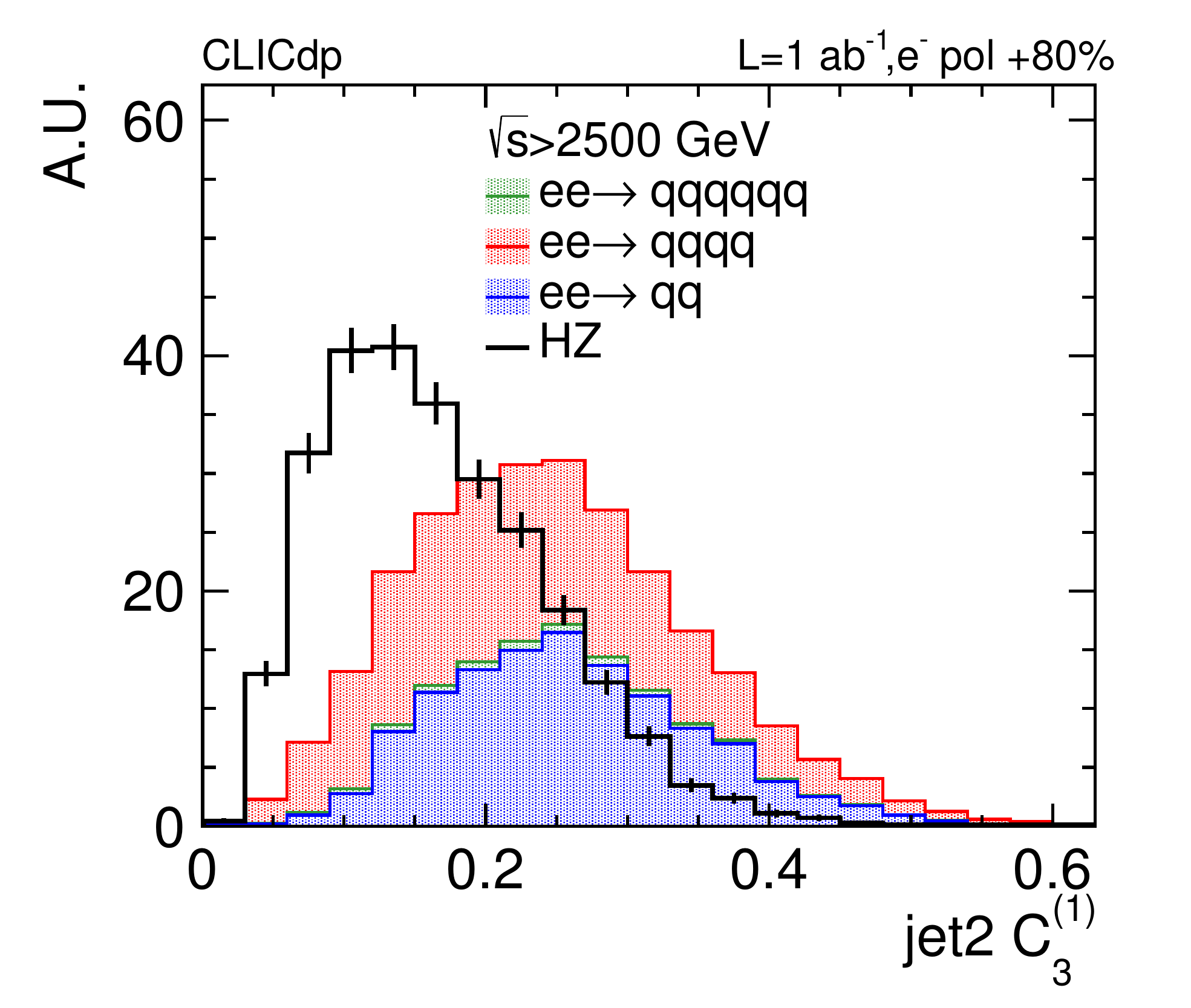}
\end{minipage}
\caption{The jet energy correlation ratio distribution $C_{2}^{(1)}$ (top) and $C_{3}^{(1)}$ (bottom) for the leading (left) and subleading (right) jet for signal and background events with positive electron beam polarisation after the preselection on jet masses.}
\label{Fig:discrimination4_polp80}
\end{figure}

\begin{figure}[htbp!]
\centering
\begin{minipage}[l]{0.48\textwidth}
\includegraphics[width=1.0\textwidth]{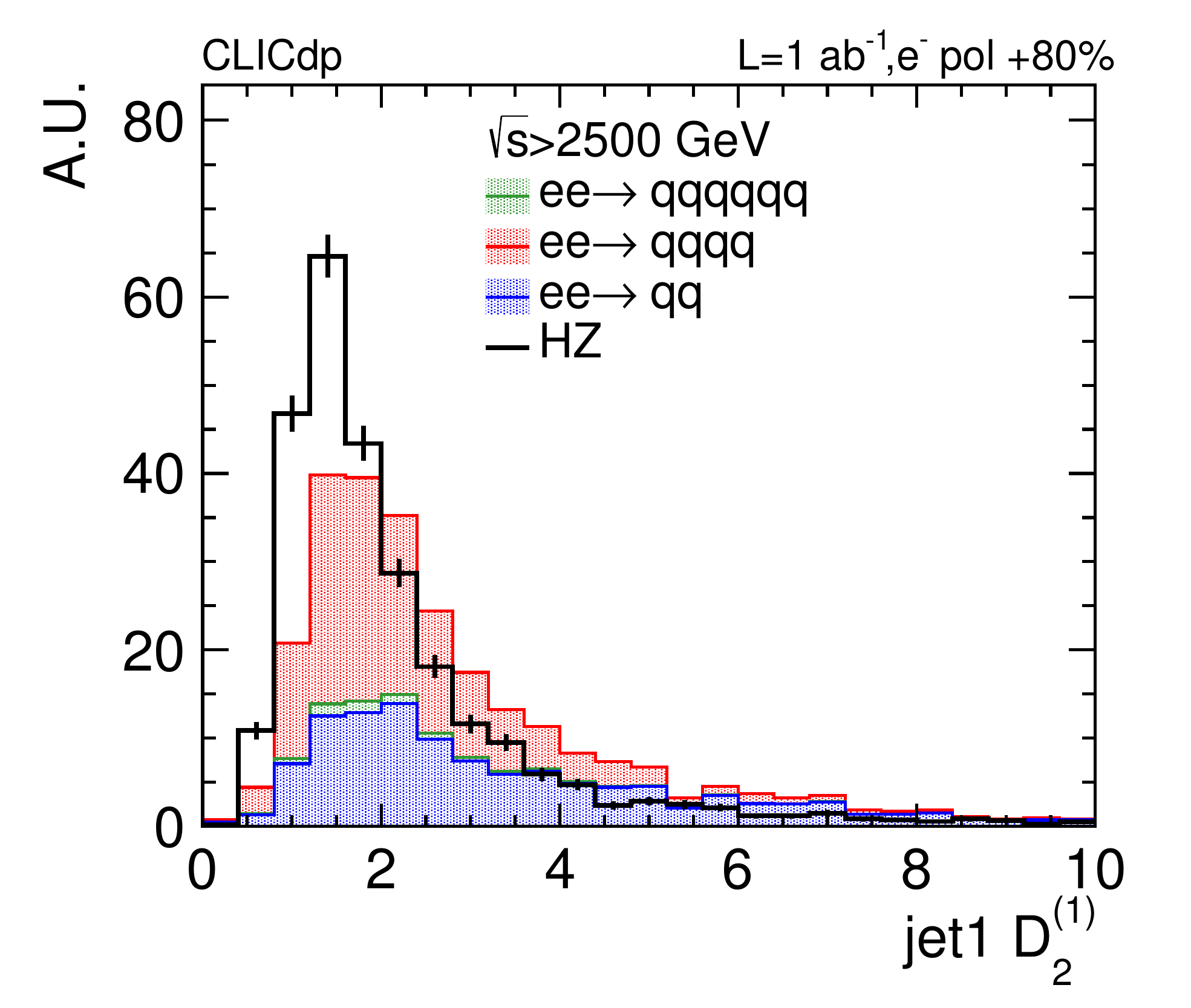}
\end{minipage}
\begin{minipage}[r]{0.48\textwidth}
\includegraphics[width=1.0\textwidth]{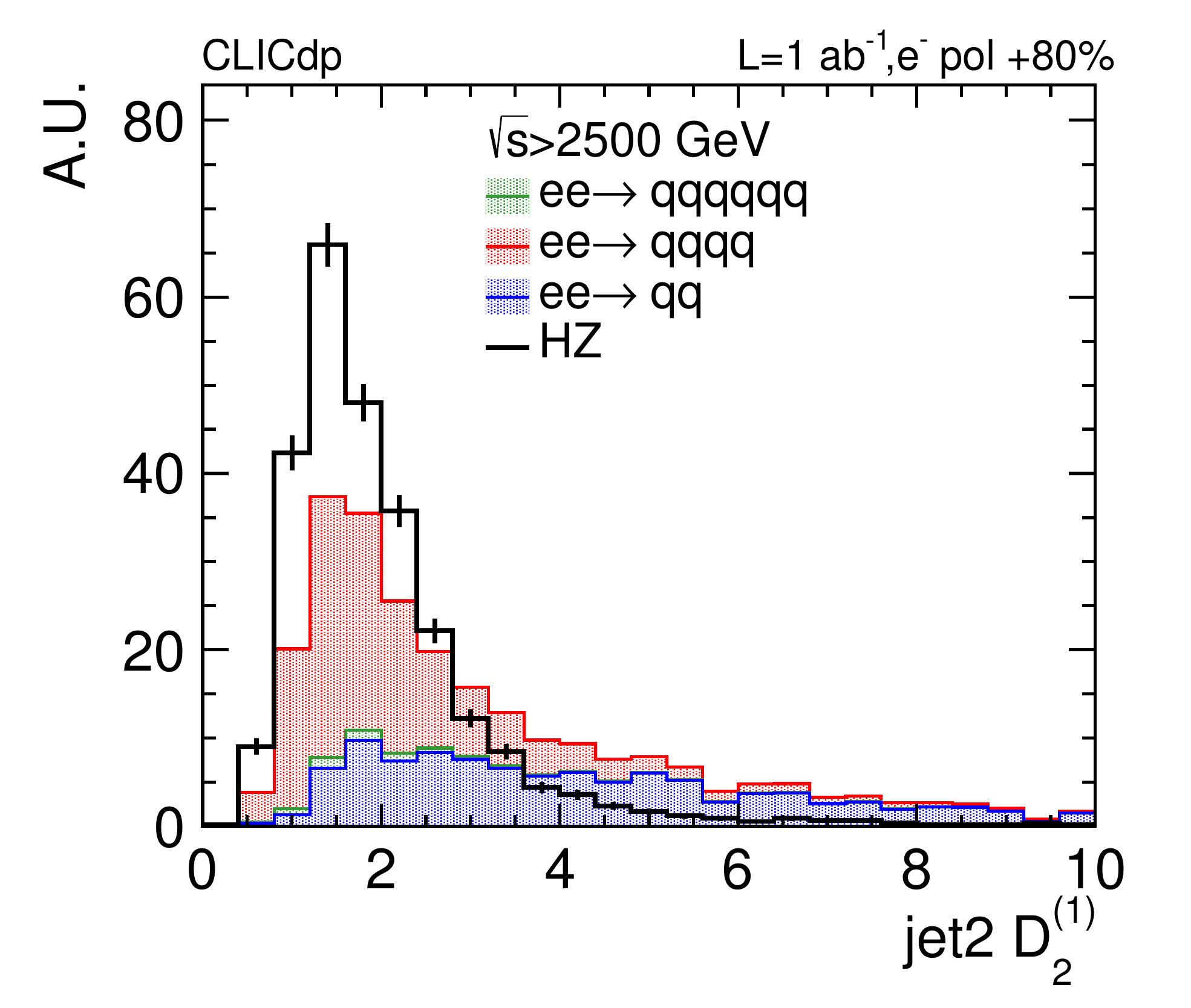}
\end{minipage}
\caption{The jet energy correlation ratio $D_{2}^{(1)}$ for the leading (left) and sub-leading jet (right) for signal and background events with positive electron beam polarisation after the preselection on jet masses.}
\label{Fig:discrimination5_polp80}
\end{figure}

\begin{figure}[htbp!]
\centering
\begin{minipage}[l]{0.31\textwidth}
\includegraphics[width=1.0\textwidth]{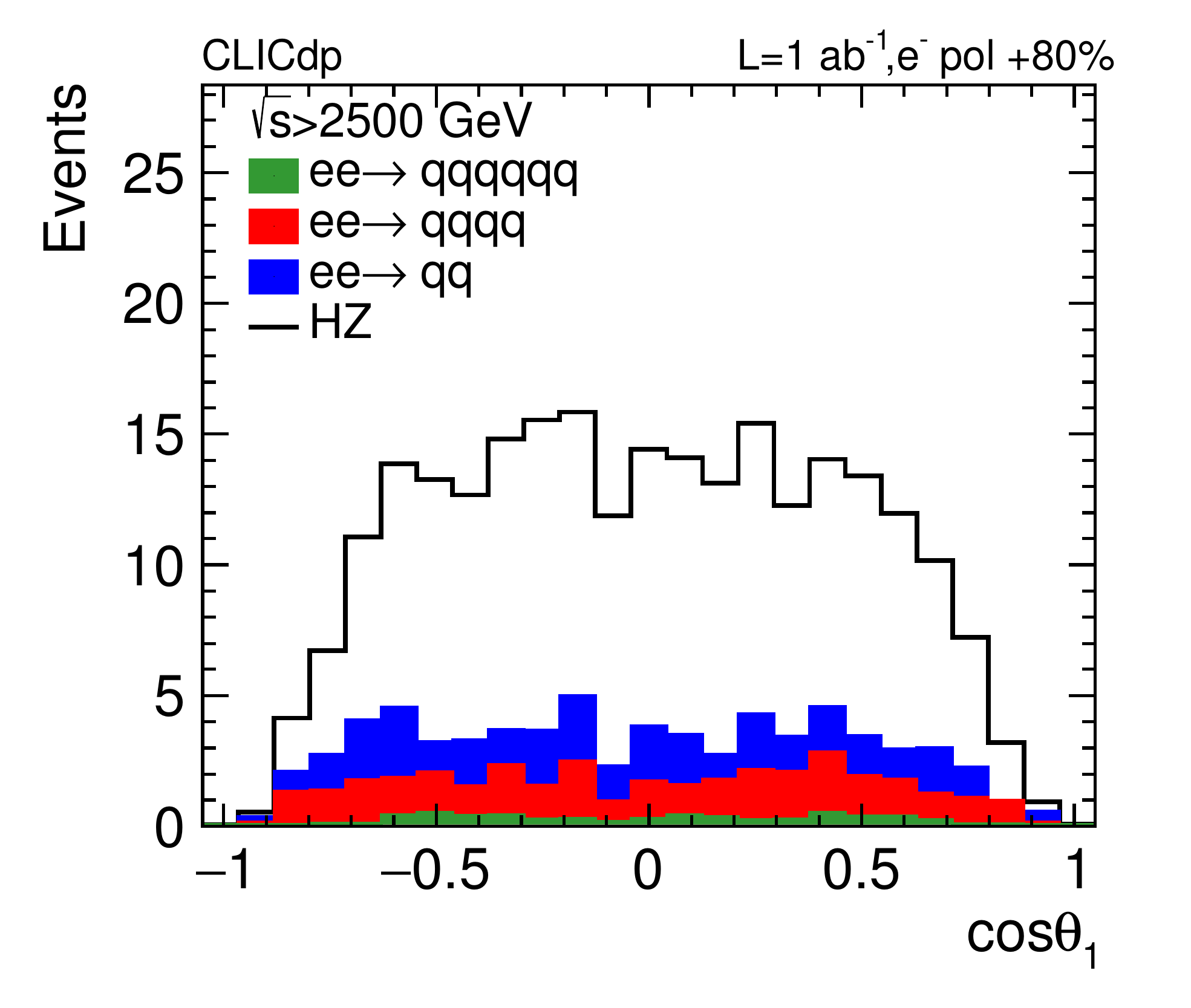}
\end{minipage}
\begin{minipage}[c]{0.31\textwidth}
\includegraphics[width=1.0\textwidth]{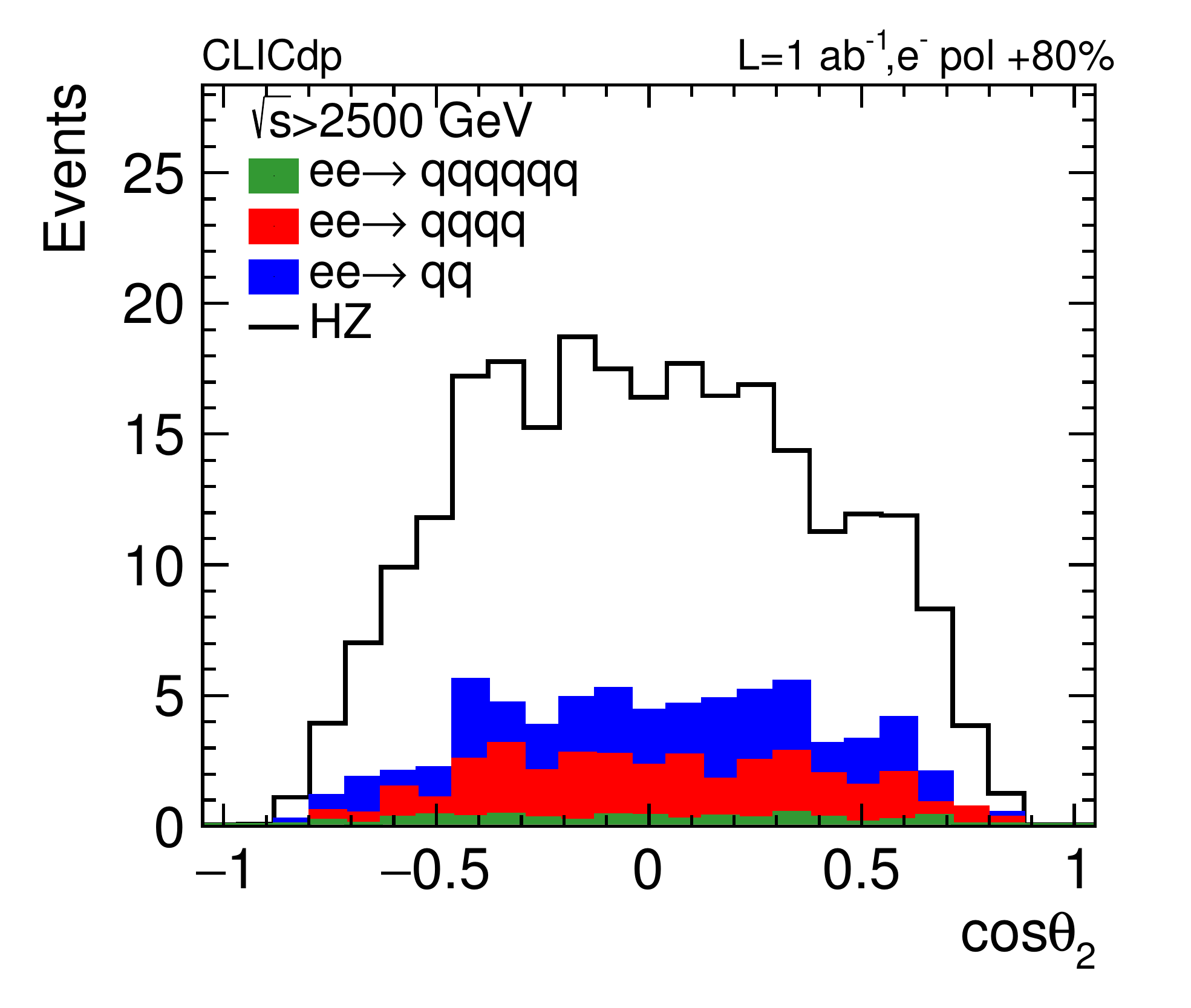}
\end{minipage}
\begin{minipage}[l]{0.31\textwidth}
\includegraphics[width=1.0\textwidth]{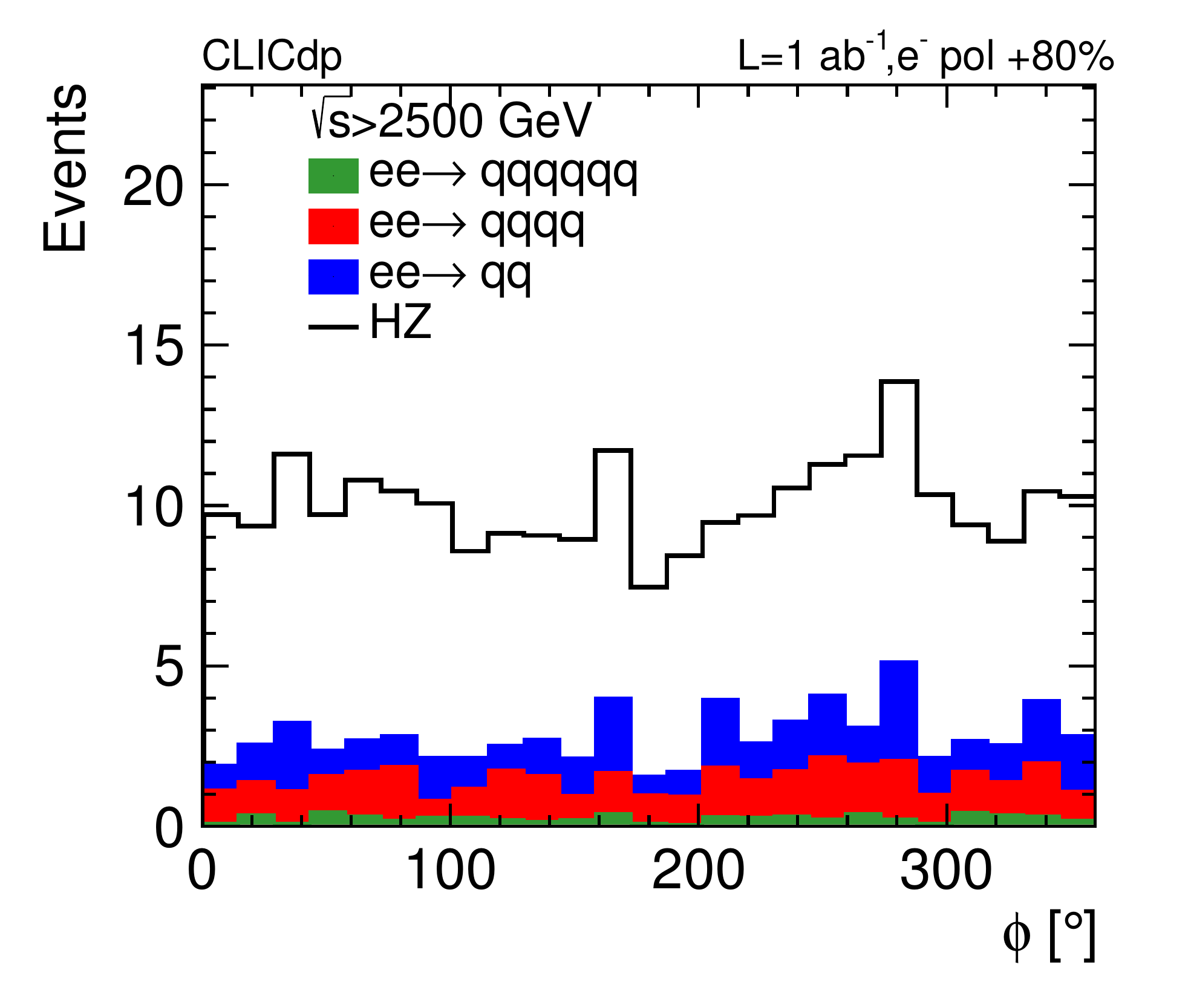}
\end{minipage}
\begin{minipage}[l]{0.31\textwidth}
\includegraphics[width=1.0\textwidth]{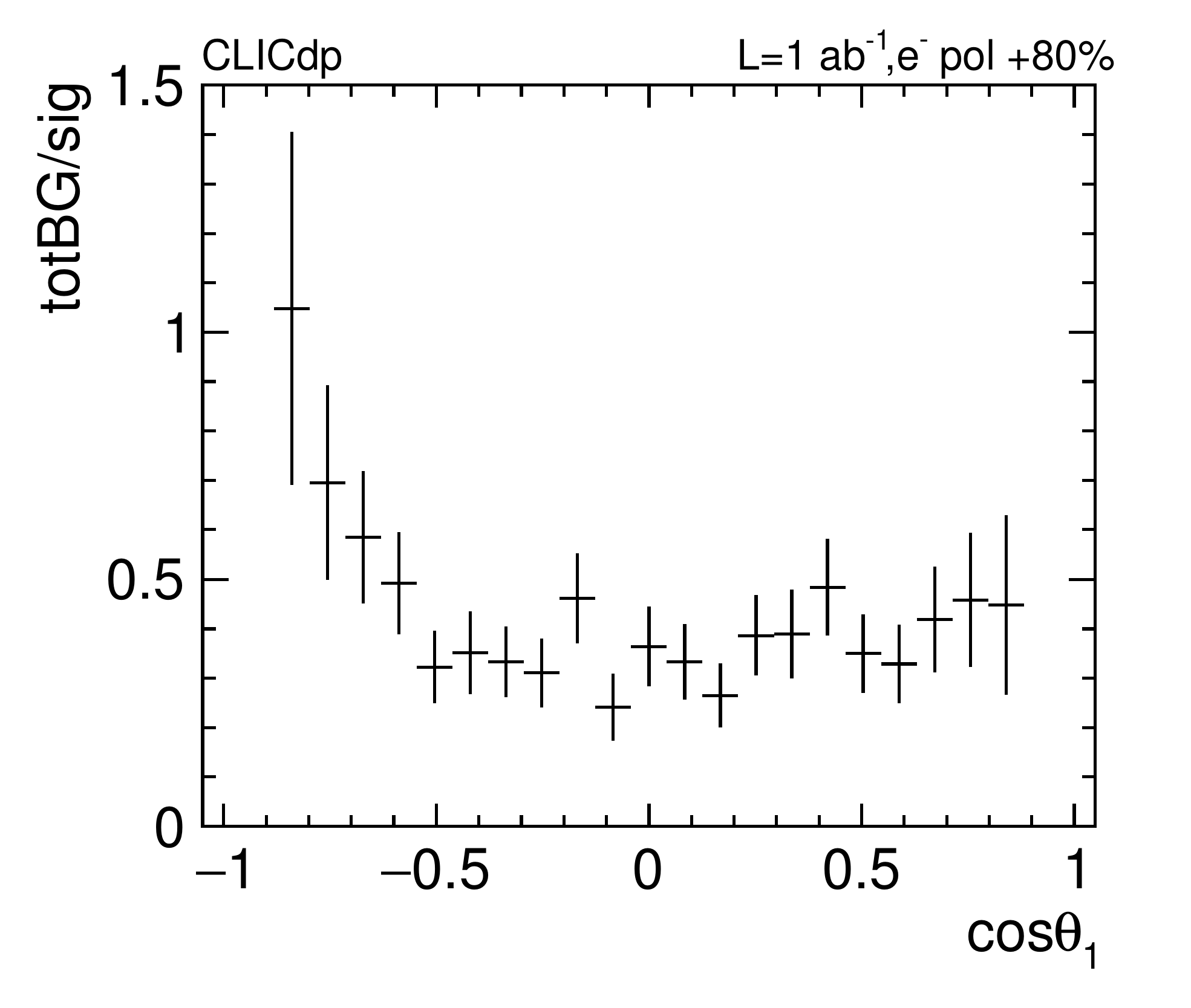}
\end{minipage}
\begin{minipage}[c]{0.31\textwidth}
\includegraphics[width=1.0\textwidth]{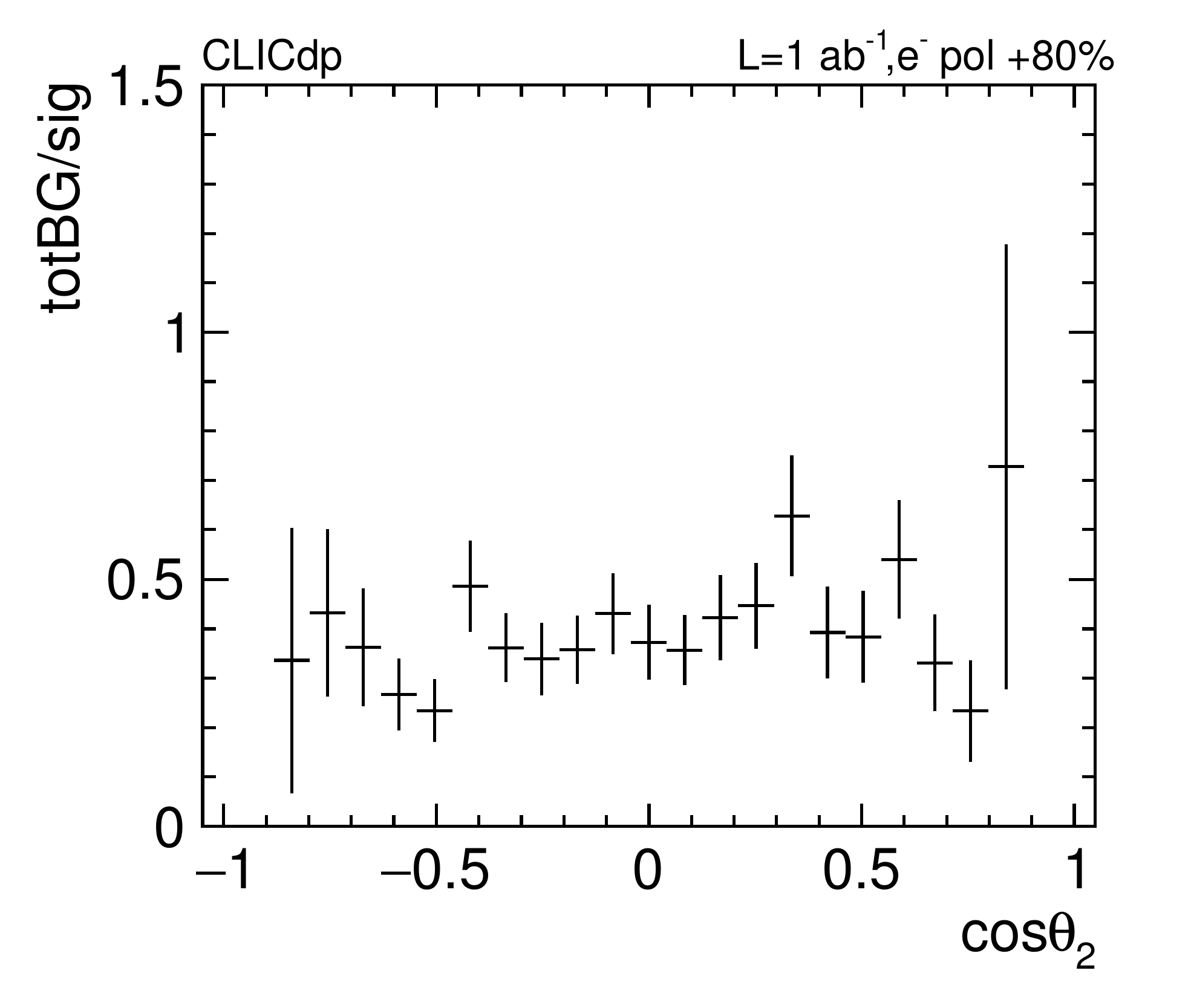}
\end{minipage}
\begin{minipage}[l]{0.31\textwidth}
\includegraphics[width=1.0\textwidth]{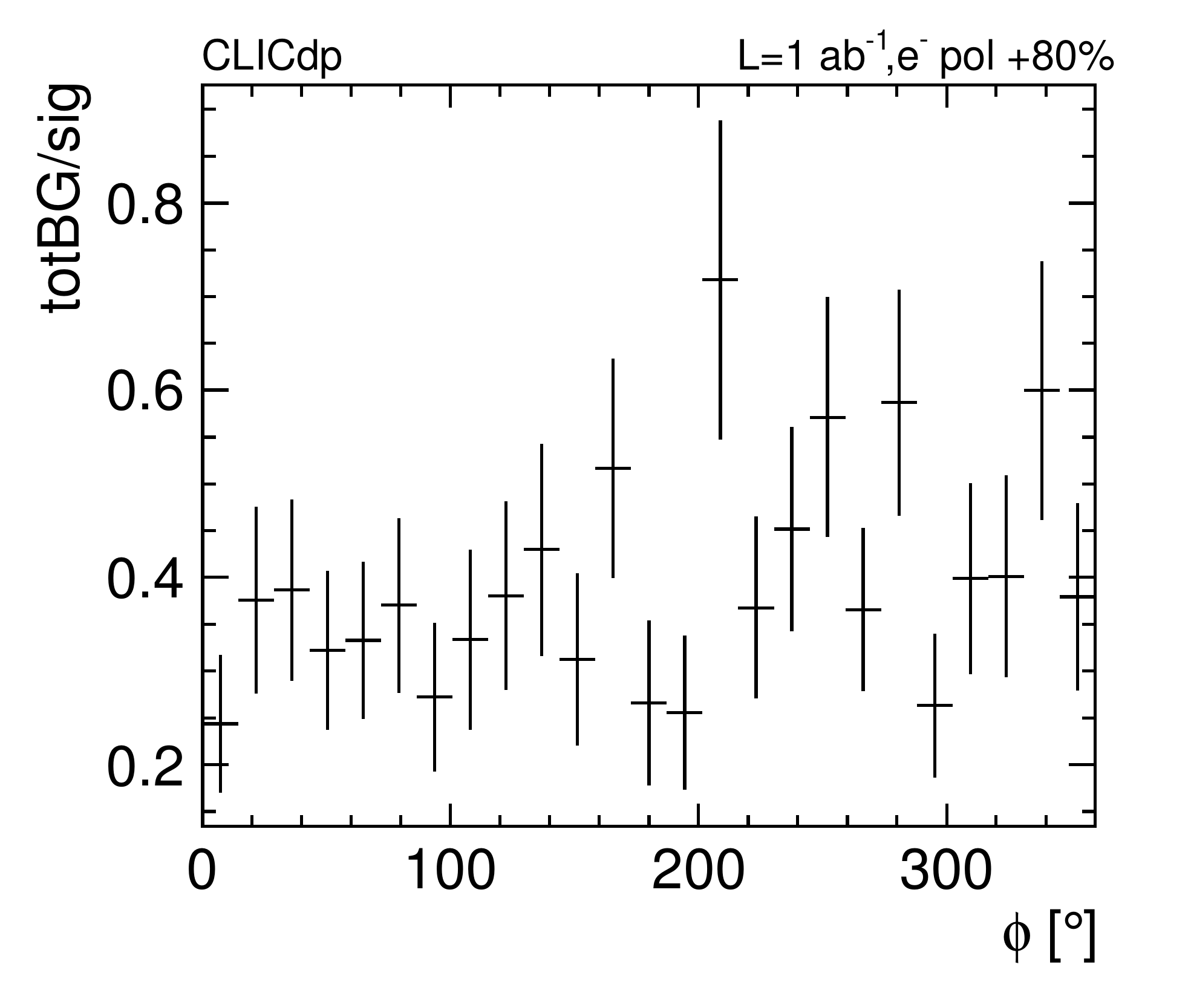}
\end{minipage}
\caption{The three reconstructed angular distributions of $\cos\theta_{1}$ (left), $\cos\theta_{2}$ (centre), and $\phi$ (right) for signal and background events with positive electron beam polarisation. The lower plots show the ratios between background and signal events after the preselection on jet masses.}
\label{Fig:angles_signal_background_polp80}
\end{figure}

\end{document}